\documentclass[12pt]{article} 
\usepackage[sectionbib]{natbib}
\usepackage{array,epsfig,fancyheadings,rotating,multirow}

\usepackage{enumerate}
\usepackage{float}
\usepackage{tcolorbox}
\usepackage{xcolor}
\usepackage{comment}

\usepackage[]{hyperref}  
\hypersetup{hidelinks}
\usepackage{sectsty, secdot}
\sectionfont{\fontsize{12}{14pt plus.8pt minus .6pt}\selectfont}
\renewcommand{\theequation}{\thesection\arabic{equation}}
\subsectionfont{\fontsize{12}{14pt plus.8pt minus .6pt}\selectfont}

\usepackage{amsmath}
\usepackage{amssymb}
\usepackage{amsfonts}
\usepackage{amsthm}

\newtheorem{theorem}{Theorem}
\newtheorem{lemma}{Lemma}

\newtheorem{proposition}{Proposition}
\theoremstyle{definition}

\newtheorem{remark}{Remark}
\DeclareMathOperator\arctanh{arctanh}

\newcommand{\bs}{\boldsymbol{s}}

\makeatletter
\def\varphiall{\@ifnextchar[{\varphiall@i}{\varphiall@i[]}}
\def\varphiall@i[#1]{\@ifnextchar[{\varphiall@ii{#1}}{\varphiall@ii{#1}[#1]}}
\def\varphiall@ii#1[#2]{\mathcal{GH}_{\delta,\beta_{#1},\gamma_{#1},a_{#2}}}
\def\hatvarphiall{\@ifnextchar[{\hatvarphiall@i}{\hatvarphiall@i[]}}
\def\hatvarphiall@i[#1]{\@ifnextchar[{\hatvarphiall@ii{#1}}{\hatvarphiall@ii{#1}[#1]}}
\def\hatvarphiall@ii#1[#2]{\widehat{\mathcal{GH}}_{\delta,\beta_{#1},\gamma_{#1},a_{#2}}}
\makeatother

\def\R{\mathbb{R}}

\begin{document}
\thispagestyle{empty}


\renewcommand{\baselinestretch}{2}

\renewcommand{\thefootnote}{}
$\ $\par


\fontsize{12}{14pt plus.8pt minus .6pt}\selectfont \vspace{0.8pc}
\centerline{\large\bf Parsimonious Compactly Supported Covariance Models }
\vspace{2pt} 
\centerline{\large\bf in the Gauss Hypergeometric Class: Identifiability, }
\vspace{2pt} 
\centerline{\large\bf Reparameterizations, and Asymptotic Properties}
\vspace{.25cm}
 \author{Bevilacqua, Christian Caama\~no-Carrillo Tarik Faouzi and Xavier Emery}
\vspace{.4cm} 
\centerline{Moreno Bevilacqua, Christian Caama\~no-Carrillo, Tarik Faouzi and Xavier Emery} 
\vspace{.4cm} 
\centerline{\it Universidad Adolfo Ib\'a\~nez,\\
Universit\`a  Ca' Foscari,\\Universidad del B\'io-B\'io
}
\vspace{2pt}
\centerline{\it 
Universidad de Santiago de  Chile,\\Universidad de Chile}
 \vspace{.55cm} \fontsize{9}{11.5pt plus.8pt minus.6pt}\selectfont


\begin{quotation}
\noindent {\it Abstract:}
We study isotropic covariance functions in the Gauss hypergeometric ($\mathcal{GH}$) class, a flexible family that encompasses the  celebrated Mat\'ern ($\mathcal{MT}$)  and Generalized Wendland (${\cal GW}$)   models.
We refine validity conditions for positive definiteness in the $\mathcal{GH}$ family and clarify two distinct sources of non-identifiability: a structural one, due to an intrinsic symmetry in the shape parameters, and the classical fixed-domain one, where different parameter values may induce equivalent Gaussian measures and only microergodic combinations are consistently estimable.

To remove the structural ambiguity, we restrict the parameter space to obtain a unique representation of the correlation function.
We then introduce a parsimonious compactly supported subclass, denoted $\mathcal{H}$, selected by maximizing the integral range. It has simple validity conditions, yields sparse covariance matrices, and admits compact-support reparameterizations that recover the Mat\'ern model as a limit case.

Under fixed-domain asymptotics, we establish consistency and asymptotic normality for likelihood-based estimation of the associated microergodic parameter, including cases where nuisance covariance parameters are estimated over compact sets.
Simulation experiments and a climate-data application illustrate finite-sample behavior and show that $\mathcal{H}$ achieves performance comparable to ${\cal GW}$ while retaining the computational advantages of compact support relative to $\mathcal{MT}$, for both Gaussian and Tukey-$h$ random fields.
\vspace{9pt}

\par
\end{quotation}\par

\def\thefigure{\arabic{figure}}
\def\thetable{\arabic{table}}

\renewcommand{\theequation}{\thesection.\arabic{equation}}

\fontsize{12}{14pt plus.8pt minus .6pt}\selectfont

\section{Introduction}\label{sec1}

Gaussian random fields (RFs) are widely used in spatial statistics, machine
learning, numerical analysis, and computer experiments
\citep{Cressie:1993,stein-book,Banerjee-Carlin-Gelfand:2004,
williams2006gaussian,wendland2004scattered,cp2003}.
For isotropic Gaussian RFs, covariance specification is central to modeling,
prediction, and uncertainty quantification. The Mat\'ern ($\mathcal{MT}$)
family is a standard choice because its smoothness parameter controls
mean-square differentiability \citep{mateporcu}, while the compactly supported
generalized Wendland (${\cal GW}$) family provides continuous smoothness
control together with sparse covariance matrices and associated computational
gains \citep{gnei02,BFFP,da1989,da2006}. Connections between these two
families have been studied in \cite{BFFP,bevilacqua2022unifying}.

In this paper, we focus on the Gauss hypergeometric ($\mathcal{GH}$)
covariance class introduced by \citet{emery2022gauss}. This highly flexible
class includes the Mat\'ern and ${\cal GW}$ families as special cases and has
been recognized as a prominent extension of the Mat\'ern model
\citep{mateporcu}. Although \citet{emery2022gauss} derived convenient
sufficient conditions for positive definiteness, a sharper description of the
admissible parameter space and a systematic inferential analysis are still
lacking. The aim of this paper is to address these issues.

Our contributions are as follows. We first sharpen the validity conditions of
the $\mathcal{GH}$ class and distinguish its structural non-identifiability,
caused by symmetry in the shape parameters, from fixed-domain
non-identifiability, under which only suitable microergodic combinations are
consistently estimable. We remove the structural ambiguity by restricting the
parameter space to obtain a unique representation of the correlation function.

We then identify a parsimonious compactly supported subclass by maximizing the
integral range, a global and threshold-free measure of dependence proportional
to the spectral density at zero
\citep{Vanmarcke:1983,Lantuejoul:2002,HristopulosZukovic:2011}.

{
The maximum-integral-range criterion is not intended as a universal
statistical optimum. Rather, it provides an intrinsic rule for selecting a
canonical subclass once smoothness and compact support have been fixed. When
the support radius is constrained by a desired sparsity level, maximizing the
integral range favors stronger large-scale dependence without enlarging the
support and hence without increasing the number of nonzero covariance
entries. Other criteria, such as matching a prescribed effective range or
spectral mass over a particular frequency band, may be preferable when
different aspects of the dependence structure are of primary interest.}

We call the resulting subclass the hypergeometric ($\mathcal H$) family. It
has a particularly simple validity characterization that, unlike that of the
${\cal GW}$ family, does not become increasingly restrictive with the spatial
dimension. It also admits a beta-type scale-mixture representation and
generalizes Euclid's hat and the classical spherical models, both widely used
in applied geostatistics
\citep{Gneiting1999,Chiles:Delfiner:2012}. We further derive a compact-support
reparameterization under which $\mathcal H$ converges to the Mat\'ern model,
analogously to the ${\cal GW}$--Mat\'ern bridge established in
\cite{bevilacqua2022unifying}.

{
From a practical perspective, $\mathcal H$ is a native compactly supported
family. Relative to Mat\'ern, it provides exact sparsity while retaining
continuous control of local regularity. Relative to ${\cal GW}$, its simpler
validity structure becomes increasingly relevant as the dimension grows.
Combined with the maximum-integral-range construction, this allows the
proposed family to retain stronger large-scale dependence under a fixed
support, and therefore under a fixed sparsity budget.

The Mat\'ern-limit reparameterization provides a complementary computational
interpretation. For comparable limiting smoothness and scale, we show that the
$\mathcal H$ family can attain shorter valid compact supports than its
${\cal GW}$ counterpart, with this difference becoming more pronounced in
higher-dimensional input spaces. Thus, the maximum-integral-range
construction has a concrete practical implication: it can either preserve
stronger dependence for a prescribed sparsity pattern or allow greater
sparsity for a comparable dependence scale, depending on the parameterization
being considered.

These advantages come with trade-offs. The expression of $\mathcal H$ depends
explicitly on the dimension and, unlike ${\cal GW}$, no general closed-form
expression is available for arbitrary dimension when the smoothness parameter
is a nonnegative integer. Thus, $\mathcal H$ should not be viewed as uniformly
preferable to ${\cal GW}$.

The confluent hypergeometric class \citep{ma2022beyond} instead combines
Mat\'ern-type local regularity with polynomial long-range decay and remains
globally supported. Covariance tapering \citep{Furrer:2006} induces sparsity
by multiplying a base covariance by a compactly supported positive definite
taper. These are alternative modeling strategies; the comparisons below do
not establish uniform superiority of $\mathcal H$ over them.
}

Finally, for fixed and known smoothness, we establish consistency and
asymptotic normality of the likelihood-based estimator of the associated
microergodic parameter, including settings in which scale and shape nuisance
parameters are estimated over compact sets.

A simulation study investigates finite-sample behavior under increasing- and
fixed-domain scenarios. {Additional experiments assess robustness
under a different sampling design, compare $\mathcal H$ with ${\cal GW}$,
Mat\'ern, and tapered Mat\'ern covariances, and consider a non-Gaussian
Tukey-$h$ setting.} A real-data analysis of climate data further illustrates
that $\mathcal H$ achieves empirical performance comparable to ${\cal GW}$
while retaining the computational advantages associated with compact support
relative to Mat\'ern.

The $\mathcal H$ model is implemented in the \texttt{GeoModels} package
\citep{geomodels}. The remainder of the paper is organized as follows.
Section~\ref{Sec2} reviews the $\mathcal{MT}$ and ${\cal GW}$ models.
Section~\ref{sec3} introduces the $\mathcal{GH}$ model and discusses
identifiability. Section~\ref{4} provides validity conditions and derives the
$\mathcal H$ subclass. Section~\ref{5} presents closed-form expressions and
compact-support reparameterizations. Section~\ref{44} establishes asymptotic
properties of the microergodic estimator. All proofs are collected in the
Supplementary Material, which also contains the simulation study and the
real-data results.

\section{Background: the Mat{\'e}rn and Generalized Wendland  correlation models}\label{Sec2}

Let $\{Z(\bs):\bs\in D\}$ be a zero-mean stationary isotropic Gaussian RF
on a bounded set $D\subset\R^d$, with
\[
{\rm Cov}\!\left(Z(\bs),Z(\bs')\right)
=\sigma^2\phi(\|\bs'-\bs\|),
\]
where $\sigma^2>0$, $\phi(0)=1$, and $\phi\in\Phi_d$, the class of continuous
radial correlation functions that are positive semidefinite on $\R^d$.

\cite{Shoe38} characterized the class $\Phi_d$ as scale mixtures
of the characteristic functions of random vectors uniformly
distributed on the spherical shell of $\R^d$, with any finite and nonnegative measure, $F$:
$$ \phi(x)= \int_{0}^{\infty} \Omega_{d}(x r) F(\text{d} r), \qquad x \ge 0,$$
with $\Omega_{d}(x)= \Gamma(d/2)(2/x)^{d/2-1}J_{d/2-1}(x)$ and $J_{\nu}$ the Bessel function of the first kind of order $\nu$.
The class $\Phi_d$ is nested, with the inclusion relation $\Phi_{1} \supset \Phi_2 \supset \ldots \supset \Phi_{\infty}$ being strict, and where $\Phi_{\infty}:= \bigcap_{d \ge 1} \Phi_d$ is the class of continuous mappings $\phi$, the radial version of which is positive semidefinite on any $d$-dimensional Euclidean space.

Fourier transforms of radial versions of members of $\Phi_d$, for a given $d$, have a simple expression, as reported in \citet{Yaglom:1987} or
\citet{stein-book}. For a member $\phi$ of the class $\Phi_d$ such that $\phi(\|\cdot\|)$ is absolutely integrable in $\mathbb{R}^d$, we define its isotropic spectral density as
\begin{equation} \label{FT}
 \widehat{\phi}(z)= \frac{z^{1-d/2}}{(2 \pi)^{\frac{d}{2}}} \int_{0}^{\infty} u^{d/2} J_{d/2-1}(uz)  \phi(u) {\rm d} u, \qquad z \ge 0.
\end{equation}

In the following we review two important examples:

\begin{enumerate}
\item \textbf{The  Mat{\'e}rn  (${\cal MT}$) family}

The Mat\'ern family is a two-parameter family of globally supported
correlation functions, defined by \citep{Matern}:
\begin{equation}\label{eq:matern}
 {\cal MT}_{\nu, a} (x) = 
 \frac{2^{1-\nu}}{\Gamma(\nu)} \left ( \frac{x}{a} \right )^{\nu} {\cal K}_{\nu}\left ( \frac{x}{a} \right ), \qquad x > 0,
\end{equation}

where ${\cal K}_{\nu}$ is the modified Bessel function of the second kind of order $\nu$ \citep[][9.6.22]{Abramowitz-Stegun:1965}. The associated spectral density, $\widehat{{\cal MT}}_{a,\nu}$, is given by
\begin{equation} \label{spectral_matern}
\widehat{{\cal MT}}_{\nu,a}(z)= \frac{\Gamma(\nu+d/2)}{\pi^{d/2} \Gamma(\nu)}
\frac{a^d}{(1+a^2z^2)^{\nu+d/2}}, \qquad z \ge 0.
\end{equation}

Arguments in \cite{stein-book} show that ${\cal MT}_{\nu,a}$ belongs to $\Phi_{\infty}$ when both the smoothness parameter $\nu$ and the scale parameter $a$ are strictly positive.
A key feature of the Mat\'ern model is the smoothness parameter $\nu$,
which determines the mean-square differentiability of the sample paths
of the associated Gaussian RF. In particular, for any integer
$m=0,1,\ldots$, the sample paths are $m$-times mean-square differentiable
(in any direction) if and only if $\nu>m$.
For $\nu=m+\tfrac12$, the correlation reduces to an exponential multiplied
by a polynomial of degree $m$.
\item \textbf{The  Generalized Wendland (${\cal GW}$) family}

The ${\cal GW}$ family is a three-parameter family of compactly supported
correlation functions, defined by \citep{gnei02,BFFP}:
\begin{equation} 
\label{WG4*}
\begin{split}
    {\cal GW}_{\kappa,\mu, a}(x) &=
	\frac{\Gamma(\kappa)\Gamma(2\kappa+\mu+1)}{\Gamma(2\kappa)\Gamma(\kappa+\mu+1)2^{\mu+1}} \left( 1- \frac{x^2}{a^2} \right)^{\kappa+\mu}_+ \\
    &\times
	    {}_2F_1\left(\frac{\mu}{2},\frac{\mu+1}{2};\kappa+\mu+1;\left(1- \frac{x^2}{a^2} \right )_+ \right), \; x \geq 0,
        \end{split}
	\end{equation}
	with $ a > 0$ a compact support parameter, $\kappa > -\frac{1}{2} $ a smoothness parameter and $\mu $ a shape parameter, 
	and where $(\cdot)_+$ stands for the positive part function and ${}_2F_1$ for the Gauss hypergeometric function \citep[][15.1.1]{Abramowitz-Stegun:1965}.
The associated spectral density is given by:
\begin{equation}\label{llkk}
\widehat{{\cal GW}}_{\kappa,\mu, a}(z)={L_{\mathrm{GW}}}a^{d}\mathstrut_1 F_2\Big(\lambda;\lambda+\frac{\mu}{2},\lambda+\frac{\mu}{2}+\frac{1}{2};-\frac{(za)^{2}}4\Big), \quad  z > 0,
\end{equation}
where ${}_1F_2$ is a generalized hypergeometric function, $\lambda=
(d+1)/2+\kappa$ and ${L_{\mathrm{GW}}}=\frac{2^{-d}\pi^{-\frac{d}{2}}\Gamma(\mu+2\kappa+1)\Gamma(2\kappa+d)\Gamma(\kappa)}{  \Gamma\left(\kappa+{d}/{2}\right)\Gamma(\mu+2\kappa+d+1)\Gamma(2\kappa) }$
is a normalization constant.

For a given $\kappa > -\frac{1}{2}$, the validity conditions for 
${\cal GW}_{\kappa,\mu,a} \in \Phi_d$ follow as a special case 
of Theorem~\ref{theopp} by setting $l = \frac{1}{2}$:
\begin{itemize}
    \item[(A)] ${\cal GW}_{\kappa,\mu,a} \in \Phi_d$ if and only if 
    $\mu \geq \kappa + \frac{d+1}{2}$, whenever $\kappa \geq 0$ or $d \geq 2$;
    
    \item[(B)] ${\cal GW}_{\kappa,\mu,a} \in \Phi_1$ if 
    $\mu \geq \frac{\sqrt{8\kappa+9}-1}{2}$, and only if 
    $\mu \geq \kappa + 1$, for $-\frac{1}{2} < \kappa < 0$.
\end{itemize}
Since $\kappa + 1 < \frac{\sqrt{8\kappa+9}-1}{2}$ for all 
$-\frac{1}{2} < \kappa < 0$, a gap exists between the necessary and 
sufficient conditions. This partially corrects \cite{bevele}, which 
claimed in case (B) that 
$\mu \geq \frac{\sqrt{8\kappa+9}-1}{2}$ 
is both necessary and sufficient for $d = 1$. Whether this lower bound 
is also necessary remains an open problem.

	

As for Mat\'ern, the ${\cal GW}_{\kappa,\mu,a}$ family provides continuous
control of the mean-square differentiability of the associated Gaussian RF
through the smoothness parameter $\kappa$. Specifically, for any integer
$m=1,2,\ldots$, the sample paths are $m$-times mean-square differentiable in
any direction if and only if $\kappa>m-\tfrac12$; for
$0\leq\kappa\leq\tfrac12$, they are not mean-square differentiable.

When $\kappa = k \in \mathbb{N}$,  the Generalized Wendland  correlation simplifies into the product of an Askey (truncated power) correlation with a polynomial of degree $k$.

Recently, \cite{bevilacqua2022unifying}
considered a compact support reparameterized version of the ${\cal GW}$ model, and proved that
   \begin{equation}   \label{cpoo}
   \lim_{\mu\to\infty} {\cal GW}_{\kappa,\mu,g(a, \mu,\kappa)}(x)={\cal MT}_{\kappa+1/2,a}(x),\quad \kappa > -0.5,
     \end{equation}
 where  $g(a, \mu,\kappa)=a  \left(\frac{\Gamma(\mu+2\kappa +1)} {\Gamma(\mu)}\right)^{\frac{1}{1+2\kappa}}$,
with uniform convergence on $[0,\infty)$.


Consequently, the reparameterized ${\cal GW}$ family can be viewed as a
compactly supported extension of Mat\'ern, with the additional parameter
$\mu$ controlling the transition from finite compact support to the globally
supported Mat\'ern limit.

\end{enumerate}

\section{The Gauss hypergeometric model and the identifiability problem
}\label{sec3}

The $\mathcal{GH}$ class in \cite{emery2022gauss} was introduced through
the following spectral density:
\begin{equation}\label{ppoon}
\widehat{{\cal GH}}_{\delta,\beta,\gamma,a}(z)=\frac{\Gamma(\delta)\Gamma(\beta-\frac{d}{2})\Gamma(\gamma-\frac{d}{2}) a^{d}}{ 2^d\pi^{\frac{d}{2}} \Gamma\left(\delta-\frac{d}{2}\right)\Gamma(\beta)\Gamma(\gamma) }\mathstrut_1 F_2\Big(\delta;\beta,\gamma;-\frac{(za)^{2}}{4}\Big), \;  z \geq 0.
\end{equation}

The spectral density in \eqref{ppoon} has the same general form as the
${\cal GW}$ density in \eqref{llkk}, but without its parameter constraints
inside ${}_1F_2$.

Using results from \cite{cho2018}, \citet{emery2022gauss} showed that, for
a given $a>0$, the following conditions:
$$(a)\ \delta>\frac{d}{2},\qquad
(b)\ 2(\beta-\delta)(\gamma-\delta)\ge\delta,\qquad
(c)\ \beta+\gamma \ge 3\delta+\frac{1}{2}\,,$$
are sufficient for the positivity of the spectral density on $[0,\infty)$.
The associated four-parameter isotropic correlation model  is given by:
\begin{equation}
    \label{emery}
    \begin{split}
    \mathcal{GH}_{\delta,\beta,\gamma,a}(x) &= \frac{\Gamma(\beta-\frac{d}{2})\Gamma(\gamma-\frac{d}{2})}
    { \Gamma(\beta-\delta + \gamma-\frac{d}{2})  \Gamma(\delta-\frac{d}{2})} \left( 1 - \frac{x^2}{a^2} \right)_+^{\beta-\delta + \gamma-\frac{d}{2}-1} \\
    &\times 
 {}_2F_1\left( \beta-\delta; \gamma-\delta; \beta-\delta + \gamma -\frac{d}{2}; \left( 1 - \frac{x^2}{a^2}\right)_+ \right), \; x \geq 0.
 \end{split}
\end{equation}
Here $a>0$ is a compact support parameter, $\delta>\frac{d}{2}$ is a smoothness parameter, and $\beta$ and $\gamma$ are two additional  shape parameters.

As in the ${\cal MT}$ and ${\cal GW}$ families, the ${\cal GH}$ model
provides continuous control of the mean-square differentiability of the
associated Gaussian RF through the smoothness parameter $\delta$.

Moreover, the ${\cal GH}$ class includes the ${\cal GW}$ and ${\cal MT}$
families as special cases \citep{emery2022gauss}. For instance:
\begin{equation}\label{ww3}
\mathcal{GH}_{\delta,\delta+\frac{\mu}{2},\delta+\frac{\mu}{2} + \frac{1}{2},a}(x)= {\cal GW}_{\kappa,\mu,a}(x),
 \end{equation}
 where  $\delta=\frac{d+1}{2}+\kappa$.

Although the ${\cal GH}$ family is highly flexible, it is \emph{not identifiable in the usual parametric sense}.
Indeed, even under increasing-domain asymptotics---where one would in principle expect all parameters to be estimable \citep{mardia1984maximum}---the parameter-to-correlation map is not injective.
In particular,
$
\mathcal{GH}_{\delta,\beta,\gamma,a}(x)=\mathcal{GH}_{\delta,\gamma,\beta,a}(x),
$
because the Gauss hypergeometric function is symmetric in its first two arguments:
$
{}_2F_1(a,b;\cdot\,;x) = {}_2F_1(b,a;\cdot\,;x).
$
Hence distinct parameter values (obtained by swapping $\beta$ and $\gamma$) induce exactly the same correlation function.
This implies that additional constraints are required.

A related inferential limitation arises under fixed-domain (infill) asymptotics, where distinct parameter values may induce equivalent Gaussian measures. The following theorem establishes the equivalence condition for two distinct Gaussian measures with $\mathcal{GH}$ correlation models that share a common smoothness parameter. Hereafter, we denote by $P(\rho_i)$, $i = 0, 1$, a zero-mean Gaussian measure with an isotropic covariance function $\rho_i$.

\begin{theorem} \label{Wv_vs_Wv}
Let $P(\sigma_i^2{\cal GH}_{\delta,\beta_i,\gamma_i,a_i})$, $i=0,1$, be two
zero-mean Gaussian measures, and let $\delta>d/2$. Assume that
\[
\frac{d+1}{2}+3\delta
<
\min(\beta_1+\gamma_1,\beta_0+\gamma_0)
\]
and
$
2(\beta_i-\delta)(\gamma_i-\delta)\ge\delta$,
$i=0,1$.

Then, for any bounded infinite set $D\subset\R^d$, $d=1,2,3$, the Gaussian measures
$P(\sigma_1^2{\cal GH}_{\delta,\beta_1,\gamma_1,a_1})$ and
$P(\sigma_0^2{\cal GH}_{\delta,\beta_0,\gamma_0,a_0})$
are equivalent on the paths of $\{Z(\bs): \bs \in D\}$, if and only if
\begin{equation}\label{condition1_iff2}
\frac{\sigma_0^2 L (\beta_0,\gamma_0)}{a_0^{2\delta-d}} =
\frac{\sigma_1^2 L(\beta_1,\gamma_1)}{a_1^{2\delta-d}},
\end{equation}
where $L(\beta_i,\gamma_i)=\dfrac{2^{2\delta-d}\Gamma(\beta_i-d/2)\Gamma(\gamma_i-d/2)}
{\Gamma(\gamma_i-\delta)\Gamma(\beta_i-\delta)}$, $i=0,1$.
\end{theorem}

The restriction to $d=1,2,3$ is typical when studying equivalence 
of Gaussian measures for isotropic models on $\mathbb{R}^d$, and 
is not specific to the ${\cal GH}$ class. Indeed, the same dimensional 
restriction appears in the equivalence results for the ${\cal MT}$
model \citep{Zhang:2004}, and stems from the integrability 
conditions required by the criteria of \citet{Stein:2004}.

An immediate consequence of Theorem~\ref{Wv_vs_Wv} is that, for fixed $\delta$, the 
  \emph{microergodic parameter} is given  by
$
\frac{\sigma^2 L(\beta,\gamma)}{a^{2\delta-d}}.
$
Because $L(\beta,\gamma)=L(\gamma,\beta)$, this quantity does not distinguish $(\beta,\gamma)$ from $(\gamma,\beta)$.
Thus, while the microergodic parameter is the relevant identifiable feature under fixed-domain asymptotics,
the individual shape parameters $\beta$ and $\gamma$ are not identifiable without an additional constraint.

\section{Fixing the identifiability problem: the Hypergeometric correlation model}\label{4}

In this section, we first refine the validity conditions of the $\mathcal{GH}$ model,
clarifying the relationship between necessary and sufficient constraints.
We then address the non-uniqueness discussed in the previous section by imposing an appropriate restriction on the parameter space, which restores a unique representation of the correlation function.
Finally, we introduce a parsimonious compactly supported subclass selected through an explicit optimality criterion.

To this end, we adopt a ${\cal GW}$-type reparameterization of the
$\mathcal{GH}_{\delta,\beta,\gamma,a}$ model and set
\begin{equation}\label{repp}
\delta=\frac{d+1}{2}+\kappa, \qquad
\beta=\delta+\frac{\mu}{2}, \qquad
\gamma=\beta+l, \qquad l\ge 0.
\end{equation}
As noted in~\eqref{ww3}, this parametrization includes the ${\cal GW}$ model as the
special case $l=\frac{1}{2}$.

The restriction $l\ge 0$ in~\eqref{repp} is equivalent to imposing the ordering
constraint $\gamma\ge\beta$ on the original $\mathcal{GH}_{\delta,\beta,\gamma,a}$
family. This removes the intrinsic label-switching ambiguity induced by the symmetry
of ${}_2F_1$ in its upper parameters, namely
$\mathcal{GH}_{\delta,\beta,\gamma,a}=\mathcal{GH}_{\delta,\gamma,\beta,a}$,
and thus yields a unique representation up to this exchange.

The following theorem improves Theorem~1 in~\cite{emery2022gauss} by providing sharp
validity conditions for the reparameterized model
$\mathcal{GH}_{\delta,\delta+\frac{\mu}{2},\delta+\frac{\mu}{2}+l,a}$.

\begin{theorem}\label{theopp}
Let  $\kappa > -\frac{1}{2}$,
$\delta = \frac{d+1}{2}+\kappa$, $a > 0$. Then:
\begin{enumerate}
\item If $0 \le l \le \frac{d}{2}+\kappa$, then
\[
\mathcal{GH}_{\delta,\delta+\frac{\mu}{2},\delta+\frac{\mu}{2}+l,a}
\in \Phi_d
\quad\text{if and only if}\quad
\mu \ge \delta - l + \frac{1}{2}.
\]
\item If $l > \frac{d}{2}+\kappa$, then
\[
\mathcal{GH}_{\delta,\delta+\frac{\mu}{2},\delta+\frac{\mu}{2}+l,a}
\in \Phi_d
\quad\text{if}\quad
\mu \ge \sqrt{2\kappa + l^2 + d + 1} - l,
\]
and only if
$
\mu \ge \delta - l + \frac{1}{2}.
$
\end{enumerate}
\end{theorem}

\begin{remark}
Tracing back the conditions of Theorem \ref{theopp} to the original parametrization $(\delta,\beta,\gamma,a)$, and assuming  $\delta> d/2$ and $\gamma\ge\beta$ we obtain:
\begin{enumerate}
\item If $0\le \gamma-\beta \le \delta-\frac12$, then
\[
\mathcal{GH}_{\delta,\beta,\gamma,a}\in\Phi_d
\quad\text{if and only if}\quad
\beta+\gamma \ge 3\delta+\frac12.
\]
\item If $\gamma-\beta > \delta-\frac12$, then
\[
\mathcal{GH}_{\delta,\beta,\gamma,a}\in\Phi_d
\quad\text{if}\quad
\beta+\gamma \ge 2\delta+\sqrt{(\gamma-\beta)^2+2\delta},
\]
and only if
$
\beta+\gamma \ge 3\delta+\frac12.
$
\end{enumerate}
In particular, on the broad regime $\gamma-\beta\le \delta-\tfrac12$ the sufficient conditions (a)--(c) in Section \ref{sec3} reduce to the single inequality $\beta+\gamma\ge 3\delta+\tfrac12$ and become sharp, since this inequality is also necessary. This sharpens the earlier validity region reported in \citet{emery2022gauss}. Outside this regime, the sufficient bound above is explicit while a residual gap with the necessary condition remains.
\end{remark}

Setting $l=\tfrac12$ in Theorem~\ref{theopp} recovers the
${\cal GW}$ validity conditions reported in Section~\ref{Sec2}, including
the remaining necessary--sufficient gap for $d=1$ and
$-\tfrac12<\kappa<0$.

Even after imposing $l\ge0$, the reparameterized model retains the two shape
parameters $(\mu,l)$. To obtain a parsimonious subclass, we select $l$ by
maximizing the integral range at fixed smoothness and support, first over the
regime where the validity boundary is exact and then over the region guaranteed
by the sufficient conditions of Theorem~\ref{theopp}.

For an isotropic correlation $\phi \in \Phi_d$, the integral range is
\begin{equation}
\label{intrang}
A=\int_{\mathbb{R}^d} \phi(\| \boldsymbol{h}\|)\, \mathrm{d}\boldsymbol{h}
= \frac{2\pi^{\frac{d}{2}}}{\Gamma(\frac{d}{2})}
\int_{0}^{\infty} x^{d-1}\phi(x)\,\mathrm{d}x.
\end{equation}

The integral range is a global, threshold-free measure of dependence
\citep{Lantu1991,Yaglom:1987,matheron1989,HristopulosZukovic:2011}. For integrable
correlations,
$
A=(2\pi)^d\widehat\phi(0),
$
so maximizing $A$ at fixed support favors low-frequency power, hence stronger
large-scale dependence, through a criterion intrinsic to the model class.

For a parametric model $\phi_{\boldsymbol{\theta}} \in \Phi_d$,
we denote the associated integral range by $A_{\boldsymbol{\theta},d}$.
For example, for the Mat\'ern correlation one has
\begin{equation}
\label{ftt}
A_{\nu,a,d}=
2^{d}\pi^{d/2}a^d\,
\frac{\Gamma\!\left(\nu+\frac{d}{2}\right)}
{\Gamma(\nu)}.
\end{equation}

The following theorem provides the integral range of
$\mathcal{GH}_{\delta,\delta+\frac{\mu}{2},\delta+\frac{\mu}{2}+l,a}$.

\begin{theorem}\label{theopkkp}
The integral range of $\mathcal{GH}_{\delta,\delta+\frac{\mu}{2},\delta+\frac{\mu}{2}+l ,a}$ is given by
\begin{equation}\label{integral range}
A_{\kappa,\mu,l,a,d}=a^d \left[
\frac{\pi^{d/2}\Gamma(\frac{d+1}{2}+ \kappa)\,
\Gamma\left(\frac{\mu+1}{2}+\kappa\right)\,
\Gamma\left(\frac{\mu+1}{2}+\kappa+l\right)}
{\Gamma(\kappa+\frac{1}{2})\,
\Gamma\left(\frac{\mu+1+d}{2}+\kappa \right)\,
\Gamma\left(\frac{\mu+1+d}{2}+\kappa+l\right)}
\right].
\end{equation}
\end{theorem}

The factor $a^d$ only rescales the integral range. Moreover, for fixed
$(\kappa,l,a,d)$, $A_{\kappa,\mu,l,a,d}\asymp\mu^{-d}$ as $\mu\to\infty$,
so the integral range vanishes as $\mu$ increases.

For the next results, \emph{admissible region} denotes the parameter set
guaranteed by Theorem~\ref{theopp}; in Regime~2 this refers to its sufficient
bound. The following theorem maximizes the integral range over $\mu$ for fixed
$l$.

\begin{theorem}\label{theopkkp2}
Let $\kappa>-\frac12$, $a>0$, and $d\in\mathbb{N}$ be fixed, and set
$\delta=\frac{d+1}{2}+\kappa$.
For each $l\ge0$, the map $\mu\mapsto A_{\kappa,\mu,l,a,d}$ is non-increasing on
its admissible domain. Hence its maximum is attained at the smallest admissible
value of $\mu$.
In particular,
\begin{enumerate}
\item if $0\le l\le d/2+\kappa$, then the maximum is attained at
$$\mu^\star(l)=\mu_1(l):=\delta-l+\frac12;$$
\item if $l>d/2+\kappa$, then the maximum is attained at
$$\mu^\star(l)=\mu_2(l):=\sqrt{2\kappa+l^2+d+1}-l.$$
\end{enumerate}
\end{theorem}

For instance, if $l=\frac{1}{2}$ and $d\ge2$  (or $d=1$ and $\kappa\ge0$), then
by \eqref{ww3}
\[
\mathcal{GH}_{\delta,\delta+\frac{\mu}{2},\delta+\frac{\mu+1}{2},a}
={\cal GW}_{\kappa,\mu,a},
\]
and the maximum integral range (over the admissible $\mu$) is attained at
$\mu=\mu_1(\frac{1}{2})=\delta=\frac{d+1}{2}+\kappa$, which coincides with the
lower bound of the ${\cal GW}$ model in this case.

\begin{figure}[h!]
\begin{center}
\begin{tabular}{cc}
\includegraphics[width=6.45cm,height=6.6cm]{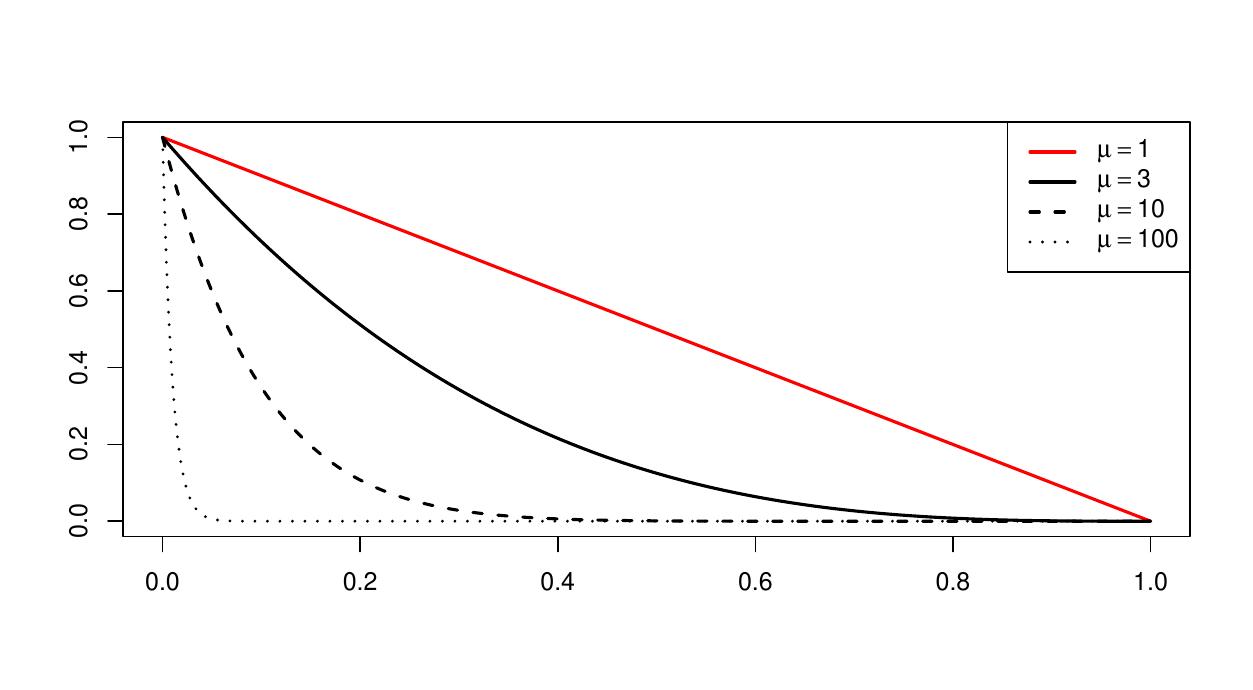}&\includegraphics[width=6.45cm,height=6.6cm]{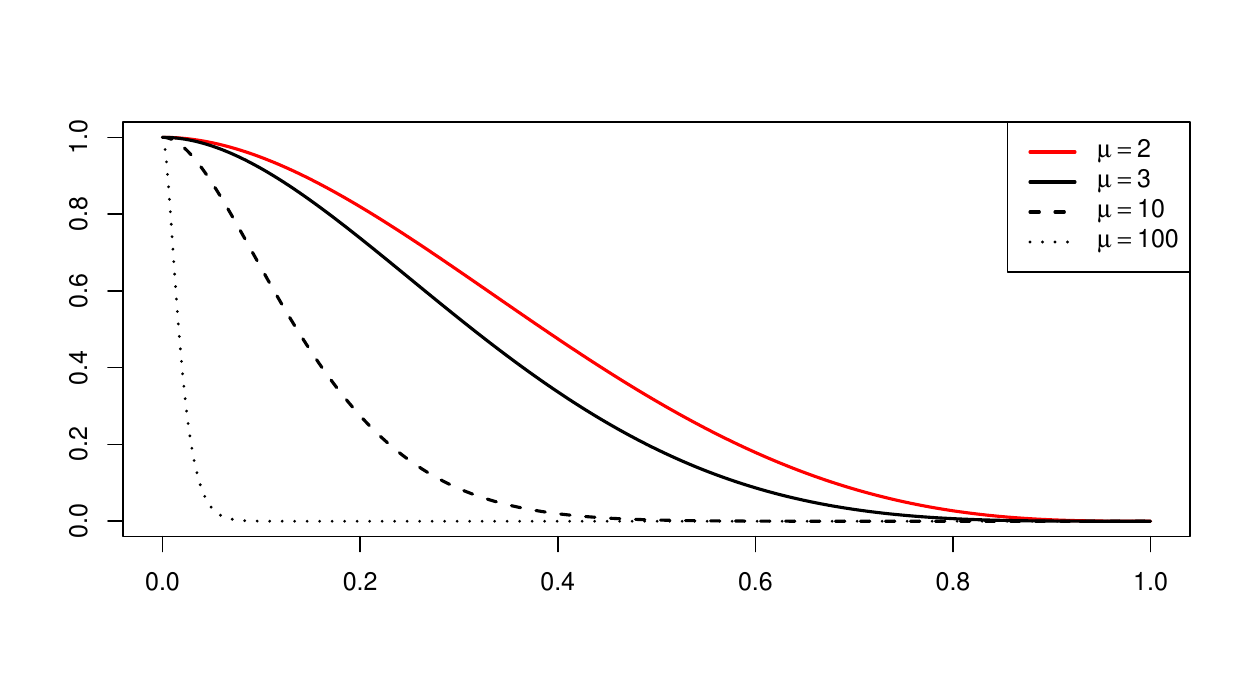}\\
\end{tabular}
\end{center}
\caption{
Left part:
${\cal GW}_{0,\mu,1}$  for $\mu=\mu_1(\frac{1}{2}),3,10,100$   with  $\delta=\frac{d+1}{2}+\kappa$ and $\kappa=0$.
Right panel: the same comparison with $\kappa=1$.
In both cases $d=1$.
}
 \label{ordinary44}
\end{figure}

As an example, consider the case $d=1$ and $\kappa=0,1$.
Figure~\ref{ordinary44} (left) depicts ${\cal GW}_{0,\mu,1}$ and Figure~\ref{ordinary44}
(right) depicts ${\cal GW}_{1,\mu,1}$ for $\mu=\mu_1(\frac{1}{2}),3,10,100$.
It is clear that, in both panels, the correlation with the largest area under
the curve is obtained for $\mu=\mu_1\!\left(\tfrac12\right)$ (red line).

The next step is to determine which $l\ge0$ yields the largest attainable
integral range, i.e., which $l$ maximizes $A_{\kappa,\mu^\star(l),l,a,d}$.

\begin{theorem}\label{theopkkp333}
Let $\kappa>-1/2$, $a>0$, and $d\in\mathbb N$ be fixed, and set
$\tau=d/2+\kappa$.
\begin{enumerate}
\item The function
$ l\mapsto A_{\kappa,\mu_1(l),l,a,d}$ is non-decreasing on $[0,\tau]$.
Consequently,
\[
\max_{0\leq l\leq\tau} A_{\kappa,\mu_1(l),l,a,d}
= A_{\kappa,1,\tau,a,d}.
\]
\item If $d\geq2$, or if $d=1$ and $\kappa\geq0$, then the function
$ l\mapsto A_{\kappa,\mu_2(l),l,a,d}$ is non-increasing on
$(\tau,\infty)$ and
\[
\lim_{l\downarrow\tau} A_{\kappa,\mu_2(l),l,a,d}
= A_{\kappa,1,\tau,a,d}.
\]
Under either of these dimensional--smoothness conditions, the largest integral
range over the full parameter region guaranteed by the sufficient conditions
of Theorem~\ref{theopp} is therefore achieved at
$(\mu,l)=(1,d/2+\kappa)$.
\end{enumerate}
For $d=1$ and $-1/2<\kappa<0$, the first assertion remains valid, but no
global maximization claim over the second regime is made.
\end{theorem}

\begin{figure}[h!]
\begin{center}
\begin{tabular}{cc}
\includegraphics[width=6.45cm,height=6.6cm]{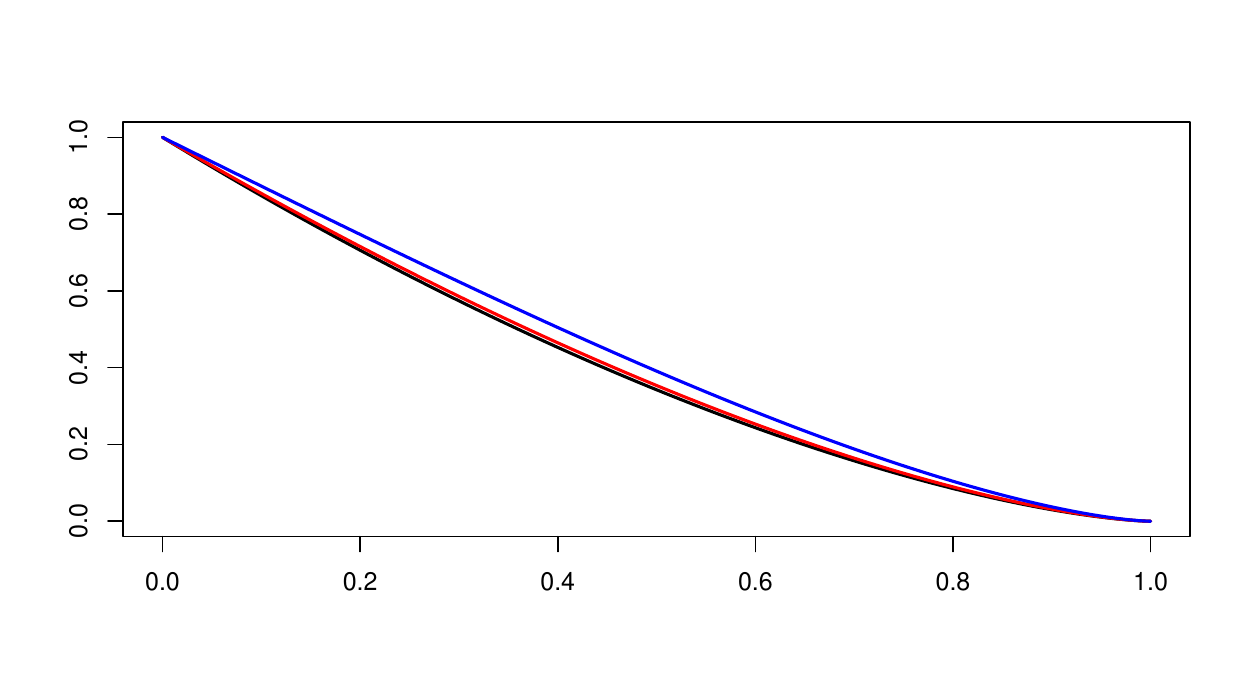}&\includegraphics[width=6.45cm,height=6.6cm]{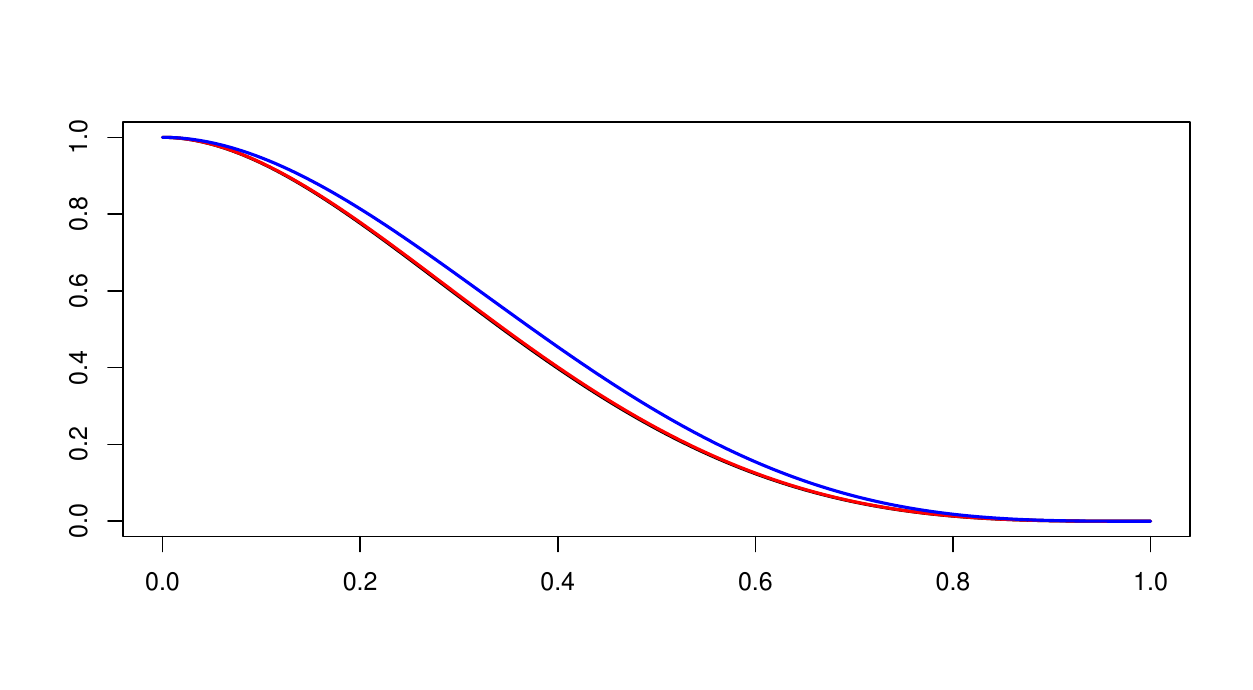}\\
\end{tabular}
\end{center}
\caption{
Left part:
$\mathcal{GH}_{\delta,\delta+\frac{\mu(l) }{2},\delta+\frac{\mu(l) }{2}+l,1}$  with
$\delta=\frac{d+1}{2}+\kappa$, $\kappa=0$,   $\mu(l) =\delta-l+\frac{1}{2}$  and $l=0, \frac{1}{2}, \frac{d}{2}+\kappa$ (black, red and blue lines  respectively).
Right panel: the same comparison with $\kappa=1$. The case
$l=\frac12$ (red line) corresponds to the ${\cal GW}_{\kappa,\mu,1}$ model.
The blue  line corresponds to the correlation model that maximizes the integral range.
In both cases $d=2$.
}
 \label{ordinary}
\end{figure}

As an example, Figure   \ref{ordinary}  compares
 $\mathcal{GH}_{\delta,\delta+\frac{\mu_1(l) }{2},\delta+\frac{ \mu_1(l)  }{2}+l,1}$
when        $l=0, \frac{1}{2}, d/2+\kappa$ (black, red and blue lines, respectively). In particular the left part depicts  the case  $\kappa=0$ and the right part
depicts  the case  $\kappa=1$ ($d=2$ in both cases).
The case $l=d/2+\kappa$ (blue line) has the largest integral range and
exceeds the maximum integral range attained by the ${\cal GW}$ model
(red line).

Theorem~\ref{theopkkp333} therefore motivates fixing $l=d/2+\kappa$.
At $\mu=1$ the selected boundary member is
$\mathcal{GH}_{\delta,\delta+1/2,2\delta,a}$, which includes the Euclid's hat
model at $\kappa=0$ and its (fractional) mont\'ee extensions for
$\kappa\ne0$ \citep{Gneiting1999,emery2022gauss,matheron65}. We define the
resulting \emph{hypergeometric correlation} by

\begin{equation}\label{venezia}
\mathcal{H}_{\kappa,\mu,a,d}(x)
:=\mathcal{GH}_{\delta,\delta+\frac{\mu}{2},
\delta+\frac{\mu+d}{2}+\kappa,a}(x),
\qquad x \ge 0.
\end{equation}
An explicit expression is given by
\begin{equation}
\label{emery2}
\begin{split}
\mathcal{H}_{\kappa,\mu,a,d}(x)
&=
\frac{\Gamma\!\left(\kappa+\frac{\mu+1}{2}\right)
      \Gamma\!\left(2\kappa+\frac{d+\mu+1}{2}\right)}
     {\Gamma\!\left(\mu+\frac{d+1}{2}+2\kappa\right)
      \Gamma\!\left(\kappa+\frac{1}{2}\right)}
\left(1-\frac{x^2}{a^2}\right)_+^{\mu+\frac{d-1}{2}+2\kappa}
\\[4pt]
&\quad\times
{}_2F_1\!\left(
\frac{\mu}{2},
\frac{\mu+d}{2}+\kappa;
\mu+\frac{d+1}{2}+2\kappa;
\left(1-\frac{x^2}{a^2}\right)_+
\right),
\;x \ge 0.
\end{split}
\end{equation}

Notably, the model $\mathcal{H}_{\kappa,\mu,a,d}$ always lies in Regime~1 of
Theorem~\ref{theopp}, where validity is completely characterized. Indeed, by
setting $l = d/2+\kappa$ in Theorem~\ref{theopp}, we obtain
$
\mathcal{H}_{\kappa,\mu,a,d} \in \Phi_d$ if and only if
$\mu \ge 1$.

\begin{figure}[h!]
\begin{center}
\begin{tabular}{cc}
\includegraphics[width=5.9cm,height=6.2cm]{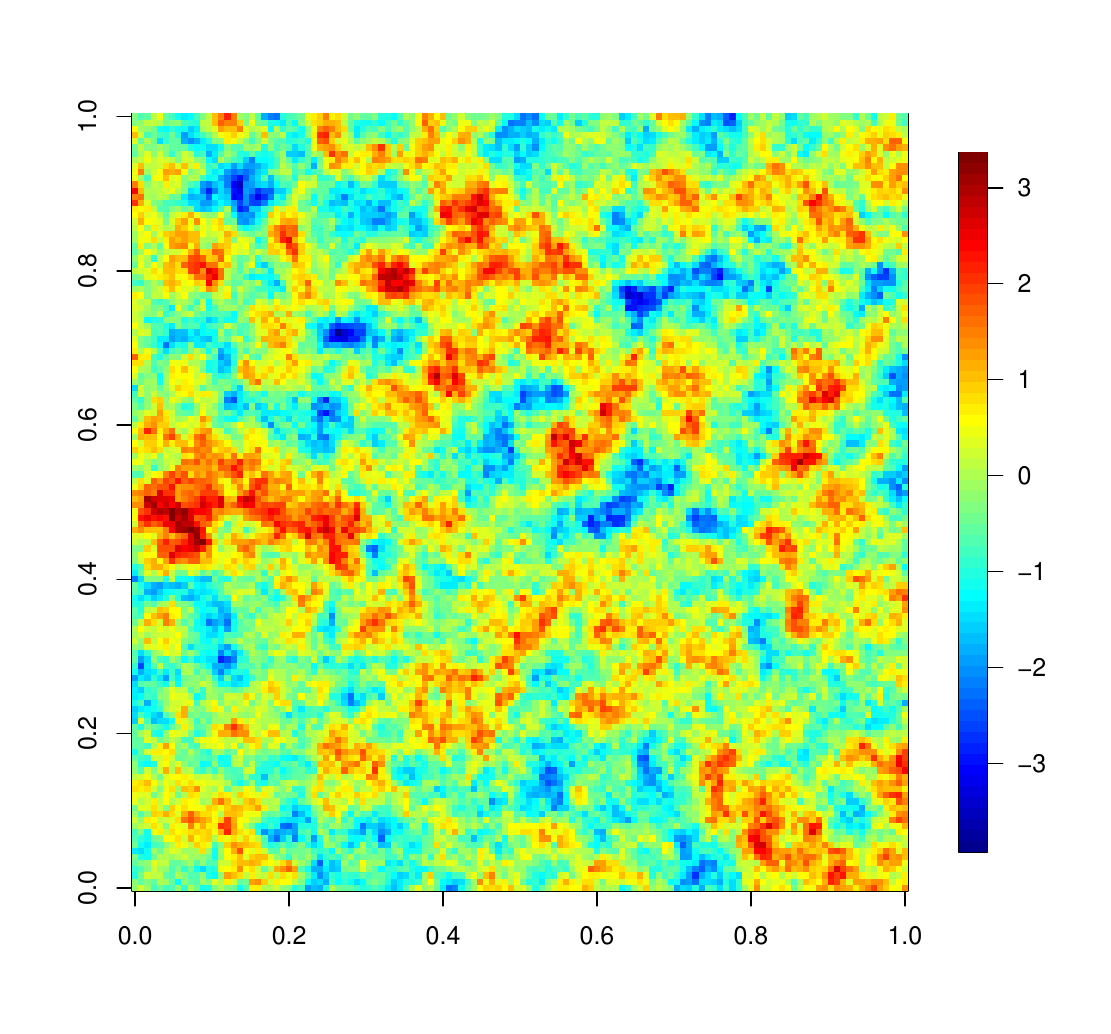}&\includegraphics[width=5.9cm,height=6.2cm]{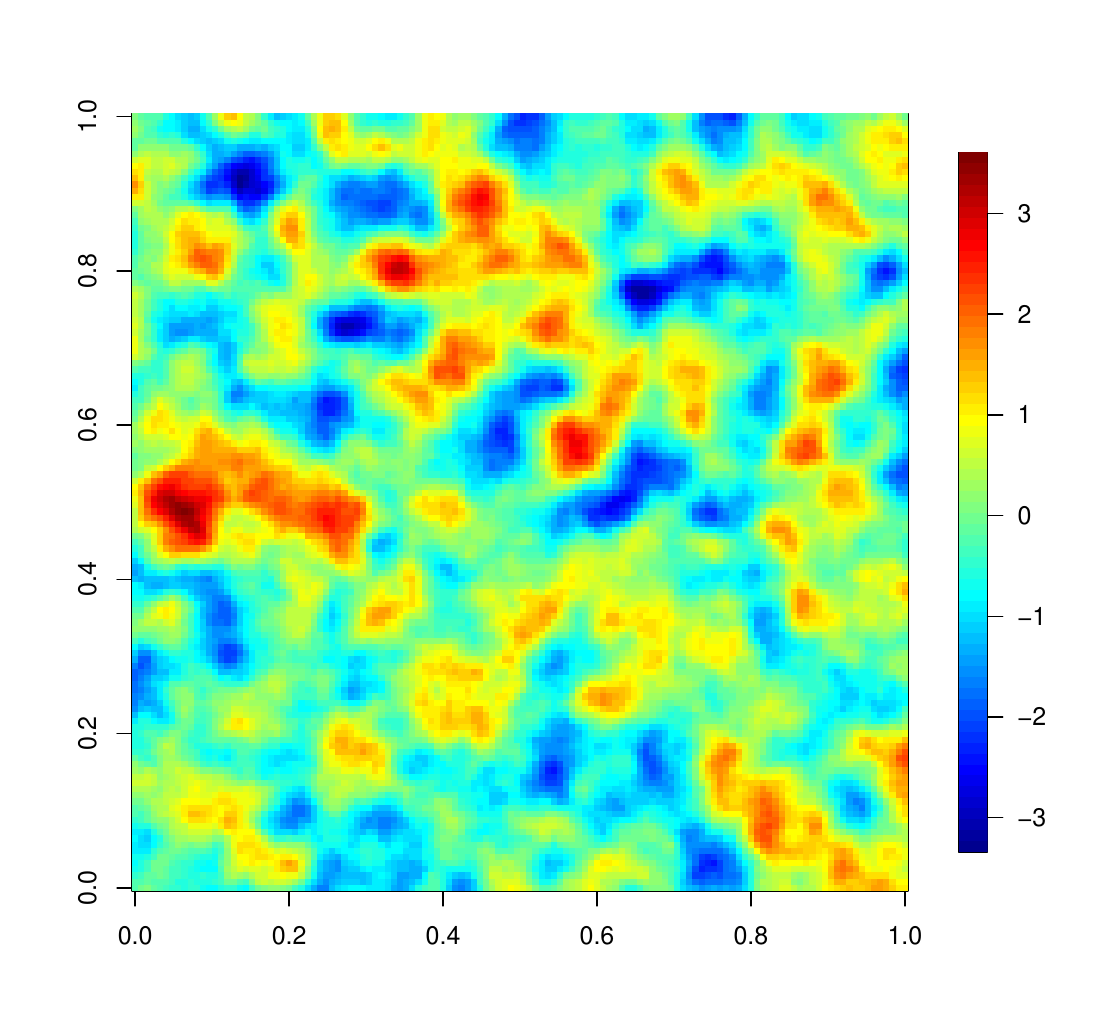}\\
\end{tabular}
\end{center}
\caption{
Two Gaussian RF realizations with
${\cal H}_{\kappa,4,0.2,2}$
correlation model, for $\kappa = 0, 1$ (from left to right).
}
 \label{figsmo}
\end{figure}

The validity bound $\mu\ge1$ is simpler than for
${\cal GW}_{\kappa,\mu,a}$ and independent of $d$, although the expression
of $\mathcal H_{\kappa,\mu,a,d}$ itself depends on dimension.

{
The maximum-integral-range construction fixes $l$, but it does not force every
fitted $\mathcal H$ model to attain the maximum. Within the selected family,
the maximum occurs at $\mu=1$, while larger values of $\mu$ progressively
reduce the integral range. Thus, $\mu$ remains free to accommodate weaker
large-scale dependence in a particular application.

Fixing $l=d/2+\kappa$ reduces covariance-shape flexibility relative to the
full identifiable $\mathcal{GH}$ family. The scale-mixture representation
below does not imply that a single $\mathcal H$ covariance can reproduce
every admissible $\mathcal{GH}$ correlation.

This distinction also has a direct modeling consequence. In the regime where
the ${\cal GW}$ validity boundary is exact, its smallest admissible value
$\mu=\kappa+(d+1)/2$ grows with dimension, whereas the $\mathcal H$ boundary
remains $\mu=1$. Together, Theorems~\ref{theopkkp2}--\ref{theopkkp333} show
that $\mathcal H$ can retain stronger large-scale dependence at the same
compact-support radius without reducing covariance sparsity.
Proposition~\ref{prop:bridge-support} gives the complementary sparsity
interpretation under the Mat\'ern-limit reparameterization.
}

The next theorem shows that ${\cal H}_{\kappa,1,a,d}$ is a scale-mixture
building block for every ${\cal H}_{\kappa,\mu,a,d}$ and for a broad subclass
of $\mathcal{GH}$.
\begin{theorem}\label{emery-antologico}
Let $a > 0$, $\kappa > -\frac{1}{2}$, $\delta = \frac{d+1}{2}+\kappa$, $\mu>1$ and
$l\ge 0$. Assume that the parameters $(\mu,l)$ belong to the admissible region
of Theorem~\ref{theopp}, so that
$\mathcal{GH}_{\delta,\delta+\frac{\mu}{2},\delta+\frac{\mu}{2}+l,a}\in\Phi_d$.
In addition, assume $\mu + 2l > 1 + d + 2\kappa$.
Then
$\mathcal{GH}_{\delta,\delta+\frac{\mu}{2},
\delta+\frac{\mu}{2}+l,a}$
admits a beta-type scale mixture representation in terms of
$\mathcal H_{\kappa,1,b,d}$ with $b=a\sqrt{uv}$:
\begin{eqnarray}
\label{emery-mitico}
\mathcal{GH}_{\delta,\delta+\frac{\mu}{2},
\delta+\frac{\mu}{2}+l,a}(x)
&=&
\frac{\Gamma\!\left(\delta+\frac{\mu-d}{2}\right)
      \Gamma\!\left(\delta+\frac{\mu-d}{2}+l\right)}
     {\Gamma\!\left(\delta+\frac{1-d}{2}\right)
      \Gamma\!\left(\delta+\frac{1}{2}+\kappa\right)
      \Gamma\!\left(\frac{\mu-1}{2}\right)
      \Gamma\!\left(\frac{\mu-1-d}{2}-\kappa+l\right)}
\nonumber\\
&&\times
\int_0^1\!\!\int_0^1
u^{\delta-\frac{d+1}{2}}
(1-u)^{\frac{\mu-3}{2}}
v^{\delta-\frac{1}{2}+\kappa}
(1-v)^{\frac{\mu-d-3}{2}-\kappa+l}
\nonumber\\
&&\times
\mathcal{H}_{\kappa,1,a\sqrt{uv},d}(x)
\,\mathrm{d}u\,\mathrm{d}v,
\qquad x \ge 0.
\end{eqnarray}
In particular, when $l=d/2+\kappa$, one obtains
\begin{eqnarray*}
\mathcal{H}_{\kappa,\mu,a,d}(x)
&=&
\frac{\Gamma\!\left(\delta+\frac{\mu-d}{2}\right)
      \Gamma\!\left(\delta+\frac{\mu}{2}+\kappa\right)}
     {\Gamma\!\left(\delta+\frac{1-d}{2}\right)
      \Gamma\!\left(\delta+\frac{1}{2}+\kappa\right)
      \Gamma^2\!\left(\frac{\mu-1}{2}\right)}
\\
&&\times
\int_0^1\!\!\int_0^1
u^{\delta-\frac{d+1}{2}}
(1-u)^{\frac{\mu-3}{2}}
v^{\delta-\frac{1}{2}+\kappa}
(1-v)^{\frac{\mu-3}{2}}
\\
&&\times
\mathcal{H}_{\kappa,1,a\sqrt{uv},d}(x)
\,\mathrm{d}u\,\mathrm{d}v,
\; x \ge 0.
\end{eqnarray*}
\end{theorem}
The representation can support spectral simulation
\citep{BevilacquaEmeryCuevasPacheco2025FastSimulationGRF}.
{
It also has a hierarchical interpretation through latent beta-distributed
scales. The theorem remains stationary, but suggests nonstationary extensions
based on spatially varying mixing laws or supports, provided positive
definiteness is preserved.
}
As for ${\cal GW}$, $\kappa$ controls mean-square smoothness.
Figure~\ref{figsmo} shows unit-variance Gaussian RF realizations with
$\mathcal H_{\kappa,4,0.2,2}$ for $\kappa=0,1$ on the unit square.

\section{Some special cases and bridges to the Mat\'ern model}\label{5}

Although $\mathcal H_{\kappa,\mu,a,d}$ generally requires numerical
evaluation of a Gauss hypergeometric function \citep{Johansson2019}, several
special cases simplify. We first consider
$\kappa=k\in\mathbb N_0$, $d=1$, and $\mu\ge1$.

\begin{theorem}\label{cff}
If $\kappa=k$ is a nonnegative integer, $d=1$, and $\mu\ge1$, then, for $x\ge0$,
\[
\mathcal{H}_{k,\mu,a,1}(x)=
\begin{cases}
\displaystyle
\frac{\sqrt{\pi}\,
\Gamma\!\left(\frac{\mu}{2}+2k+1\right)}
{\Gamma\!\left(k+\tfrac12\right)}
\sum_{n=0}^{k}
\frac{(n+k)!}
     {2^{2n} n!(k-n)!\,
      \Gamma\!\left(n+\frac{\mu}{2}+k+1\right)}
\\[0.7em]\displaystyle\qquad\qquad\times
\left(\frac{x}{a}\right)^{k-n}
\left(1-\frac{x}{a}\right)^{\mu+2k+2n},
& 0\le x<a,\\[1.1em]
0, & x\ge a.
\end{cases}
\]
\end{theorem}

General closed forms for arbitrary $d>1$, $\mu\ge1$, and integer $\kappa$
are more difficult to obtain. Nevertheless, additional simplifications
can be derived using known identities of the Gauss hypergeometric function.
For instance, when $d$ is odd, $\mu\ge1$, and $\kappa=0$,
the following closed-form expression holds.

\begin{theorem}\label{cff1}
Let $d$ be an odd integer and $\mu\ge1$.
If $\kappa=0$, then
\[
\mathcal{H}_{0,\mu,a,d}(x)
=
C\left(1-\frac{x}{a}\right)_+^{\mu+\frac{d-1}{2}}
\sum_{n=0}^{\frac{d-1}{2}}
\frac{\left(\frac{1-d}{2}\right)_n
      \left(\frac{d+1}{2}\right)_n
      \left(1-\frac{x}{a}\right)_+^n}
     {2^n n!\,
      \Gamma\!\left(n+\frac{d-1}{2}+\mu+1\right)},
\]
where
$
C=
\frac{2^{\mu+\frac{d-1}{2}}
\Gamma\!\left(\frac{\mu+1}{2}\right)
\Gamma\!\left(\frac{\mu+d+1}{2}\right)}
{\sqrt{\pi}}.
$
\end{theorem}

Additional closed-form expressions are reported in the Supplementary Material, including spherical models \citep{Chiles:Delfiner:2012} that are widely used in geostatistical practice and are recovered as special cases of the proposed hypergeometric family.

The Mat\'ern model can be recovered as a limiting case of 
$\mathcal{H}_{\kappa,\mu,a,d}$ under an appropriate compact-support 
reparameterization.
To establish this result, we first characterize the equivalence between the Gaussian measures 
$P(\sigma_0^2 \mathcal{MT}_{\nu,\alpha})$ 
and 
$P(\sigma_1^2 \mathcal{H}_{\kappa,\mu,a,d})$.
The proof is provided in the Supplementary Material.

\begin{theorem}\label{ThmX}
Let $P(\sigma_0^2{\cal MT}_{\nu,\alpha})$ and
$P(\sigma_1^2{\cal H}_{\kappa,\mu,a,d})$
be two zero-mean Gaussian measures, and let
$\kappa>-\frac{1}{2}$ , $\nu=\kappa+\frac{1}{2}$ and $\mu>1+d/2$ .
For any bounded infinite set $D\subset\mathbb{R}^d$,
$d=1,2,3$, the Gaussian measures
$P(\sigma_0^2{\cal MT}_{\nu,\alpha})$
and
$P(\sigma_1^2{\cal H}_{\kappa,\mu,a,d})$
are equivalent on the paths of $\{Z(\bs): \bs \in D\}$,
if and only if
\begin{equation}\label{cafu}
\frac{\sigma_0^2}{\alpha^{2\nu}}
=
\frac{\sigma_1^2}{a^{2\kappa+1}}
\,F(\kappa,\mu,d),
\end{equation}
where
\begin{equation}\label{ka4}
F(\kappa,\mu,d)
=
\frac{2^{2\kappa+1}
\Gamma\!\left(\frac{\mu+1}{2}+\kappa\right)
\Gamma\!\left(\frac{\mu+d+1}{2}+2\kappa\right)}
{\Gamma\!\left(\frac{\mu}{2}\right)
 \Gamma\!\left(\frac{\mu+d}{2}+\kappa\right)}.
\end{equation}
\end{theorem}

Assuming a common variance and imposing the relation $\nu = \kappa + \frac{1}{2}$,
the equivalence condition~(\ref{cafu}) is equivalently expressed as
\[
a = \alpha\, F(\kappa,\mu,d)^{\frac{1}{1+2\kappa}}
:= b(\alpha,\mu,\kappa,d).
\]

{
Although the Gaussian-measure equivalence result above assumes
$\mu>1+d/2$, the mapping $b(\alpha,\mu,\kappa,d)$ is well defined throughout
the validity region $\mu\ge1$. We use it below as a Mat\'ern-limit
reparameterization of the full $\mathcal H$ family.
}

Using this compact-support reparameterization and
following the same  arguments in \cite{bevilacqua2022unifying},
one obtains
\begin{equation}\label{popi}
\lim_{\mu\to\infty}
{\cal H}_{\kappa,\mu,b(\alpha,\mu,\kappa,d),d}(x)
=
{\cal MT}_{\kappa+\frac{1}{2},\alpha}(x),
\quad \kappa>-\frac{1}{2},
\end{equation}
with uniform convergence on $[0,\infty)$.

Thus the reparameterized $\mathcal H$ family is a compactly supported
extension of Mat\'ern: for fixed $(\kappa,\alpha)$, $\mu$ controls the support
and hence the sparsity--approximation trade-off, while \eqref{popi} recovers
Mat\'ern as $\mu\to\infty$. This is analogous to the result obtained for the
${\cal GW}$ family by \cite{bevilacqua2022unifying}; see
Equation~\eqref{cpoo}.

{
The bridge also quantifies a practical sparsity difference between $\mathcal H$
and ${\cal GW}$. For a fixed limiting Mat\'ern scale $\alpha$ and regularity
$\nu=\kappa+1/2$, the following result compares the shortest supports available
at the respective validity boundaries.

\begin{proposition}\label{prop:bridge-support}
Fix $\alpha>0$ and $\kappa>-1/2$, and let $p=1+2\kappa$. The support functions
$b(\alpha,\mu,\kappa,d)$ and $g(\alpha,\mu,\kappa)$ are strictly increasing in
$\mu$. If $d\ge2$, or if $d=1$ and $\kappa\ge0$, their minimum valid values
therefore satisfy
\begin{equation}\label{eq:bridge-support-order}
 b(\alpha,1,\kappa,d)
 \leq
 g\!\left(\alpha,\kappa+\frac{d+1}{2},\kappa\right),
\end{equation}
with equality only when $d=1$ and $\kappa=0$. Moreover, for fixed $\kappa$ and
$d\to\infty$,
\begin{align}
 b(\alpha,1,\kappa,d)
 &\sim
 \alpha\sqrt{2d}\left\{\frac{\Gamma(\kappa+1)}{\sqrt{\pi}}\right\}^{1/p},
 \label{eq:H-bridge-asymptotic}\\
 g\!\left(\alpha,\kappa+\frac{d+1}{2},\kappa\right)
 &\sim \frac{\alpha d}{2}.
 \label{eq:GW-bridge-asymptotic}
\end{align}
Consequently, the ratio of the minimum $\mathcal H$ and ${\cal GW}$ bridge supports is
of order $d^{-1/2}$.
\end{proposition}

Thus, within the two Mat\'ern-limit bridges, $\mathcal H$ can start from a
shorter valid compact support while preserving the same limiting Mat\'ern scale
and smoothness. For a fixed sampling design this can translate into more exact
zeros in the covariance matrix, and the distinction becomes increasingly
pronounced in higher-dimensional input spaces. The statement concerns support
and sparsity within the bridge parameterizations; it does not imply uniform
predictive superiority of $\mathcal H$ over ${\cal GW}$.
}

\section{Asymptotic properties of the maximum likelihood estimator
of the Hypergeometric model}\label{44}

We study the asymptotic properties of likelihood-based estimators under
fixed-domain (infill) asymptotics for the model
$\sigma^2\mathcal{H}_{\kappa,\mu,a,d}$, with $\kappa$ fixed and known.

Using Theorem~\ref{Wv_vs_Wv} and the parameterization~\eqref{repp} with $l=d/2+\kappa$,
the equivalence condition for two Gaussian measures
$P(\sigma_i^2\mathcal{H}_{\kappa,\mu_i,a_i,d})$, $i=0,1$, sharing the same $\kappa$,
for $d=1,2,3$ and $\min(\mu_0,\mu_1)> 1+d/2$, is
\begin{equation}\label{conditionPO}
\frac{\sigma_0^2\,F(\kappa,\mu_0,d)}{a_0^{2\kappa+1}}
=
\frac{\sigma_1^2\,F(\kappa,\mu_1,d)}{a_1^{2\kappa+1}}.
\end{equation}

Hence, under fixed-domain (infill) asymptotics, the individual parameters
$\sigma^2$, $\mu$ and $a$ are not consistently identifiable; only the
\emph{microergodic parameter}
\begin{equation}\label{micro}
\theta(\sigma^2,\mu,a)
:=
\frac{\sigma^2\,F(\kappa,\mu,d)}{a^{2\kappa+1}}
\end{equation}
is consistently estimable.
In our fixed-domain limit theory we assume $\mu > 1+d/2$ for $d=1,2,3$,
which is the condition ensuring equivalence of Gaussian measures (see  Theorem~\ref{Wv_vs_Wv}).
This requirement is stronger than the mere validity condition for the
$\mathcal H$ model, which is simply $\mu \ge 1$.

Let $D\subset\mathbb{R}^d$ be a bounded domain and let
$S_n=\{\mathbf{s}_1,\ldots,\mathbf{s}_n\}\subset D$ be an infill design, i.e.\
$S_n$ becomes dense in $D$ as $n\to\infty$. Let
$\mathbf{Z}_n=(Z(\mathbf{s}_1),\ldots,Z(\mathbf{s}_n))^\top$
be a realization of a zero-mean stationary Gaussian RF with covariance function
$\sigma^2\mathcal{H}_{\kappa,\mu,a,d}(\|\cdot\|)$.
Write
\[
R_n(\mu,a)
:=
\bigl[\mathcal{H}_{\kappa,\mu,a,d}
(\|\mathbf{s}_i-\mathbf{s}_j\|)\bigr]_{i,j=1}^n
\]
for the $n\times n$ correlation matrix, and
\begin{equation}\label{eq:17}
\ell_n(\sigma^2,\mu,a)
:=
-\frac{1}{2}
\Bigl(
n\log(2\pi\sigma^2)
+\log\lvert R_n(\mu,a)\rvert
+\frac{1}{\sigma^2}
\mathbf{Z}_n^\top R_n(\mu,a)^{-1}\mathbf{Z}_n
\Bigr)
\end{equation}
for the Gaussian log-likelihood. Throughout this section $\kappa$ is fixed and
known, while $\mu$ and $a$ may be either fixed or estimated as specified below.

For any $(\mu,a)$, the profile ML estimator of $\sigma^2$ is
\begin{equation}\label{hoceini}
\hat{\sigma}_n^2(\mu,a)
:=
\frac{\mathbf{Z}_n^\top R_n(\mu,a)^{-1}\mathbf{Z}_n}{n},
\end{equation}
and the corresponding plug-in estimator of the microergodic parameter~\eqref{micro} is
\begin{equation}\label{thetahat}
\hat{\theta}_n(\mu,a)
:=
\theta\!\bigl(\hat{\sigma}_n^2(\mu,a),\mu,a\bigr)
=
\frac{\hat{\sigma}_n^2(\mu,a)\,F(\kappa,\mu,d)}{a^{2\kappa+1}}.
\end{equation}
The profile log-likelihood obtained by substituting~\eqref{hoceini} into~\eqref{eq:17} is
\begin{equation}\label{eq:prof}
\mathcal{PL}_n(\mu,a)
:=
-\frac{1}{2}
\Bigl(
n\log\bigl(2\pi\hat{\sigma}_n^2(\mu,a)\bigr)
+\log\lvert R_n(\mu,a)\rvert
+n
\Bigr).
\end{equation}

When $\mu$ is fixed, $F(\kappa,\mu,d)$ is known, so the microergodic
parameter is equivalently $\sigma^2/a^{2\kappa+1}$ up to this constant. We keep
the notation \eqref{micro} throughout.

The following result establishes strong consistency and asymptotic normality of
$\hat{\theta}_n(\mu,a)$ when both $\mu$ and $a$ are treated as fixed.
The proof is provided in the Supplementary Material.

\begin{theorem}\label{theo10}
Let  $\{Z(\bs): \bs \in D\}$,  $ D\subset\mathbb{R}^d$, $d=1,2,3$, be a zero-mean Gaussian RF
with covariance $\sigma_0^2\,\mathcal{H}_{\kappa,\mu,a_0,d}$, where $\kappa$ and $\mu >  1+d/2$
are fixed and known, and $(\sigma_0^2,a_0)\in(0,\infty)^2$.
Then, for any fixed $a>0$, as $n\to\infty$:
\begin{enumerate}
\item[\rm(i)]
$\hat{\theta}_n(\mu,a)\xrightarrow{a.s.}\theta_0
:=\sigma_0^2\,F(\kappa,\mu,d)/a_0^{2\kappa+1}$;
\item[\rm(ii)]
$\sqrt{n}\,\bigl(\hat{\theta}_n(\mu,a)-\theta_0\bigr)
\xrightarrow{\mathcal{D}}
\mathcal{N}\bigl(0,\,2\theta_0^2\bigr)$.
\end{enumerate}
\end{theorem}

We now estimate $a$ by ML over the compact set $I=[a_L,a_U]$:
\begin{equation}\label{ahat_fixed}
\hat{a}_n
:=
\arg\max_{a\in I}\,\mathcal{PL}_n(\mu,a),
\quad 0<a_L<a_U<\infty,
\end{equation}
Uniform control of $\hat\theta_n(\mu,a)$ over $a\in I$, established in the
Supplementary Material, yields the following result.

\begin{theorem}\label{theo11_new}
Under the same assumptions as Theorem~\ref{theo10}, and $a_0\in I$,
let $\hat a_n$ be as in~\eqref{ahat_fixed}. Then, as $n\to\infty$:
\begin{enumerate}
\item[\rm(i)]
$\hat{\theta}_n(\mu,\hat{a}_n)\xrightarrow{p}\theta_0$;
\item[\rm(ii)]
$\sqrt{n}\bigl(\hat{\theta}_n(\mu,\hat{a}_n)-\theta_0\bigr)
\xrightarrow{\mathcal{D}} \mathcal{N}(0,2\theta_0^2)$.
\end{enumerate}
\end{theorem}

We finally allow $\mu$ to be unknown and estimate $(\mu,a)$ jointly by ML over the
compact set $J\times I$, where $J=[\mu_L,\mu_U]$ and $\mu_L> 1+d/2$:
\begin{equation}\label{joint_mle}
(\hat{\mu}_n,\hat{a}_n)
:=
\arg\max_{(\mu,a)\in J\times I}\,\mathcal{PL}_n(\mu,a).
\end{equation}
The same uniform argument over $J\times I$ gives:

\begin{theorem}\label{theo12_new}
Under the same fixed-domain setting as Theorem~\ref{theo10}, let the true
covariance be $\sigma_0^2\mathcal H_{\kappa,\mu_0,a_0,d}$, where $\kappa$ is
fixed and known,
$(\sigma_0^2,\mu_0,a_0)\in(0,\infty)\times J\times I$, and
$\mu_0>1+d/2$. Let $(\hat{\mu}_n,\hat{a}_n)$ be as in~\eqref{joint_mle}.
Then, as $n\to\infty$:
\begin{enumerate}
\item[\rm(i)]
$\hat{\theta}_n(\hat{\mu}_n,\hat{a}_n)\xrightarrow{p}\theta_0$;
\item[\rm(ii)]
$\sqrt{n}\,\bigl(\hat{\theta}_n(\hat{\mu}_n,\hat{a}_n)-\theta_0\bigr)
\xrightarrow{\mathcal{D}}
\mathcal{N}\bigl(0,\,2\theta_0^2\bigr)$.
\end{enumerate}
\end{theorem}

Theorems~\ref{theo11_new}--\ref{theo12_new} show that estimating nuisance
covariance parameters over compact sets does not alter the fixed-domain limit
theory. Their proofs use uniform spectral envelopes to obtain a direct
sandwich for the profile quadratic forms, followed by the usual chi-square
limit at the true covariance parameters. This avoids the stronger $\mu$
restriction required by the parameter-monotonicity argument used for
generalized Wendland models in \citet{BFFP}.

{
The fixed-domain theory assumes that $\kappa$ is known. This is not specific to
the proposed family: the classical Mat\'ern fixed-domain likelihood theory for
the microergodic parameter likewise treats the smoothness parameter as fixed
\citep{Zhang:2004}. Allowing $\kappa$ to vary changes both the high-frequency
spectral decay and the microergodic normalization. Profiling over a plausible
grid of $\kappa$ values, or selecting among them by a holdout predictive score,
is a practical model-selection strategy. Subsequent plug-in uncertainty
assessments treat the selected smoothness as fixed; the present theorems do
not account for smoothness-selection uncertainty or establish post-selection
coverage. A general joint likelihood theory for the microergodic parameter
and unknown smoothness remains an important topic for future work.
}

\section{Concluding Remarks}\label{sec8}

We revisited the Gauss hypergeometric covariance class
\citep{emery2022gauss}, sharpening its validity conditions and separating
structural non-identifiability from fixed-domain equivalence. An ordering
constraint removes the former, while the latter identifies the relevant
microergodic parameter.

Within this formulation, the proposed $\mathcal H$ family provides a
parsimonious compactly supported subclass selected by an integral-range
criterion. It combines simple validity conditions, continuous smoothness,
scale-mixture representations, and links to Mat\'ern through compact-support
reparameterizations. {In the regime where the ${\cal GW}$ validity boundary is exact,
the dimension-free validity boundary of $\mathcal H$ allows stronger
large-scale dependence at a fixed support, while the Mat\'ern-bridge comparison
shows that $\mathcal H$ can attain shorter valid supports, particularly as
dimension grows.} For fixed smoothness, its microergodic parameter remains consistent and
asymptotically normal when scale and shape nuisance parameters are estimated
over compact sets. The simulations and climate application show competitive fit
and prediction together with sparse covariance computations.

Open problems include the remaining validity gap, fixed-domain inference with
unknown smoothness, and scalable estimation, as well as anisotropic,
nonstationary, and multivariate extensions. The model ${\cal H}$ is implemented in the
\texttt{GeoModels} package \citep{geomodels}.


\section*{Acknowledgements}
This work was funded by the National Agency for Research and Development of Chile, under grants ANID FONDECYT 1240308 (M. Bevilacqua), ANID FONDECYT 1250862 (C. Caamaño-Carrillo) and ANID PIA AFB230001 (X. Emery).

\par


\bibhang=1.7pc
\bibsep=2pt
\fontsize{9}{14pt plus.8pt minus .6pt}\selectfont
\renewcommand\bibname{\large \bf References}
\expandafter\ifx\csname
natexlab\endcsname\relax\def\natexlab#1{#1}\fi
\expandafter\ifx\csname url\endcsname\relax
  \def\url#1{\texttt{#1}}\fi
\expandafter\ifx\csname urlprefix\endcsname\relax\def\urlprefix{URL}\fi

  \bibliographystyle{chicago}      
  \bibliography{sample}   

\vskip .65cm
\noindent
Moreno Bevilacqua,  Facultad de Ingenier\'ia y Ciencias, Universidad Adolfo Ib\'a\~nez, Vi\~na del Mar, Chile.
Dipartimento di Scienze Ambientali, Informatica e Statistica, Universit\`a  Ca' Foscari, Venezia, Italy.
\vskip 2pt
\noindent
E-mail: (moreno.bevilacqua@uai.cl)
\vskip 2pt

\noindent
Christian Caama\~no-Carrillo, Departamento de Estad\'istica, Universidad del B\'io-B\'io, Concepci\'on, Chile.
\vskip 2pt
\noindent
E-mail: (chcaaman@ubiobio.cl)
\vskip 2pt

\noindent
Tarik Faouzi
Departamento de Matem\'atica y Ciencia de la Computaci\'on, Universidad de Santiago de  Chile, Santiago, Chile
\vskip 2pt
\noindent
E-mail: (tarik.faouzi@usach.cl)
\vskip 2pt

\noindent
Xavier Emery
Department of Mining Engineering, Universidad de Chile, Santiago, Chile.
\vskip 2pt
\noindent
E-mail: (xemery@uchile.cl)
\vskip 2pt


\clearpage

\renewcommand{\baselinestretch}{2}

\markright{ \hbox{\footnotesize\rm Statistica Sinica: Supplement
}\hfill\\[-13pt]
\hbox{\footnotesize\rm
}\hfill }

\markboth{\hfill{\footnotesize\rm Moreno Bevilacqua, Christian Caama\~no-Carrillo Tarik Faouzi and Xavier Emery} \hfill}
{\hfill {\footnotesize\rm Covariance Models
in the Gauss Hypergeometric Class} \hfill}

\renewcommand{\thefootnote}{}
$\ $\par \fontsize{12}{14pt plus.8pt minus .6pt}\selectfont


 \centerline{\large\bf Parsimonious Compactly Supported Covariance Models}
\vspace{2pt}
 \centerline{\large\bf in the Gauss Hypergeometric Class: Identifiability,}
\vspace{2pt}
 \centerline{\large\bf Reparameterizations, and Asymptotic Properties}
\vspace{.25cm}
 \author{Bevilacqua, Christian Caama\~no-Carrillo Tarik Faouzi and Xavier Emery}
\vspace{.4cm}
\centerline{Moreno Bevilacqua$^{1,2}$, Christian Caama\~no-Carrillo$^{3}$, Tarik Faouzi$^{4}$ and Xavier Emery$^{5}$} 
\vspace{.4cm}
 \centerline{\it $^{1}$Universidad Adolfo Ib\'a\~nez, \\  $^{2}$Universitá  Ca’ Foscari, \\ $^{3}$Universidad del B\'io-B\'io}
\vspace{2pt}
\centerline{\it 
$^{4}$Universidad de Santiago de  Chile, \\$^{5}$Universidad de Chile}
\vspace{.55cm}
 \centerline{\bf Supplementary Material}
\vspace{.55cm}
\fontsize{9}{11.5pt plus.8pt minus .6pt}\selectfont
\par

\setcounter{section}{0}
\setcounter{equation}{0}
\setcounter{figure}{0}
\setcounter{table}{0}
\setcounter{lemma}{0}
\setcounter{corollary}{0}
\setcounter{proposition}{0}
\setcounter{definition}{0}
\setcounter{example}{0}
\setcounter{remark}{0}
\def\theequation{S\arabic{section}.\arabic{equation}}
\def\thesection{S\arabic{section}}

\fontsize{12}{14pt plus.8pt minus .6pt}\selectfont

\section{A simulation study}\label{sec6}

We study the finite-sample behavior of ML estimation under the proposed model
$\sigma^2 \mathcal{H}_{\kappa,\mu,a,d}$.
We generate $500$ independent realizations of a zero-mean Gaussian random field
on $[0,1]^2$ ($d=2$) with $\sigma^2=1$, $\kappa\in\{0,1,2\}$, and $\mu\in\{4,8\}$.
Each realization is simulated via Cholesky decomposition at $n=2500$ locations
randomly sampled uniformly over the unit square.

Throughout, we treat $\kappa$ as known and focus on ML estimation of
$\sigma^2$, $a$, $\mu$, and the microergodic quantity
$\sigma^2F(\kappa,\mu,d)/a^{2\kappa+1}$.
To create a gradual transition from increasing-domain to fixed-domain behavior
while keeping the sampling design fixed, we follow
\cite{Zhang:Zimmerman:2005,Shaby:Kaufmann:2013} and vary the scale parameter $a$.
Specifically, we consider $a\in\{0.1,0.3,0.6,0.9\}$.
Here $a=0.1$ corresponds to weak dependence relative to the unit-square domain
and thus approximates an increasing-domain regime, whereas $a=0.9$
induces strong dependence and is much closer to a fixed-domain setting.

Table~\ref{tabs19_new} reports the empirical bias and standard deviation (in parentheses)
of the $500$ ML estimates $\widehat{\sigma}^2_i$, $\widehat{a}_i$, $\widehat{\mu}_i$ and
$\widehat{\sigma}^2_iF(\kappa,\widehat{\mu}_i,d)/\widehat{a}_i^{2\kappa+1}$, $i=1,\ldots,500$.
The main pattern is clear: the individual covariance parameters
$\sigma^2$, $a$, and $\mu$ become more variable as $a$ increases,
whereas the microergodic parameter behaves in the opposite way.

\begin{table}[H]
\begin{center}
\caption{Bias and standard deviation (in parentheses) of ML estimates
$\widehat{\sigma}^2_i$, $\widehat{a}_i$,$\widehat{\mu}_i$, and $\widehat{\sigma}^2_iF(\kappa,\widehat{\mu}_i,d)/\widehat{a}_i^{2\kappa+1}$ for $i=1,\ldots,500$ when estimating a zero-mean Gaussian RF with covariance
$\sigma^2\mathcal{H}_{\kappa,\mu,a,2}$ on $[0,1]^2$ ($n=2500$ locations), with $\sigma^2=1$, $\kappa\in\{0,1,2\}$, $\mu\in\{4,8\}$, and $a\in\{0.1,0.3,0.6,0.9\}$.}\label{tabs19_new}
\begingroup
\renewcommand{\arraystretch}{0.78}
\scalebox{0.70}{
\begin{tabular}{cc|cc|cc|cc|}
\cline{3-8}
 &  & \multicolumn{2}{c|}{$\kappa=0$} & \multicolumn{2}{c|}{$\kappa=1$} & \multicolumn{2}{c|}{$\kappa=2$} \\ \hline
\multicolumn{1}{|c|}{} &  & \multicolumn{1}{c|}{$\mu=4$} & \multicolumn{1}{c|}{$\mu=8$} & \multicolumn{1}{c|}{$\mu=4$} & \multicolumn{1}{c|}{$\mu=8$} & \multicolumn{1}{c|}{$\mu=4$} & \multicolumn{1}{c|}{$\mu=8$}  \\ \hline

\multicolumn{1}{|c|}{\multirow{4}{*}{$\widehat{\sigma}^2$}} & \multicolumn{1}{c|}{\multirow{2}{*}{$a=0.1$}}  
& 0.00299 & 0.00107 & 0.00433 & 0.00184 & 0.00408 & 0.00212 \\
\multicolumn{1}{|c|}{} &
& (0.04138) & (0.03205) & (0.04914) & (0.03759) & (0.04794) & (0.03889) \\ \cline{2-8}

\multicolumn{1}{|c|}{} & \multicolumn{1}{c|}{\multirow{2}{*}{$a=0.3$}} 
& 0.01575 & 0.00504 & 0.01504 & 0.00770 & 0.01173 & 0.00726 \\
\multicolumn{1}{|c|}{} &
& (0.10407) & (0.05976) & (0.13271) & (0.08844) & (0.12980) & (0.09514) \\ \cline{2-8}

\multicolumn{1}{|c|}{} & \multicolumn{1}{c|}{\multirow{2}{*}{$a=0.6$}}  
& 0.06435 & 0.01181 & 0.04389 & 0.01385 & 0.03073 & 0.01553 \\
\multicolumn{1}{|c|}{} &
& (0.26849) & (0.11312) & (0.25614) & (0.16863) & (0.23655) & (0.18127) \\ \cline{2-8}

\multicolumn{1}{|c|}{} & \multicolumn{1}{c|}{\multirow{2}{*}{$a=0.9$}}  
& 0.12097 & 0.02437 & 0.09481 & 0.01848 & 0.05189 & 0.02400 \\
\multicolumn{1}{|c|}{} &
& (0.38703) & (0.17580) & (0.41788) & (0.24205) & (0.32244) & (0.25633) \\ \hline

\multicolumn{1}{|c|}{\multirow{4}{*}{$\widehat{a}$}} & \multicolumn{1}{c|}{\multirow{2}{*}{$a=0.1$}} 
& 0.02090 & 0.08978 & 0.00276 & 0.01635 & 0.00137 & 0.00612 \\
\multicolumn{1}{|c|}{} &
& (0.10465) & (0.20471) & (0.01397) & (0.06865) & (0.00827) & (0.02363) \\ \cline{2-8}

\multicolumn{1}{|c|}{} & \multicolumn{1}{c|}{\multirow{2}{*}{$a=0.3$}}   
& 0.81440 & 0.32342 & 0.02153 & 0.05872 & 0.00511 & 0.01092 \\
\multicolumn{1}{|c|}{} &
& (2.75327) & (0.75816) & (0.10263) & (0.26379) & (0.04194) & (0.06661) \\ \cline{2-8}

\multicolumn{1}{|c|}{} & \multicolumn{1}{c|}{\multirow{2}{*}{$a=0.6$}}  
& 3.91933 & 2.29317 & 0.55575 & 0.45367 & 0.04601 & 0.07026 \\
\multicolumn{1}{|c|}{} &
& (8.37251) & (4.53301) & (2.64211) & (1.50576) & (0.22104) & (0.33102) \\ \cline{2-8}

\multicolumn{1}{|c|}{} & \multicolumn{1}{c|}{\multirow{2}{*}{$a=0.9$}}  
& 7.46177 & 4.50776 & 1.66114 & 1.55866 & 0.16050 & 0.22433 \\
\multicolumn{1}{|c|}{} &
& (14.27313) & (7.73620) & (5.73980) & (4.14495) & (0.77112) & (0.80880) \\ \hline

\multicolumn{1}{|c|}{\multirow{4}{*}{$\widehat{\mu}$}} & \multicolumn{1}{c|}{\multirow{2}{*}{$a=0.1$}}  
& 1.00129 & 8.12410 & 0.17091 & 1.81217 & 0.10804 & 0.81383 \\
\multicolumn{1}{|c|}{} &
& (5.12648) & (18.51370) & (0.88371) & (7.52792) & (0.66348) & (3.13054) \\ \cline{2-8}

\multicolumn{1}{|c|}{} & \multicolumn{1}{c|}{\multirow{2}{*}{$a=0.3$}} 
& 12.17228 & 9.21399 & 0.42205 & 1.98131 & 0.12180 & 0.42181 \\
\multicolumn{1}{|c|}{} &
& (41.02159) & (21.40586) & (2.01811) & (8.89166) & (0.96416) & (2.54729) \\ \cline{2-8}

\multicolumn{1}{|c|}{} & \multicolumn{1}{c|}{\multirow{2}{*}{$a=0.6$}} 
& 29.22591 & 32.41524 & 5.42300 & 7.57716 & 0.55245 & 1.34734 \\
\multicolumn{1}{|c|}{} &
& (61.35913) & (63.46363) & (25.61297) & (24.98706) & (2.54928) & (6.27535) \\ \cline{2-8}

\multicolumn{1}{|c|}{} & \multicolumn{1}{c|}{\multirow{2}{*}{$a=0.9$}} 
& 36.69791 & 42.83158 & 10.93102 & 17.38350 & 1.29228 & 2.84537 \\
\multicolumn{1}{|c|}{} &
& (68.21306) & (72.15506) & (37.54848) & (45.95964) & (6.09577) & (10.03183) \\ \hline

\multicolumn{1}{|c|}{\multirow{4}{*}{$\frac{\widehat{\sigma}^2F(\kappa,\widehat{\mu},d)}{\widehat{a}^{2\kappa+1}}$}} & \multicolumn{1}{c|}{\multirow{2}{*}{$a=0.1$}}  
& 0.02847 & -0.15330 & 567.68210 & 7656.58100 & $1.40740\times10^{7}$ & $4.32344\times10^{8}$ \\
\multicolumn{1}{|c|}{} &
& (1.79328) & (5.08178) & (12695.11000) & (107860.90000) & ($2.18681\times10^{8}$) & ($3.79118\times10^{9}$) \\ \cline{2-8}

\multicolumn{1}{|c|}{} & \multicolumn{1}{c|}{\multirow{2}{*}{$a=0.3$}}  
& 0.00880 & -0.04215 & -0.59747 & -0.93738 & -2994.48500 & -48566.15000 \\
\multicolumn{1}{|c|}{} &
& (0.45226) & (0.95325) & (296.09960) & (1651.58800) & (448665.40000) & (4813284.00000) \\ \cline{2-8}

\multicolumn{1}{|c|}{} & \multicolumn{1}{c|}{\multirow{2}{*}{$a=0.6$}} 
& 0.00331 & -0.01499 & 0.66440 & -0.59339 & 165.89200 & -325.76500 \\
\multicolumn{1}{|c|}{} &
& (0.21584) & (0.43426) & (32.70379) & (167.22660) & (11788.42000) & (115466.10000) \\ \cline{2-8}

\multicolumn{1}{|c|}{} & \multicolumn{1}{c|}{\multirow{2}{*}{$a=0.9$}} 
& 0.00148 & -0.01088 & 0.05882 & -0.61039 & -4.95065 & -62.59252 \\
\multicolumn{1}{|c|}{} &
& (0.14205) & (0.28237) & (9.47927) & (46.79391) & (1482.45500) & (13819.55000) \\ \hline

\end{tabular}}
\endgroup
\end{center}
\end{table}

Table~\ref{tabs19_new} shows the expected fixed-domain pattern. As $a$ increases,
the individual estimates $\widehat a$ and $\widehat\mu$ become markedly more
variable (with $\widehat{\sigma}^2$ comparatively more stable), whereas the bias
and dispersion of the microergodic estimator
\[
\widehat{\theta}
=
\frac{\widehat{\sigma}^2F(\kappa,\widehat{\mu},d)}
{\widehat{a}^{2\kappa+1}}
\]
decrease. This contrast becomes more pronounced for larger $\kappa$, reflecting
the factor $a^{-(2\kappa+1)}$ in the identifiable combination.

{
To make the change in scale across the fixed-domain configurations easier to
read, Figure~\ref{fig:microergodic-log} displays the absolute bias and empirical
standard deviation of $\widehat{\theta}$ on a logarithmic vertical scale. The
plot uses the same Monte Carlo results reported in Table~\ref{tabs19_new} and
makes clear the strong reduction in both quantities as the support radius
increases.

\begin{figure}[H]
\centering
\begin{tabular}{cc}
\includegraphics[width=0.47\textwidth]{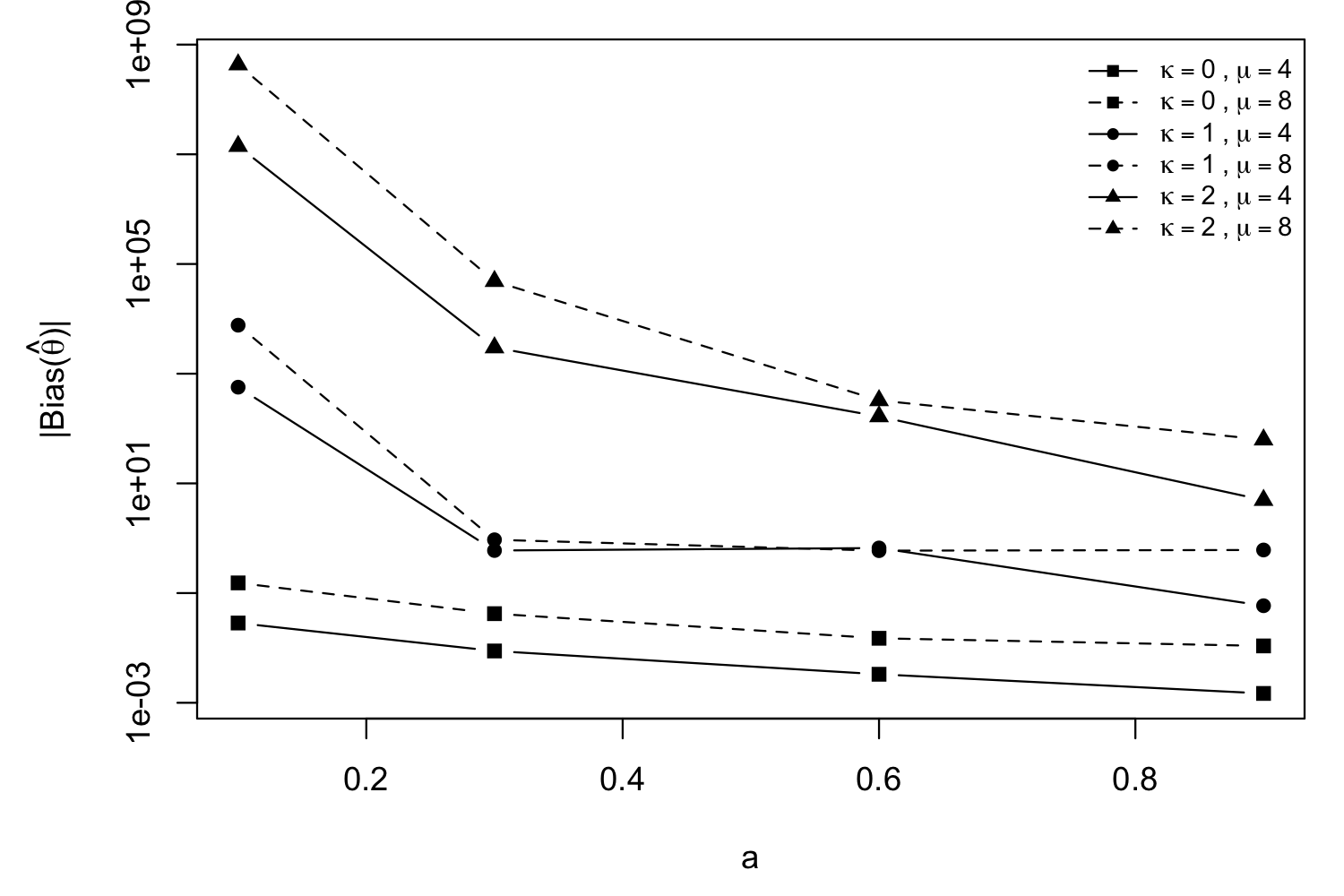} &
\includegraphics[width=0.47\textwidth]{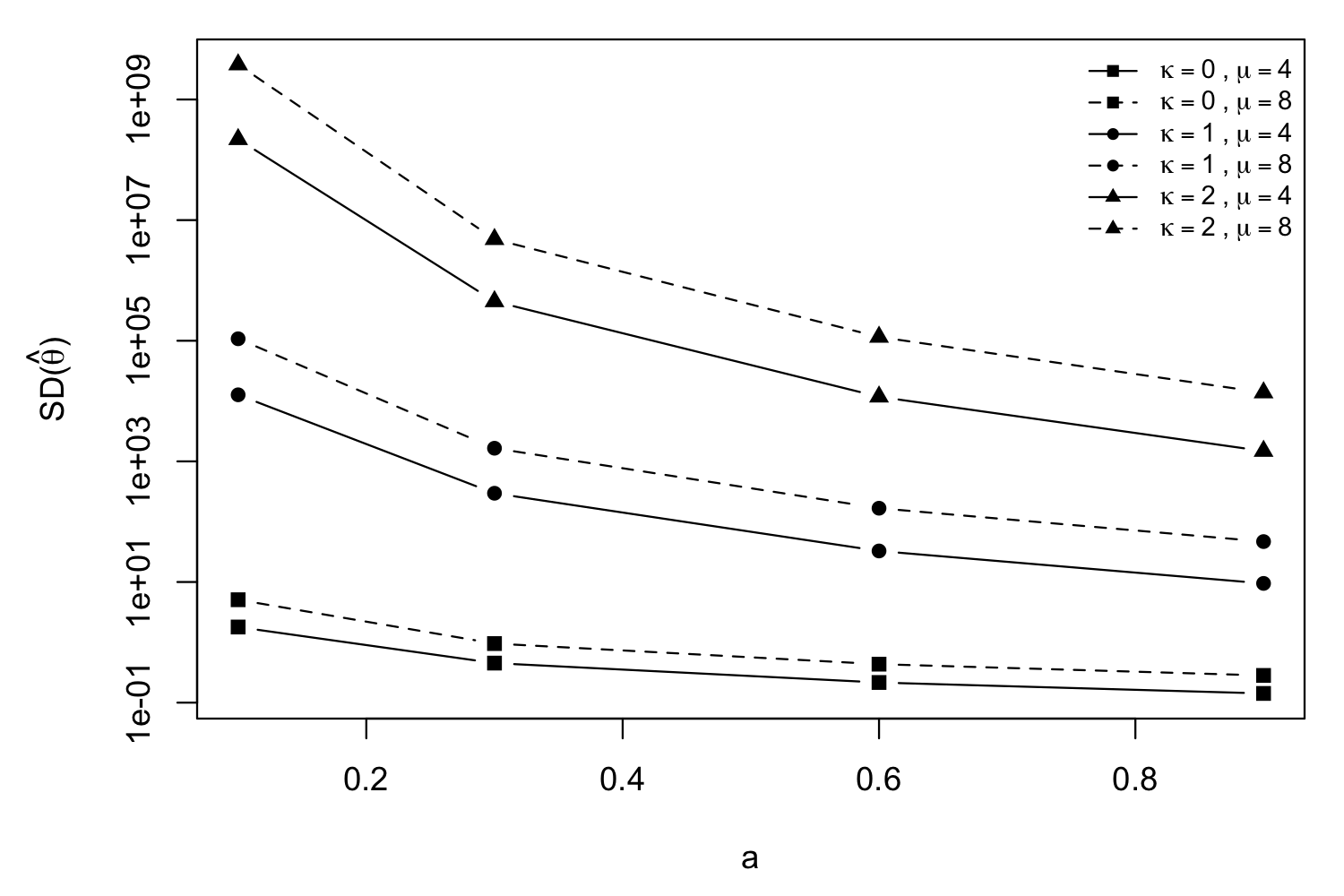}
\end{tabular}
\caption{Absolute bias (left) and empirical standard deviation (right) of the
microergodic estimator $\widehat{\theta}$ as functions of the support radius
$a$, shown on logarithmic vertical scales for the six $(\kappa,\mu)$
configurations in Table~\ref{tabs19_new}.}
\label{fig:microergodic-log}
\end{figure}
}

\begin{figure}[h!]
\begin{center}
\begin{tabular}{cc}
\includegraphics[width=5.9cm,height=6.2cm]{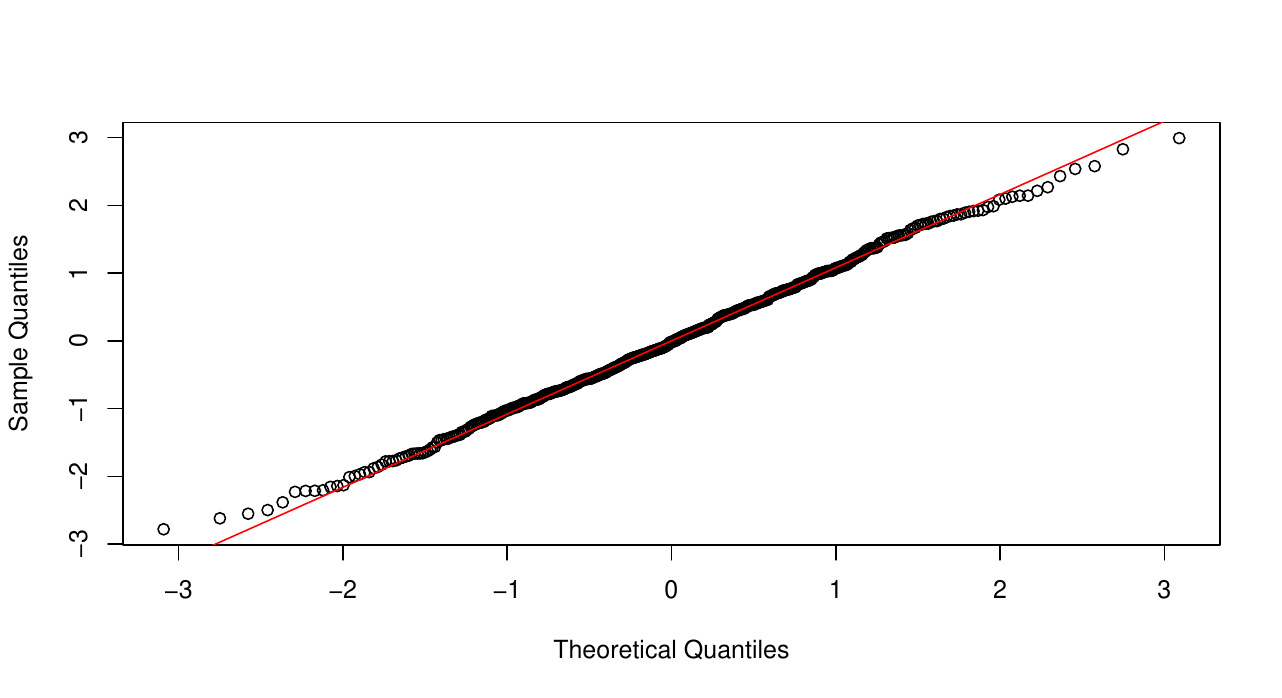}&\includegraphics[width=5.9cm,height=6.2cm]{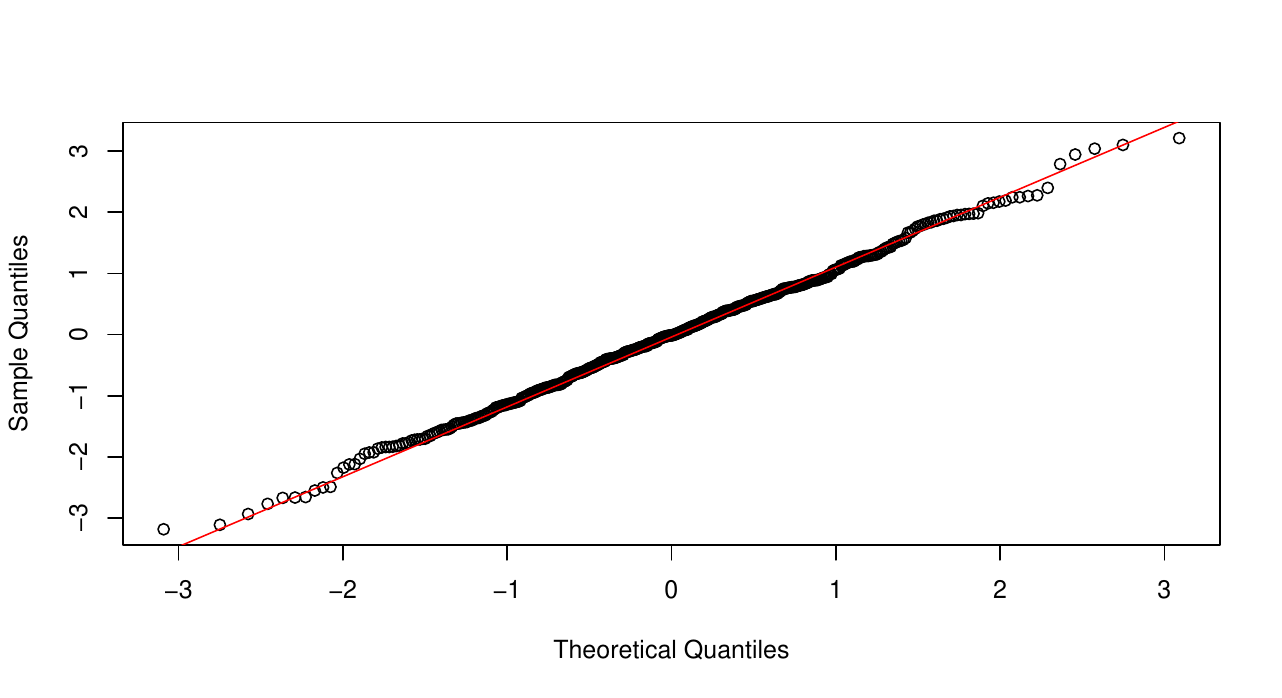}\\
($\mu=4$)&($\mu=8$)
\end{tabular}
\end{center}
\caption{Normal QQ-plots of standardized microergodic estimates in the fixed-domain  setting ($a=0.9$) with $\kappa=1$ and $\sigma^2=1$, comparing $\mu=4$ (left) and $\mu=8$ (right).
The standardized statistics are $Z_i=\sqrt{n}\,(\widehat{\theta}_i-\theta_0)/(\sqrt{2}\,\theta_0)$,
where $\widehat{\theta}_i=\widehat{\sigma}^2_iF(\kappa,\widehat{\mu}_i,d)/\widehat{a}_i^{2\kappa+1}$, $\theta_0=\sigma^2 F(\kappa,\mu,d)/a^{2\kappa+1}$, and $n=2500$.}
 \label{boxplots}
\end{figure}

Figure~\ref{boxplots} reports normal QQ-plots of the standardized
microergodic estimates for $a=0.9$ and $\kappa=1$, with $\mu=4$ and $8$.
The points align reasonably well with the standard normal reference line,
supporting the fixed-domain Gaussian approximation at this sample size.

\subsection{Additional robustness and comparative simulations}
\label{sec:additional-simulations}

{
We complement the preceding fixed-domain experiment with two targeted
simulations designed to assess robustness beyond the original sampling scheme
and to compare the proposed model with alternative covariance constructions.
Both experiments use $200$ Monte Carlo replications.

First, we consider a different spatial design. Starting from a regular
$41\times 41$ grid on $[0,1]^2$, each coordinate is independently perturbed
by a uniform random variable on $[-0.01,0.01]$, giving $n=1681$ locations.
Zero-mean Gaussian RFs are generated from
$\mathcal H_{\kappa,\mu,a,2}$ with
$\sigma^2=1$, $\kappa=1$, $\mu=4$, and compact-support radius $a=0.2$.
We compare $\mathcal H$, generalized Wendland (${\cal GW}$), Mat\'ern, and
tapered Mat\'ern covariance models. For the Mat\'ern specifications we fix
$\nu=\kappa+1/2=1.5$.

The $\mathcal H$ and ${\cal GW}$ support radii and the taper range are all
set equal to $0.2$. 
The tapered Mat\'ern covariance is obtained by multiplying the Mat\'ern
correlation by the Wendland $C^2$ taper
\[
T(r)=(1-r)^4_+(1+4r),
\]
which corresponds to the generalized Wendland correlation
${\cal GW}_{1,3,1}(r)$, where $r=x/0.2$, so that $T(r)=0$ for
$r\geq1$. Equivalently, as a function of the original distance $x$,
the taper is ${\cal GW}_{1,3,0.2}(x)$.

For $\mathcal H$ and ${\cal GW}$, the support radius and smoothness are fixed
at the common values specified above, and the remaining covariance parameters
are estimated by full maximum likelihood. For the dense Mat\'ern model,
$\nu=1.5$ is fixed and the remaining covariance parameters are estimated by
maximum likelihood, whereas for the tapered Mat\'ern model the Mat\'ern scale
is optimized with the variance profiled out. Prediction is assessed through
exact Gaussian leave-one-out quantities, conditional on the covariance
parameters estimated from the complete realization; the covariance parameters
are not re-estimated within each leave-one-out fit.

\begin{table}[H]
\centering
\caption{Additional Gaussian simulation on the jittered-grid design
($n=1681$, 200 replications). Entries are Monte Carlo means with empirical
standard deviations in parentheses, except for structural sparsity, which is
fixed by the common sampling design and support radius. The data-generating
covariance is $\mathcal H_{1,4,0.2,2}$ with $\sigma^2=1$.}
\label{tab:additional-gaussian}
\resizebox{\textwidth}{!}{
\begin{tabular}{lccccccc}
\hline
Model & RMSE & MAE & Log score & CRPS & Coverage$_{0.95}$ &
Structural zeros (\%) & Cholesky time (s)\\
\hline
$\mathcal H$
& 0.24610 (0.00633)
& 0.19132 (0.00475)
& -0.03429 (0.02486)
& 0.13535 (0.00332)
& 0.94985 (0.00486)
& 90.012
& 0.03291 (0.00230)\\

${\cal GW}$
& 0.24610 (0.00632)
& 0.19132 (0.00474)
& -0.03425 (0.02477)
& 0.13535 (0.00332)
& 0.95039 (0.00486)
& 90.012
& 0.03294 (0.00228)\\

Mat\'ern
& 0.24651 (0.00617)
& 0.19159 (0.00462)
& -0.03166 (0.02366)
& 0.13558 (0.00322)
& 0.95707 (0.00413)
& 0
& 0.72921 (0.02917)\\

Tapered Mat\'ern
& 0.24639 (0.00651)
& 0.19155 (0.00492)
& -0.03273 (0.02623)
& 0.13552 (0.00345)
& 0.94502 (0.00492)
& 90.012
& 0.03274 (0.00123)\\
\hline
\end{tabular}}
\end{table}

Table~\ref{tab:additional-gaussian} shows that the four covariance models have
essentially indistinguishable predictive performance under the alternative
sampling design. The main computational difference is between the sparse
models and the dense Mat\'ern covariance. By construction, $\mathcal H$,
${\cal GW}$, and tapered Mat\'ern have the same structural sparsity pattern,
with about $90\%$ zero off-diagonal covariance entries, and require about
$0.033$ seconds for Cholesky factorization, compared with about $0.73$ seconds
for the dense Mat\'ern covariance. Thus, in this experiment, the dense
Mat\'ern factorization is roughly $22$ times slower, without a corresponding
improvement in the predictive scores. Among the sparse alternatives,
$\mathcal H$, ${\cal GW}$, and tapered Mat\'ern exhibit nearly identical
prediction accuracy and factorization cost. This supports a competitive,
rather than uniformly superior, interpretation of the proposed model.

Second, we examine a non-Gaussian setting using the same jittered-grid design.
We generate Tukey-$h$ RFs with $\mathcal H_{1,4,0.2,2}$ correlation,
$\sigma^2=1$, and $h=0.15$. The Tukey-$h$ likelihood implemented in
\texttt{GeoModels} is fitted directly, with $\kappa=1$ fixed and
$a$, $\mu$, $\sigma^2$, and $h$ estimated jointly. Table~\ref{tab:tukey-sim}
reports the resulting Monte Carlo summaries.

\begin{table}[H]
\centering
\caption{Tukey-$h$ simulation on the jittered-grid design
($n=1681$, 200 replications).}
\label{tab:tukey-sim}
\begin{tabular}{lrrrr}
\hline
Parameter & True value & Mean & Bias & SD\\
\hline
$a$          & 0.20 & 0.20448 &  0.00448 & 0.04214\\
$\mu$        & 4.00 & 4.11748 &  0.11748 & 1.27105\\
$\sigma^2$   & 1.00 & 1.01364 &  0.01364 & 0.09124\\
$h$          & 0.15 & 0.14982 & -0.00018 & 0.01878\\
\hline
\end{tabular}
\end{table}

The Tukey-$h$ experiment shows small empirical biases for all four estimated
parameters. In particular, the estimator of the tail parameter is accurately
centered at its true value, while $\mu$ is the most variable of the estimated
covariance parameters. We use this experiment only as a finite-sample
robustness check for the non-Gaussian likelihood; the Gaussian fixed-domain
limit theory developed in the paper is not invoked for the Tukey-$h$ model.
}

\section{A real data application}\label{sec7}

\begin{figure}[h!]
\begin{center}
\begin{tabular}{cc}
\includegraphics[width=6.0cm,height=5.2cm]{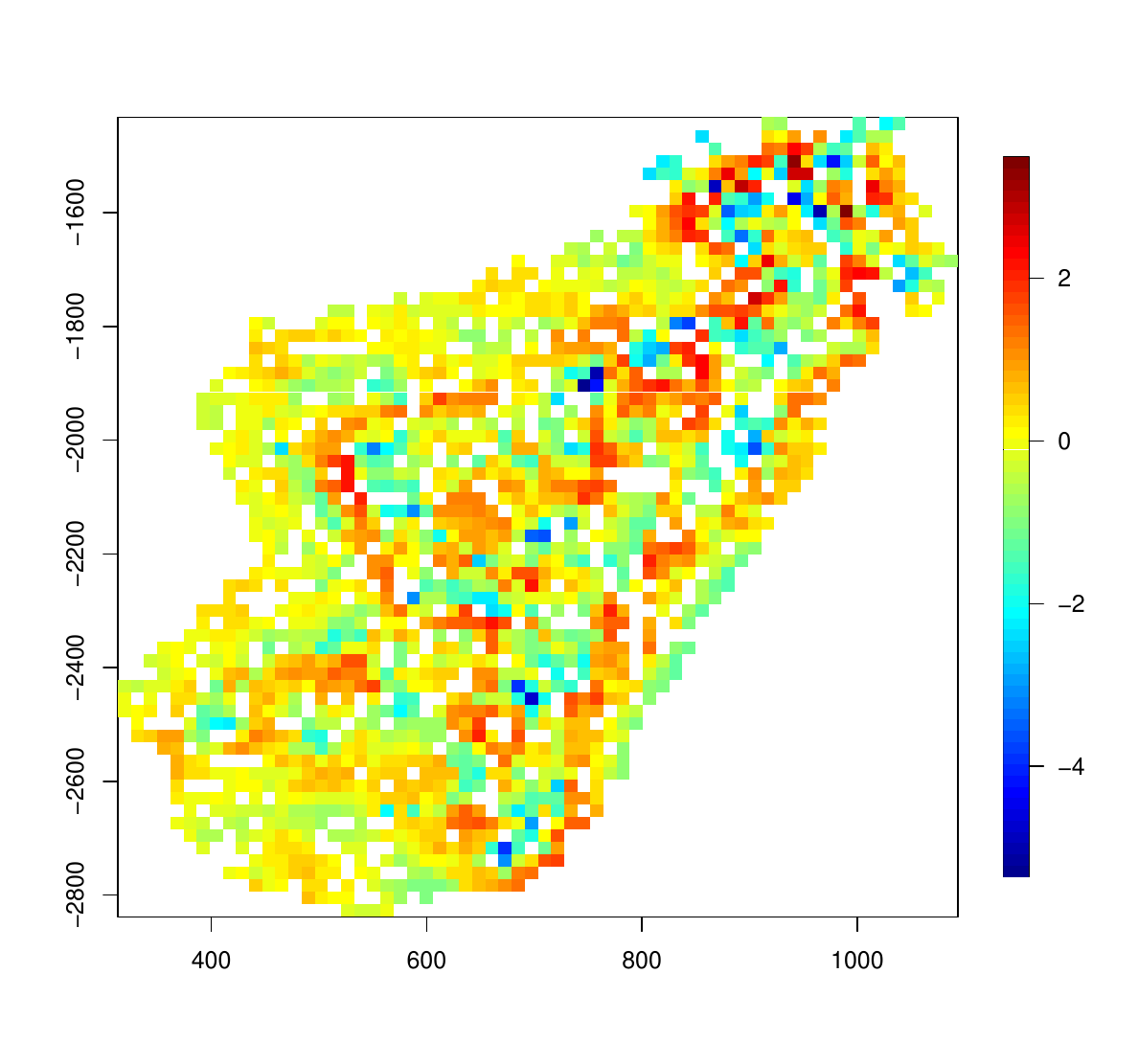}
&
\includegraphics[width=6.0cm,height=5.2cm]{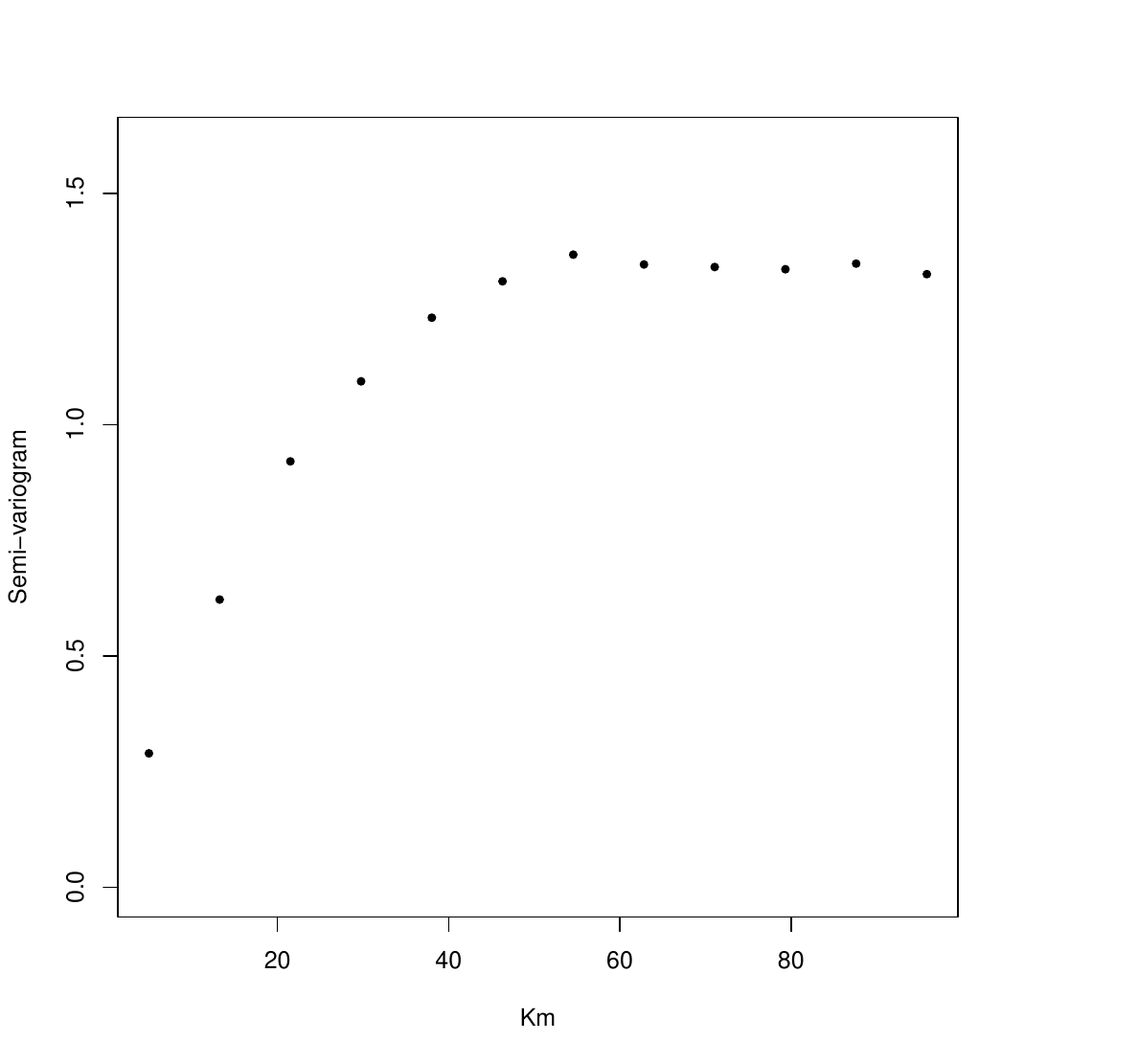}
\end{tabular}
\end{center}
\caption{
Spatial map of the GAM-detrended residuals and corresponding empirical
semivariogram for the Madagascar September mean-temperature data.
}
\label{caso}
\end{figure}

We analyze September mean temperature over Madagascar using WorldClim
\citep{FickHijmans2017}, available in \textsf{R} through \texttt{geodata}
\citep{geodata}. We restrict the analysis to locations south of
$13^\circ$S, yielding 725,330 grid cells, from which we randomly sample
3,500 locations using a fixed seed. Coordinates are projected to UTM.
Following \cite{Li:Zhang:2011}, we remove large-scale spatial variation
by spline regression and model the residuals as a zero-mean RF.
Figure~\ref{caso} shows the residual map and empirical semivariogram.

{
The application is intended as an illustrative comparison of covariance
models rather than as an exhaustive analysis of the complete WorldClim
field. The results below are therefore conditional on this reproducibly
selected 3,500-site subset and are not interpreted as establishing
subsample-invariant parameter estimates or model rankings.
}

{
Stationarity and isotropy are used as parsimonious working assumptions
for comparing covariance families. Directional empirical semivariograms
at $0^\circ$, $45^\circ$, $90^\circ$, and $135^\circ$ have similar
overall shapes, ranges, and sill levels (Figure~\ref{fig:dirvario}),
providing no clear evidence of substantial residual anisotropy in the
analyzed subsample. This graphical diagnostic does not rule out weaker
anisotropy or residual nonstationarity.
}

{
\begin{figure}[h!]
\centering
\includegraphics[width=0.66\textwidth]{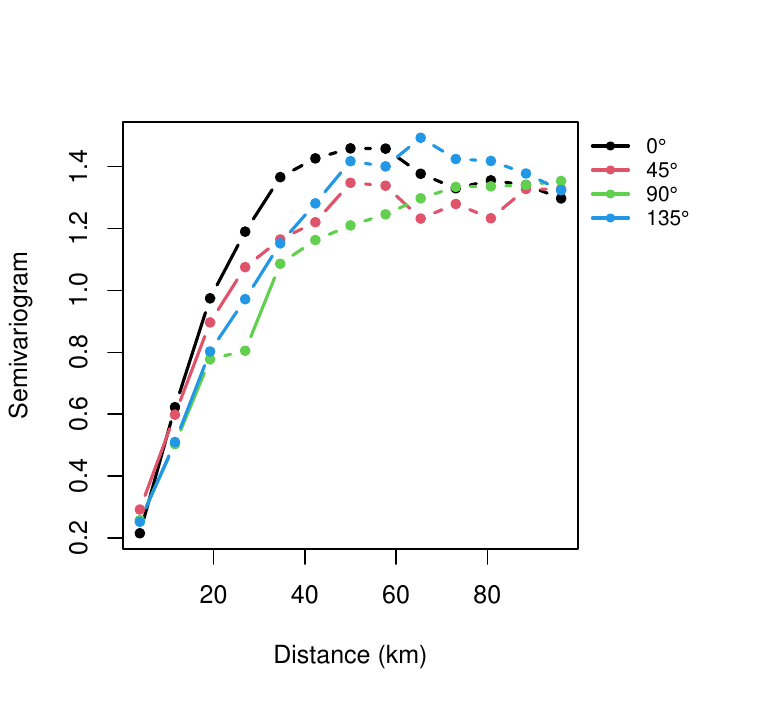}
\caption{
Directional empirical semivariograms of the GAM-detrended residuals
at $0^\circ$, $45^\circ$, $90^\circ$, and $135^\circ$.
}
\label{fig:dirvario}
\end{figure}
}

Let \(Z(\mathbf{s})\) be a standard Gaussian RF. We consider the Gaussian
model \(Y(\mathbf{s})=\sigma Z(\mathbf{s})\) and the Tukey-$h$
transformation \citep{Xu03072017,CAA},
\[
Y(\mathbf{s})
=
\sigma Z(\mathbf{s})
\exp\!\left\{\frac{hZ^2(\mathbf{s})}{2}\right\},
\qquad h\in[0,1/2).
\]
For the correlation of \(Z\), we compare
${\cal MT}_{\nu,a}$, ${\cal GW}_{\kappa,\mu,a}$, and
${\cal H}_{\kappa,\mu,a,2}$. Parameters are estimated by ML, with the
Tukey-$h$ likelihood evaluated through the Lambert-$W$ representation
in \cite{CAA}.

Table~\ref{tabsApli} summarizes the fitted models, repeated train--test
prediction, covariance sparsity, and Cholesky-factorization times.
Point-prediction errors are averages over 100 common 85/15 train--test
splits with residual-field parameters re-estimated in each training sample.
{
All predictive assessments condition on the GAM detrending fitted to the
full 3,500-site subset. They therefore assess prediction of this fixed
residual field, not the complete trend-plus-dependence procedure with the
GAM refitted within each training sample.
}

\begin{table}[htp!]
\centering
\caption{
Fitted parameters, likelihood criteria, repeated 85/15 train--test
prediction errors, covariance sparsity, and Cholesky times for the
Madagascar application. For ${\cal MT}$ the smoothness parameter is
$\nu$; for ${\cal GW}$ and $\mathcal H$ it is $\kappa$.
}
\label{tabsApli}
\resizebox{\linewidth}{!}{
\begin{tabular}{lrrrrrrrrrr}
\hline
Model
& $\widehat{\sigma}^2$
& $\widehat{\nu}/\widehat{\kappa}$
& $\widehat a$
& $\widehat\mu$
& $\widehat h$
& AIC
& RMSE
& MAE
& Zero (\%)
& Time
\\
\hline
Gaussian ${\cal MT}$
&1.2397&0.6115&15.4107&--&--&7995.96&0.67556&0.42031&0.00&6.647\\
Gaussian ${\cal GW}$
&1.2411&0.0363&89.6386&4.6270&--&7992.50&0.67495&0.41992&96.14&0.047\\
Gaussian $\mathcal H$
&1.2407&0.0371&91.8375&4.3004&--&7992.60&0.67491&0.41989&95.97&0.053\\
Tukey-$h$ ${\cal MT}$
&0.7402&0.5806&17.8254&--&0.1780&7593.69&0.67486&0.42078&0.00&6.826\\
Tukey-$h$ ${\cal GW}$
&0.7389&-0.0007&76.6675&3.2895&0.1797&7584.01&0.67411&0.42042&97.13&0.032\\
Tukey-$h$ $\mathcal H$
&0.7387&-0.0011&78.3018&2.9617&0.1796&7584.24&0.67406&0.42036&97.01&0.032\\
\hline
\end{tabular}
}
\end{table}

{
Smoothness estimates should be compared on a common regularity scale:
Mat\'ern uses $\nu$, whereas ${\cal GW}$ and $\mathcal H$ use $\kappa$,
with equivalent regularity $\nu=\kappa+1/2$. The resulting values are
$0.6115$, $0.5363$, and $0.5371$ for the Gaussian fits and
$0.5806$, $0.4993$, and $0.4989$ for the Tukey-$h$ fits. Thus all
models indicate a relatively rough residual field.

The Tukey-$h$ specifications are strongly favored by AIC, while
${\cal GW}$ and $\mathcal H$ give virtually identical likelihood and
prediction results. Point-prediction differences among all six models
are small: the Tukey-$h$ compactly supported models have slightly smaller
RMSE, whereas the Gaussian compactly supported models have slightly
smaller MAE. Both compactly supported families yield approximately
$96$--$97\%$ structural zeros and reduce Cholesky-factorization time by
more than two orders of magnitude relative to Mat\'ern.
}

{
To assess the predictive distribution beyond RMSE and MAE, we additionally
computed plug-in leave-one-out CRPS, logarithmic score, 95\% interval
score and coverage, and PIT diagnostics, also keeping the full-sample
residual-field ML estimates fixed. This additional diagnostic does not
propagate parameter or detrending uncertainty. For Tukey-$h$, the
conditional distribution is computed
on the latent Gaussian scale and transformed through the fitted Tukey-$h$
map. Results are reported in Table~\ref{tab:madaprobscores}.
}

\begin{table}[htp!]
\centering
\caption{
Distribution-sensitive plug-in leave-one-out scores. LogS is the negative
log predictive density; IS and coverage refer to 95\% predictive intervals.
Lower values are better for CRPS, LogS, and IS.
}
\label{tab:madaprobscores}
\begin{tabular}{lrrrrr}
\hline
Model & CRPS & LogS & IS & Coverage & PIT mean\\
\hline
Gaussian ${\cal MT}$      &0.32608&0.93646&4.09138&0.9311&0.5086\\
Gaussian ${\cal GW}$      &0.32585&0.93714&4.08831&0.9306&0.5086\\
Gaussian $\mathcal H$     &0.32585&0.93714&4.08845&0.9306&0.5086\\
Tukey-$h$ ${\cal MT}$     &0.32142&0.87703&4.19259&0.9329&0.5071\\
Tukey-$h$ ${\cal GW}$     &0.32146&0.87742&4.19094&0.9317&0.5074\\
Tukey-$h$ $\mathcal H$    &0.32146&0.87744&4.19095&0.9317&0.5074\\
\hline
\end{tabular}
\end{table}

{
The Tukey-$h$ models improve both CRPS and logarithmic score relative to
their Gaussian counterparts, although their interval scores are slightly
larger and 95\% coverage remains mildly below nominal. PIT means are close
to $0.5$ for all models. The PIT standard deviation increases from about
$0.237$ to $0.246$ under Tukey-$h$, toward the uniform reference
$1/\sqrt{12}\simeq0.289$, but this does not establish predictive calibration.
Thus the likelihood preference for Tukey-$h$ is reflected in important
aspects of distributional prediction, but not in every predictive criterion.

Across both point and distributional prediction, ${\cal GW}$ and
$\mathcal H$ remain practically indistinguishable. The application
therefore supports $\mathcal H$ as a competitive compactly supported
alternative rather than demonstrating empirical superiority over
${\cal GW}$; its distinguishing advantages remain the structural
properties developed in the paper.
}

\begin{figure}[h!]
\begin{center}
\begin{tabular}{cc}
\includegraphics[width=5.9cm,height=5.2cm]{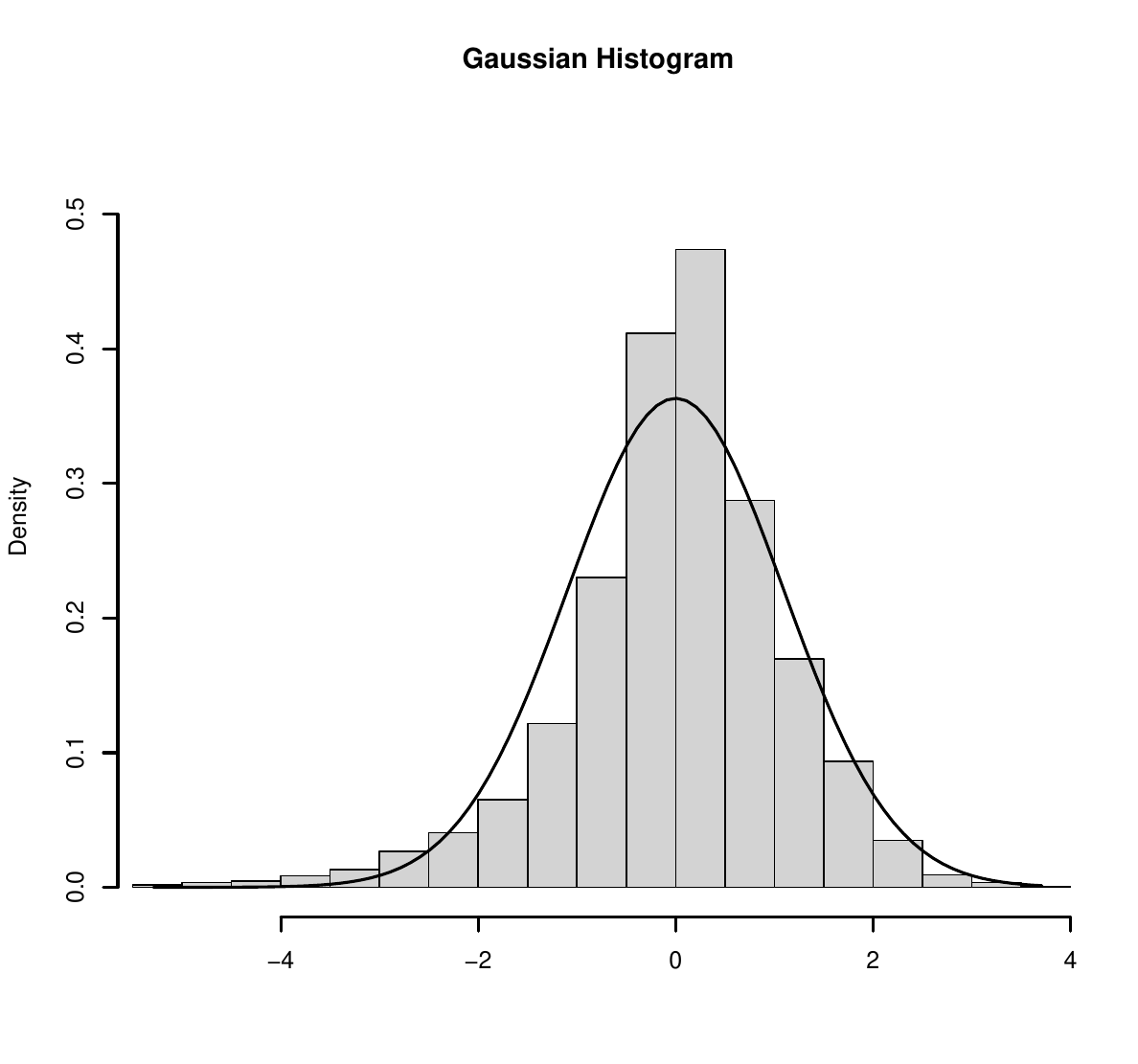}
&
\includegraphics[width=5.9cm,height=5.2cm]{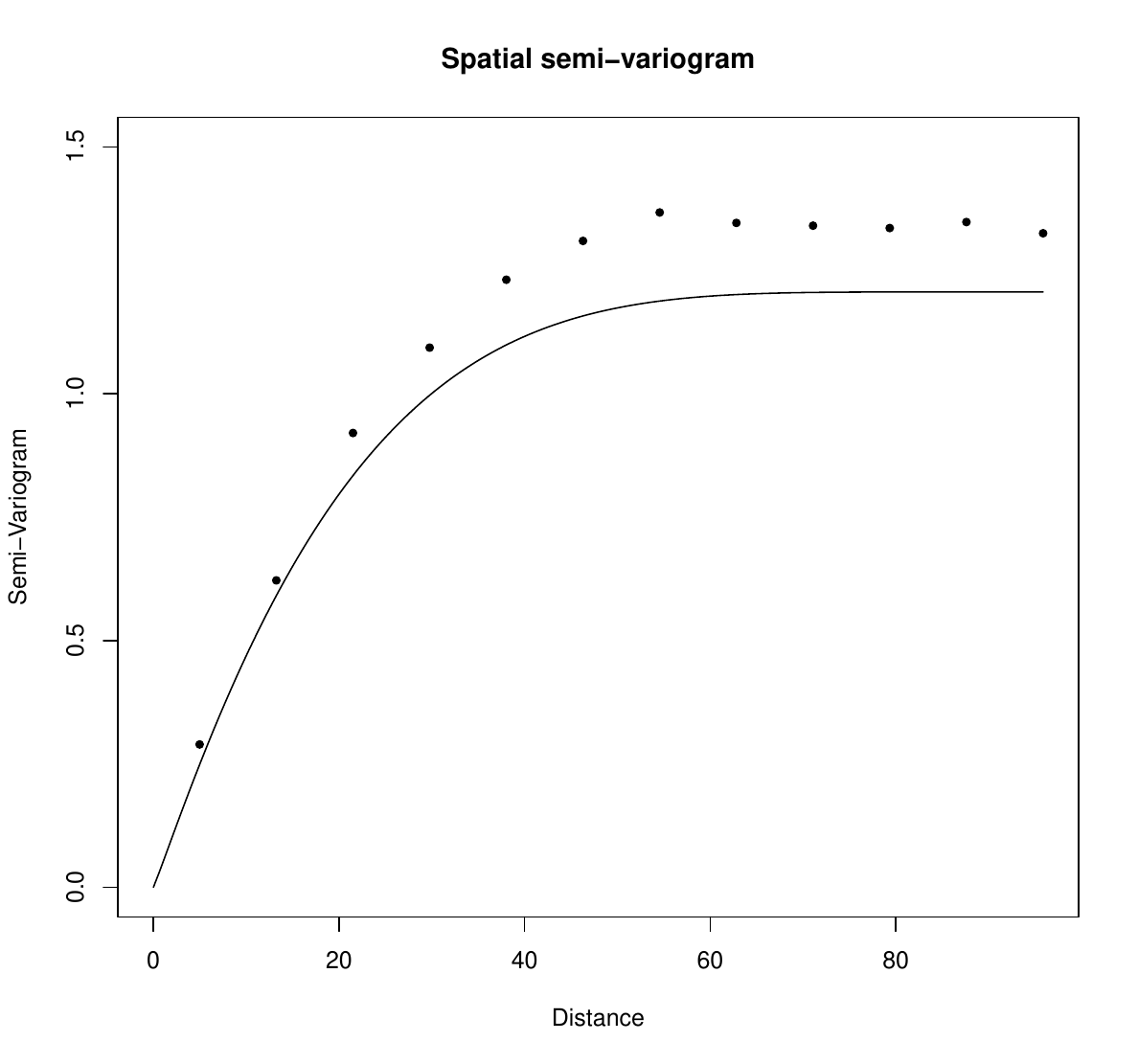}
\\
\includegraphics[width=5.9cm,height=5.2cm]{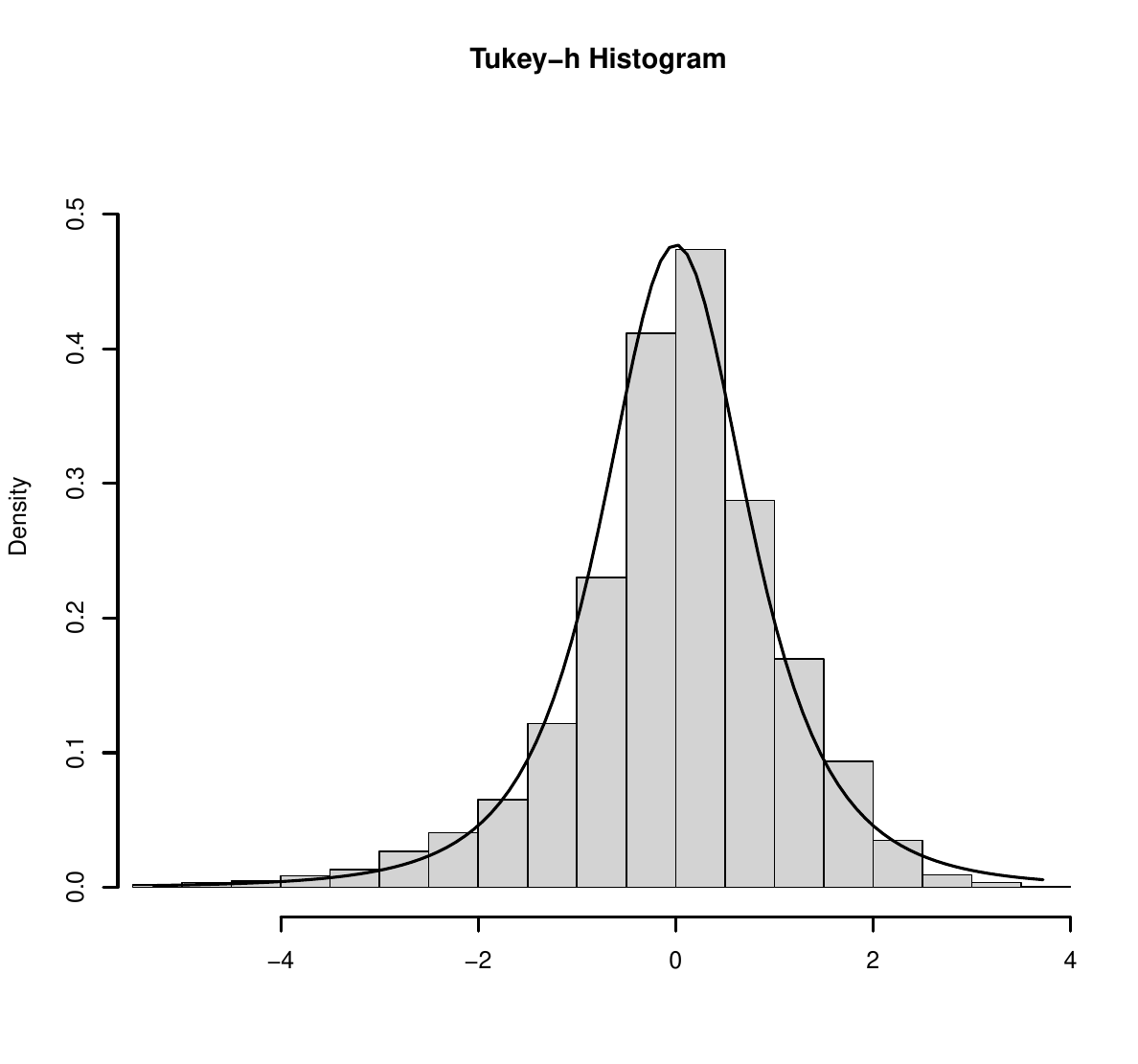}
&
\includegraphics[width=5.9cm,height=5.2cm]{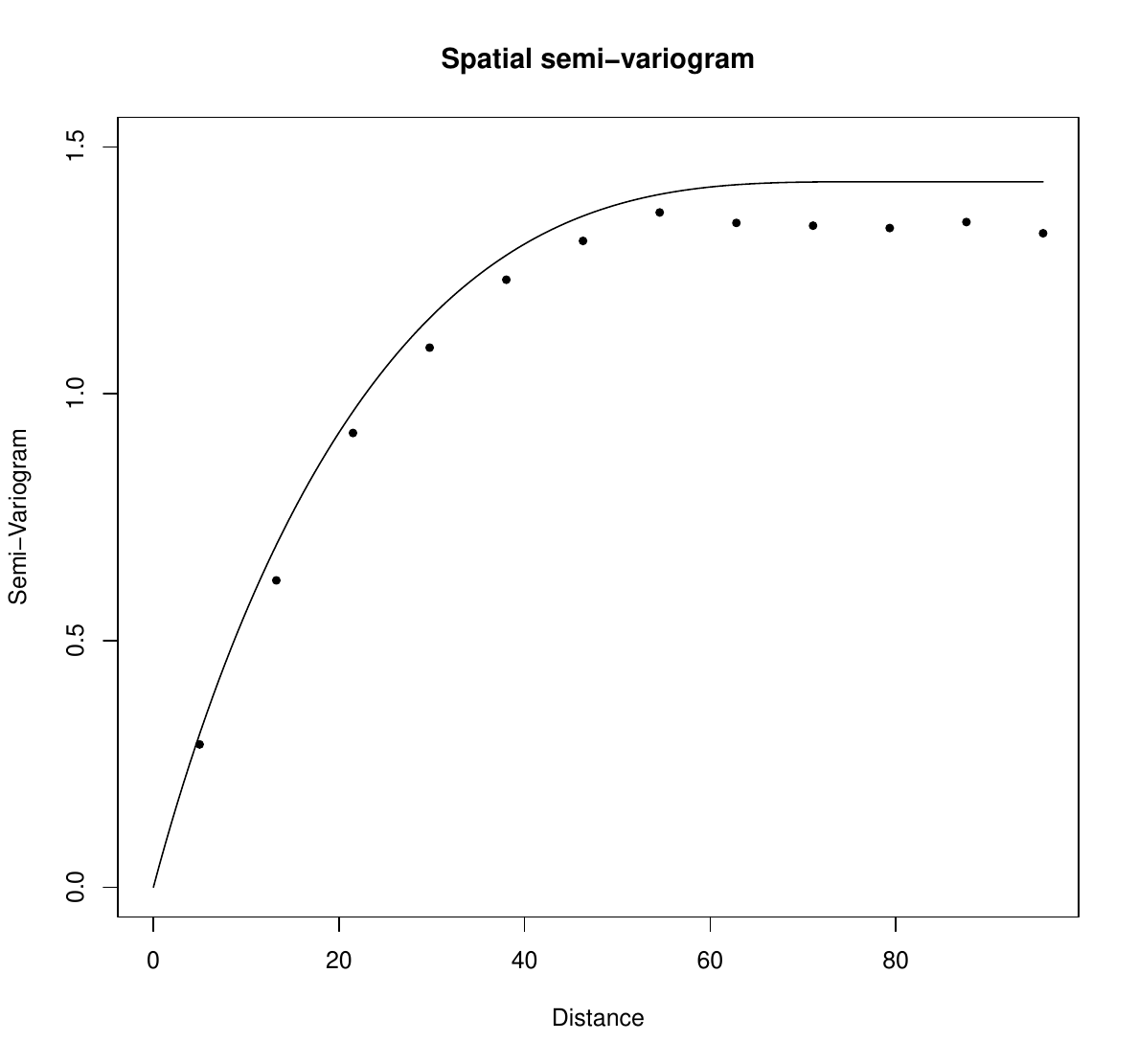}
\end{tabular}
\end{center}
\caption{
First row: normalized histogram versus fitted density and empirical
versus fitted semivariogram for the Gaussian RF with
$\sigma^2\mathcal H_{\kappa,\mu,a,2}$ covariance.
Second row: corresponding plots for the Tukey-$h$ RF.
}
\label{ordinary99}
\end{figure}

\section{Proof of Theorems}
\subsection{Proof of Theorem~\ref{Wv_vs_Wv}}

A preliminary Lemma is needed for the proof of Theorem \ref{Wv_vs_Wv}.
The following result can be  obtained from the asymptotic expansion of the 
generalized hypergeometric function ${}_1F_2$ for large argument, 
as given in \citet{Luke:1969}.

\begin{lemma}\label{the3}
Let $\widehat{{\cal GH}}_{\delta,\beta,\gamma,a}(z)$ be defined as in
\eqref{ppoon}. Assume $\delta>d/2$,
$2(\beta-\delta)(\gamma-\delta)\geq\delta$, and
$2(\beta+\gamma)\geq6\delta+1$. Set
\[
P_{\delta,\beta,\gamma,d}
:=
\frac{\Gamma(\delta)\Gamma(\beta-d/2)\Gamma(\gamma-d/2)}
{2^d\pi^{d/2}\Gamma(\delta-d/2)\Gamma(\beta)\Gamma(\gamma)}.
\]
Then, as $z\to\infty$,
\begin{align*}
\widehat{{\cal GH}}_{\delta,\beta,\gamma,a}(z)
&=
P_{\delta,\beta,\gamma,d}\,a^d
\Big[
 c_3(az)^{-2\delta}\{1+\mathcal O(z^{-2})\}\\
&\hspace{3.3cm}
+c_4(az)^{\frac12+\delta-\beta-\gamma}
 \{\cos(az+c_5)+\mathcal O(z^{-1})\}
\Big],
\end{align*}
where
\[
c_3=
\frac{2^{2\delta}\Gamma(\beta)\Gamma(\gamma)}
{\Gamma(\beta-\delta)\Gamma(\gamma-\delta)},
\qquad
c_4=
\frac{\pi^{-1/2}\Gamma(\beta)\Gamma(\gamma)}
{2^{\frac12+\delta-\beta-\gamma}\Gamma(\delta)},
\]
and
\[
c_5=\frac{\pi}{2}\left(\frac12+\delta-\beta-\gamma\right).
\]
In particular, if
$\beta+\gamma>3\delta+(d+1)/2$, then
\begin{equation}\label{eq:GH-tail-corrected}
\widehat{{\cal GH}}_{\delta,\beta,\gamma,a}(z)
=
K_{\delta,d}^{\star}L(\beta,\gamma)a^{d-2\delta}z^{-2\delta}
\{1+\mathcal O(z^{-2})\}
+\mathcal O\!\left(z^{\frac12+\delta-\beta-\gamma}\right),
\end{equation}
where
\[
K_{\delta,d}^{\star}
:=\frac{\Gamma(\delta)}{\pi^{d/2}\Gamma(\delta-d/2)},
\qquad
L(\beta,\gamma)
:=\frac{2^{2\delta-d}\Gamma(\beta-d/2)\Gamma(\gamma-d/2)}
{\Gamma(\beta-\delta)\Gamma(\gamma-\delta)}.
\]
Consequently,
$\widehat{{\cal GH}}_{\delta,\beta,\gamma,a}(z)\asymp z^{-2\delta}$.
\end{lemma}

\begin{proof}[Proof of Theorem~\ref{Wv_vs_Wv}]
Let
\[
\rho_i(\boldsymbol h)=\sigma_i^2\varphi_i(\|\boldsymbol h\|),
\qquad
\varphi_i(x)={\cal GH}_{\delta,\beta_i,\gamma_i,a_i}(x),
\qquad i=0,1,
\]
and let $f_i(z)=\sigma_i^2\widehat\varphi_i(z)$ be the corresponding spectral
densities. Define
\[
\lambda_i:=\frac12+\delta-\beta_i-\gamma_i
\]
and
\[
A_i:=
\sigma_i^2K_{\delta,d}^{\star}L(\beta_i,\gamma_i)
a_i^{d-2\delta}.
\]
The assumptions of the theorem imply the assumptions of Lemma~\ref{the3} and
$\lambda_i+2\delta<-d/2$. Hence
\begin{equation}\label{eq:tail-f-corrected}
f_i(z)
=A_i z^{-2\delta}\{1+\mathcal O(z^{-2})\}
+\mathcal O(z^{\lambda_i}),
\qquad
f_i(z)\sim A_i z^{-2\delta}.
\end{equation}
In particular, $f_0(z)z^{2\delta}$ is bounded away from zero and infinity for
all sufficiently large $z$.

For sufficiency, assume~\eqref{condition1_iff2}. Since
$K_{\delta,d}^{\star}$ is common to the two models, this condition is
equivalent to $A_0=A_1$. It follows from~\eqref{eq:tail-f-corrected} that
\[
\frac{f_1(z)-f_0(z)}{f_0(z)}
=\mathcal O(z^{-2})
 +\mathcal O(z^{\lambda_0+2\delta})
 +\mathcal O(z^{\lambda_1+2\delta}).
\]
Therefore
\[
\int_c^\infty z^{d-1}
\left(\frac{f_1(z)-f_0(z)}{f_0(z)}\right)^2\,dz<\infty
\]
for $d=1,2,3$: the contribution of the first term is bounded by a constant
times $\int_c^\infty z^{d-5}\,dz$, and the other two contributions are
integrable because $\lambda_i+2\delta<-d/2$. The standard sufficient spectral
criterion for stationary Gaussian measures on bounded domains
\citep{Ibragimov-Rozanov:1978,Stein:2004} then gives
$P(\rho_0)\equiv P(\rho_1)$.

For necessity, suppose~\eqref{condition1_iff2} fails and define
\[
\widetilde\sigma_1^2
:=
\sigma_0^2
\frac{L(\beta_0,\gamma_0)}{L(\beta_1,\gamma_1)}
\frac{a_1^{2\delta-d}}{a_0^{2\delta-d}}.
\]
Then
\[
\frac{\sigma_0^2L(\beta_0,\gamma_0)}{a_0^{2\delta-d}}
=
\frac{\widetilde\sigma_1^2L(\beta_1,\gamma_1)}{a_1^{2\delta-d}},
\]
so the sufficiency part yields
\[
P(\rho_0)
\equiv
P\!\left(\widetilde\sigma_1^2
 {\cal GH}_{\delta,\beta_1,\gamma_1,a_1}\right).
\]
Because~\eqref{condition1_iff2} fails,
$\widetilde\sigma_1^2\neq\sigma_1^2$. The latter Gaussian measure and
$P(\rho_1)$ have the same correlation operator and covariance operators that
differ by a non-unit scalar multiple. The covariance operator has infinite
rank (its spectral density is positive for all sufficiently large frequencies
by~\eqref{eq:tail-f-corrected}); hence the standard dichotomy for
infinite-dimensional Gaussian measures implies that these two measures are
mutually singular. If $P(\rho_0)$ and $P(\rho_1)$ were equivalent,
transitivity would contradict this singularity. Thus equivalence necessarily
implies~\eqref{condition1_iff2}.
\end{proof}

\subsection{Proof of Theorem \ref{theopp}}

\begin{proof}
Consider the model 
$\mathcal{GH}_{\delta,\delta+\frac{\mu}{2},\delta+\frac{\mu}{2}+l,a}$
in dimension $d$, where $\delta=\frac{d+1}{2}+\kappa$, 
$\kappa>-\frac12$.

The model $\mathcal{GH}_{\delta,\beta,\gamma,a}$ is invariant under
interchanging $\beta$ and $\gamma$. Under the present parameterization this
exchange is $(\mu,l)\mapsto(\mu+2l,-l)$, with $\delta$ and $a$ unchanged.
Thus imposing $l\ge0$, equivalently $\gamma\ge\beta$, removes the
label-switching ambiguity without losing any correlation function. We
henceforth assume $l\ge0$.

\noindent\textbf{Sufficiency.}
Under $\beta=\delta+\frac{\mu}{2}$, $\gamma=\delta+\frac{\mu}{2}+l$,
the sufficient conditions of Section~\ref{sec3} become:
\begin{enumerate}[(a)]
\item $\kappa>-\frac12$,
\item $\mu(\mu+2l)\ge d+2\kappa+1$,
\item $\mu+l\ge \frac{d}{2}+\kappa+1$.
\end{enumerate}
Condition~(b) is equivalent to the quadratic inequality
\[
\mu^2+2l\mu-(d+2\kappa+1)\ge 0,
\]
whose positive branch gives
\[
\mu\ge -l+\sqrt{l^2+d+2\kappa+1}.
\]
Only the positive branch is admissible since condition~(c) 
implies $\mu>-l$ when $l\ge0$.
For each fixed $l$, the sufficient conditions therefore reduce to
\[
\mu\ge
\max\!\Big\{\frac{d}{2}+\kappa+1-l,\;
-l+\sqrt{l^2+d+2\kappa+1}\Big\}.
\]
A direct algebraic comparison shows that
\[
\frac{d}{2}+\kappa+1-l
\ge
-l+\sqrt{l^2+d+2\kappa+1}
\quad\Longleftrightarrow\quad
0\le l\le\frac{d}{2}+\kappa.
\]
Hence condition~(c) is binding when $0\le l\le\frac{d}{2}+\kappa$,
giving $\mu\ge\delta-l+\frac12$,
and condition~(b) is binding when $l>\frac{d}{2}+\kappa$,
giving $\mu\ge\sqrt{2\kappa+l^2+d+1}-l$.
This yields exactly the two regimes in points~1 and~2.

\noindent\textbf{Necessity.}
Assume $\mathcal{GH}_{\delta,\beta,\gamma,a}\in\Phi_d$.
Since the Gamma prefactors in the spectral density are positive 
whenever $\beta>\delta$ and $\gamma>\delta$, nonnegativity 
of the spectral density requires
\[
{}_1F_2\!\left(\delta;\beta,\gamma;-\frac{(za)^2}{4}\right)
\ge 0
\quad\text{for all }z\ge0.
\]
By \citet[Theorem~3.1]{cho2018}, this implies 
$\beta>\delta$, $\gamma>\delta$, and 
$\beta+\gamma\ge3\delta+\frac12$.
The third condition, under the parameterization 
$\beta=\delta+\frac{\mu}{2}$, $\gamma=\delta+\frac{\mu}{2}+l$,
gives
\[
\mu+l\ge\frac{d}{2}+\kappa+1,
\quad\text{i.e.,}\quad
\mu\ge\delta-l+\frac12,
\]
which is the necessary lower bound for $\mu$ in both regimes.
In Regime~1 ($0\le l\le\frac{d}{2}+\kappa$), this coincides 
with the sufficient condition in point~1, so the result is 
an if and only if.
In Regime~2 ($l>\frac{d}{2}+\kappa$), this provides a 
necessary lower bound that is strictly weaker than the 
sufficient condition in point~2; whether the gap can be 
closed remains an open question.

This completes the proof.
\end{proof}

\subsection{Proof of Theorem \ref{theopkkp}}
\begin{proof}
Since $\phi(\|\cdot\|)$ is compactly supported it is absolutely integrable and,
by the spectral representation associated with \eqref{FT},
\[
A=(2\pi)^d\,\widehat{\phi}(0).
\]
For $\phi(x)=\mathcal{GH}_{\delta,\beta,\gamma,a}(x)$, the spectral density is
\eqref{ppoon}. Evaluating at $z=0$ and using ${}_1F_2(\cdot;\,0)=1$ gives
\[
\widehat{{\cal GH}}_{\delta,\beta,\gamma,a}(0)
=\frac{\Gamma(\delta)\Gamma(\beta-\frac{d}{2})\Gamma(\gamma-\frac{d}{2}) a^{d}}
{ 2^d\pi^{\frac{d}{2}} \Gamma\left(\delta-\frac{d}{2}\right)\Gamma(\beta)\Gamma(\gamma) }.
\]
Hence
\[
A=(2\pi)^d\widehat{{\cal GH}}_{\delta,\beta,\gamma,a}(0)
=\pi^{d/2}a^d\,
\frac{\Gamma(\delta)\Gamma(\beta-\frac{d}{2})\Gamma(\gamma-\frac{d}{2})}
{\Gamma\left(\delta-\frac{d}{2}\right)\Gamma(\beta)\Gamma(\gamma)}.
\]
Substituting $\delta=\frac{d+1}{2}+\kappa$, $\beta=\delta+\frac{\mu}{2}$,
$\gamma=\beta+l$ yields \eqref{integral range}.
\end{proof}

\subsection{Proof of Theorem \ref{theopkkp2}}
\begin{proof}
Since $A_{\kappa,\mu,l,a,d}=a^{d}\,A_{\kappa,\mu,l,1,d}$, the factor $a^{d}$ is
irrelevant for monotonicity and maximizers in $\mu$. Hence, we set $a=1$.

Fix $l\ge0$ and consider
\[
LA_{\kappa,d}(\mu,l):=\log\!\big(A_{\kappa,\mu,l,1,d}\big),
\]
where
\[
A_{\kappa,\mu,l,1,d}=  
\frac{\pi^{d/2}\Gamma\!\left(\frac{d+1}{2}+ \kappa\right)\,
\Gamma\!\left(\frac{\mu+1}{2}+\kappa\right)\,
\Gamma\!\left(\frac{\mu+1}{2}+\kappa+l\right)}
{\Gamma\!\left(\kappa+\frac12\right)\,
\Gamma\!\left(\frac{\mu+1}{2}+\kappa+\frac d2\right)\,
\Gamma\!\left(\frac{\mu+1}{2}+\kappa+l+\frac d2\right)}.
\]

Let $\psi=\Gamma'/\Gamma$ be the digamma function and set
\[
x=\frac{\mu+1}{2}+\kappa \qquad \Big(\text{so } \frac{dx}{d\mu}=\frac12\Big).
\]
On the validity region of the model (hence in particular on the admissible
domain in $\mu$), the Gamma factors above are well-defined and positive, so
$x>0$ and $x+l>0$. Differentiating with respect to $\mu$ gives
\[
\frac{d}{d\mu}LA_{\kappa,d}(\mu,l)
=\frac12\Big[\psi(x)-\psi\!\left(x+\tfrac{d}{2}\right)
+\psi(x+l)-\psi\!\left(x+l+\tfrac{d}{2}\right)\Big].
\]
Since $\psi$ is strictly increasing on $(0,\infty)$ and $d/2>0$, each difference
$\psi(x)-\psi(x+d/2)$ and $\psi(x+l)-\psi(x+l+d/2)$ is $\le 0$. Therefore
\[
\frac{d}{d\mu}LA_{\kappa,d}(\mu,l)\le 0
\quad\text{for all admissible }\mu,
\]
hence $\mu\mapsto A_{\kappa,\mu,l,a,d}$ is non-increasing on its admissible
domain. Consequently, its maximum is attained at the smallest admissible value
of $\mu$.

We now identify this smallest admissible value using Theorem~\ref{theopp}.
If $0\le l\le d/2+\kappa$, Theorem~\ref{theopp} provides the exact validity
condition $\mu\ge \mu_1(l):=\delta-l+\frac12$, hence $\mu^\star(l)=\mu_1(l)$.

If $l>d/2+\kappa$, Theorem~\ref{theopp} guarantees validity for all
$\mu\ge \mu_2(l):=\sqrt{2\kappa+l^2+d+1}-l$. In this regime, we take as
\emph{admissible domain} the parameter range ensured by Theorem~\ref{theopp}
(i.e., $\mu\ge\mu_2(l)$), and by monotonicity the maximum over this admissible
domain is attained at $\mu^\star(l)=\mu_2(l)$.
\end{proof}

\subsection{Proof of Theorem \ref{theopkkp333}}
\begin{proof}
Set $q=d/2$, $\alpha=\kappa+1/2>0$, and
$\tau=d/2+\kappa=q+\alpha-1/2$. As in the proof of
Theorem~\ref{theopkkp2}, the factor $a^d$ is irrelevant.

For $0\leq l\leq\tau$, put $\mu_1(l)=\tau+1-l$ and
\[
u(l)=\frac{\mu_1(l)+1}{2}+\kappa.
\]
Up to an additive constant independent of $l$,
\[
F_1(l)
=
\log\Gamma(u(l))+\log\Gamma(u(l)+l)
-\log\Gamma(u(l)+q)-\log\Gamma(u(l)+l+q).
\]
Writing $x=q/2+3\kappa/2+1$, one has
$u(l)=x-l/2$ and $u(l)+l=x+l/2$. Hence
\[
F_1'(l)
=
\frac12\left\{
\Delta_{l/2}(x)-\Delta_{l/2}(x+q)
\right\},
\qquad
\Delta_h(t):=\psi(t+h)-\psi(t-h).
\]
The digamma function is concave, so $\Delta_h$ is non-increasing in $t$.
Therefore $F_1'(l)\geq0$, proving the first assertion.

Consider now $l>\tau$. Put
\[
\mu=\mu_2(l)=\sqrt{l^2+2(q+\alpha)}-l\in(0,1).
\]
The identity $\mu(\mu+2l)=2(q+\alpha)$ gives
\[
l=\frac{q+\alpha}{\mu}-\frac{\mu}{2}.
\]
Define
\[
u=\alpha+\frac{\mu}{2},
\qquad
v=u+l=\alpha+\frac{q+\alpha}{\mu},
\]
and, up to a constant independent of $\mu$,
\[
F(\mu)=
\log\Gamma(u)+\log\Gamma(v)
-\log\Gamma(u+q)-\log\Gamma(v+q).
\]
Let
\[
g_q(t):=\psi(t+q)-\psi(t)>0.
\]
Differentiation gives
\begin{equation}\label{eq:Fmu-corrected}
F'(\mu)
=-\frac12 g_q(u)+\frac{q+\alpha}{\mu^2}g_q(v).
\end{equation}
Since $dl/d\mu=-(q+\alpha)/\mu^2-1/2<0$, it suffices to prove
$F'(\mu)>0$.

Suppose first that $d\geq2$, so $q\geq1$. The integral representation
\[
g_q(t)=\int_0^\infty e^{-tx}
\frac{1-e^{-qx}}{1-e^{-x}}\,dx
\]
shows that $t\mapsto t g_q(t)$ is non-decreasing: indeed,
$K_q(x)=(1-e^{-qx})/(1-e^{-x})$ is non-increasing for $q\geq1$, and after the
change of variable $y=tx$,
\[
tg_q(t)=\int_0^\infty e^{-y}K_q(y/t)\,dy.
\]
Since $v>u$, it follows that $g_q(v)\geq (u/v)g_q(u)$. Therefore
\[
F'(\mu)
\geq
g_q(u)\left[-\frac12+\frac{(q+\alpha)u}{\mu^2v}\right].
\]
Moreover,
\[
2(q+\alpha)u-\mu^2v
=\alpha\{2(q+\alpha)-\mu^2\}>0,
\]
so $F'(\mu)>0$.

Finally, let $d=1$ and $\kappa\geq0$, so $q=1/2$ and
$\alpha\geq1/2$. From
\[
g_{1/2}(t)=2\int_0^1\frac{x^{2t-1}}{1+x}\,dx
\]
and the inequalities
$1/2<1/(1+x)<1/(2\sqrt{x})$ for $0<x<1$, we obtain
\[
\frac{1}{2t}<g_{1/2}(t)<\frac{1}{2(t-1/4)},
\qquad t>1/4.
\]
Using these bounds in~\eqref{eq:Fmu-corrected}, with
$B=\alpha+1/2$, yields
\[
F'(\mu)
>
-\frac{1}{4(u-1/4)}+\frac{B}{2\mu^2v}.
\]
The right-hand side is positive because
\[
2B(u-1/4)-\mu^2v
=
\frac{(2\alpha-1)(4\alpha+1)+4\alpha(1-\mu^2)}{4}>0.
\]
Thus $F'(\mu)>0$ also in this case. Since $d\mu/dl<0$, the second-regime
integral range is non-increasing in $l$. Continuity and
$\mu_2(l)\to1$ as $l\downarrow\tau$ give the stated boundary limit and the
global conclusion under the stated conditions.
\end{proof}

\subsection{Proof of Theorem \ref{emery-antologico}}

\begin{proof}
Let
\[
\alpha_U=\kappa+1,
\qquad
\beta_U=\frac{\mu-1}{2},
\qquad
\alpha_V=\frac d2+2\kappa+1,
\qquad
\beta_V=\frac{\mu-d-1}{2}-\kappa+l.
\]
The assumptions $\mu>1$ and $\mu+2l>1+d+2\kappa$ ensure that all four
parameters are positive. The prefactor in~\eqref{emery-mitico} is exactly
\[
\{B(\alpha_U,\beta_U)B(\alpha_V,\beta_V)\}^{-1},
\]
so the right-hand side is
\[
\mathbb E\!\left[
\mathcal H_{\kappa,1,a\sqrt{UV},d}(x)
\right],
\]
where $U$ and $V$ are independent beta random variables with the parameters
above.

All functions involved are compactly supported and integrable, so it suffices
to compare their spectral densities. For fixed $z$, the power series below is
uniformly absolutely convergent for $(u,v)\in[0,1]^2$, and the beta weights are
integrable; hence termwise integration is justified. Write
$\delta=(d+1)/2+\kappa$ and let $P_H$ denote the spectral prefactor of
$\mathcal H_{\kappa,1,1,d}$. Then
\begin{align*}
&\mathbb E\!\left[
\widehat{\mathcal H}_{\kappa,1,a\sqrt{UV},d}(z)
\right]\\
&\quad=
P_Ha^d
\sum_{j=0}^\infty
\frac{(\delta)_j}{(\delta+\frac12)_j(2\delta)_j}
\frac{(-a^2z^2/4)^j}{j!}
\mathbb E(U^{d/2+j})\mathbb E(V^{d/2+j}).
\end{align*}
The beta moments are
\[
\mathbb E(U^{d/2+j})
=
M_U\frac{(\delta+\frac12)_j}{(\delta+\frac\mu2)_j},
\qquad
M_U=
\frac{\Gamma(\delta+\frac12)\Gamma(\delta+\frac\mu2-\frac d2)}
{\Gamma(\delta+\frac\mu2)\Gamma(\delta+\frac12-\frac d2)},
\]
and
\[
\mathbb E(V^{d/2+j})
=
M_V\frac{(2\delta)_j}{(\delta+\frac\mu2+l)_j},
\qquad
M_V=
\frac{\Gamma(2\delta)\Gamma(\delta+\frac\mu2+l-\frac d2)}
{\Gamma(\delta+\frac\mu2+l)\Gamma(2\delta-\frac d2)}.
\]
The factors $(\delta+1/2)_j$ and $(2\delta)_j$ cancel, giving
\[
P_HM_UM_Va^d
{}_1F_2\!\left(
\delta;\delta+\frac\mu2,\delta+\frac\mu2+l;
-\frac{a^2z^2}{4}
\right).
\]
A direct cancellation of Gamma factors shows that $P_HM_UM_V$ equals the
spectral prefactor of
$\mathcal{GH}_{\delta,\delta+\mu/2,\delta+\mu/2+l,a}$ in~\eqref{ppoon}.
Thus the spectral densities, and hence the covariance functions, coincide.
The specialization $l=d/2+\kappa$ gives the second identity.
\end{proof}

\subsection{Proof of Theorem \ref{cff}}


If  $d=1$ and $\kappa=k$ is a non-negative integer, using 
formula 7.3.1.70 of \cite{prud} and applying the identity $$P_a^b(x)=P^b_{-a-1}(x),$$ where $P_a^b(x)$ denotes
the associated Legendre function (of the first kind) with degree $a$ and order $b$
(see \citet[8.3.1]{Abramowitz-Stegun:1965}), one can rewrite (\ref{emery2}) as
\begin{eqnarray}\label{rep_emery2}
 \mathcal{H}_{k,\mu,a,1}(x)&=&   
 \frac{\Gamma(\frac{\mu}{2}+2k+1)}{\Gamma(k+\frac{1}{2})} \sqrt{\pi} \left( 1 - \frac{x^2}{a^2}\right)^{\frac{\mu}{2} + k}\left(\frac{x}{a}\right)^{k}   P^{-\frac{\mu}{2} - k}_k\left( \frac{a^2+x^2}{2ax}\right).\nonumber\\
\end{eqnarray}
The claim of the Theorem then derives from the following result, which is valid for $a$ being a non-negative integer and $z>1$:
\begin{equation}\label{klo}
P^{-b}_{a}(z) = \frac{(z - 1)^{b/2}}{(z + 1)^{b/2}}\sum_{j=0}^{a} \frac{(j + a)! (0.5(z - 1))^{j}}{ j! \Gamma(j + b+1)  (a - j)!}.
\end{equation}

\subsection{Proof of Theorem \ref{cff1}}
\begin{proof}
For $\mu \geq 1$ and $\kappa=0$, applying formula 7.3.1.41 from \cite{prud} to Equation (\ref{emery2}), one gets
\begin{equation*}
\mathcal{H}_{0,\mu,a,d}(x)= \frac{\Gamma(\frac{1+\mu}{2}) \Gamma(\frac{1+\mu+d}{2})}{\sqrt{\pi}} 2^{\mu+\frac{d-1}{2}} \left(1-\frac{x^2}{a^2}\right)_+^{\frac{\mu}{2}+\frac{d-1}{4}} \, P_{\frac{d-1}{2}}^{\frac{1-d}{2}-\mu}\left(\frac{x}{a}\right).
\end{equation*}
The claim of the Theorem stems from the fact that, for $m$ being an integer, the associated Legendre function of degree $m$ is a terminating series:
\begin{equation}
\label{legterm}
    P_{m}^{\eta}(z)=\left(\frac{1+z}{1-z}\right)^{\frac{\eta}{2}}\sum\limits_{n=0}^{m}\frac{(-m)_n(m+1)_n}{n!\Gamma(n-\eta+1)}\left(\frac{1-z}{2}\right)^{n},
\quad 0<z<1.
\end{equation}

\end{proof}

\subsection{Proof of Theorem \ref{ThmX}}

\begin{proof}
Set
\[
\delta=\frac{d+1}{2}+\kappa,
\qquad
\beta=\delta+\frac\mu2,
\qquad
\gamma=\delta+\frac\mu2+\frac d2+\kappa,
\qquad
\nu=\kappa+\frac12.
\]
Then $2\delta=2\nu+d$. By Lemma~\ref{the3}, the spectral density of the
$\mathcal H$ covariance satisfies
\begin{equation}\label{eq:H-tail-MT-proof}
\widehat{\mathcal H}_{\kappa,\mu,a,d}(z)
=
C_{\nu,d}F(\kappa,\mu,d)a^{-2\nu}z^{-2\nu-d}
\{1+\mathcal O(z^{-2})\}
+\mathcal O(z^{-d-2\kappa-\mu}),
\end{equation}
where
\[
C_{\nu,d}=\frac{\Gamma(\nu+d/2)}{\pi^{d/2}\Gamma(\nu)}.
\]
Indeed, in this parameterization
$L(\beta,\gamma)=F(\kappa,\mu,d)$ and
$K_{\delta,d}^{\star}=C_{\nu,d}$.
The Mat\'ern spectral density has the expansion
\begin{equation}\label{eq:MT-tail-proof}
\widehat{\mathcal{MT}}_{\nu,\alpha}(z)
=
C_{\nu,d}\alpha^{-2\nu}z^{-2\nu-d}
\{1+\mathcal O(z^{-2})\}.
\end{equation}

Assume~\eqref{cafu}. The leading terms in
$\sigma_1^2\widehat{\mathcal H}$ and
$\sigma_0^2\widehat{\mathcal{MT}}$ then coincide. Dividing the difference by
the Mat\'ern spectral density and using~\eqref{eq:H-tail-MT-proof}--
\eqref{eq:MT-tail-proof} gives
\[
\frac{\sigma_1^2\widehat{\mathcal H}_{\kappa,\mu,a,d}(z)
-\sigma_0^2\widehat{\mathcal{MT}}_{\nu,\alpha}(z)}
{\sigma_0^2\widehat{\mathcal{MT}}_{\nu,\alpha}(z)}
=
\mathcal O(z^{-2})+\mathcal O(z^{1-\mu}).
\]
Consequently,
\[
\int_c^\infty z^{d-1}
\left(
\frac{\sigma_1^2\widehat{\mathcal H}_{\kappa,\mu,a,d}(z)
-\sigma_0^2\widehat{\mathcal{MT}}_{\nu,\alpha}(z)}
{\sigma_0^2\widehat{\mathcal{MT}}_{\nu,\alpha}(z)}
\right)^2dz<\infty
\]
for $d=1,2,3$ and $\mu>1+d/2$. Notice that no oscillatory cancellation is
used: the second term is bounded absolutely by a constant times $z^{1-\mu}$.
The sufficient spectral equivalence criterion therefore proves equivalence.

For necessity, suppose~\eqref{cafu} fails and choose
$\widetilde\sigma_0^2$ so that
\[
\frac{\widetilde\sigma_0^2}{\alpha^{2\nu}}
=
\frac{\sigma_1^2F(\kappa,\mu,d)}{a^{2\kappa+1}}.
\]
The sufficiency part implies that
$P(\widetilde\sigma_0^2\mathcal{MT}_{\nu,\alpha})$ is equivalent to the
$\mathcal H$ measure. Since
$\widetilde\sigma_0^2\neq\sigma_0^2$, the two Mat\'ern measures with the same
correlation and different variances are mutually singular on the
infinite-dimensional path space. Transitivity therefore excludes equivalence
between the original Mat\'ern measure and the $\mathcal H$ measure. This proves
the necessity of~\eqref{cafu}.
\end{proof}

\subsection{Proof of Proposition~\ref{prop:bridge-support}}
{
\begin{proof}
Put $p=1+2\kappa>0$ and $q=\kappa+1/2=p/2$. From the definition of $b$,
\[
\left\{\frac{b(\alpha,\mu,\kappa,d)}{\alpha}\right\}^{p}
=F(\kappa,\mu,d).
\]
Using $\psi=\Gamma'/\Gamma$, differentiation gives
\begin{align*}
\frac{\partial}{\partial\mu}\log F(\kappa,\mu,d)
=\frac12\Bigg[&\psi\!\left(\frac\mu2+q\right)
               -\psi\!\left(\frac\mu2\right)\\
&+\psi\!\left(\frac{\mu+d}{2}+\kappa+q\right)
               -\psi\!\left(\frac{\mu+d}{2}+\kappa\right)\Bigg]>0,
\end{align*}
because $q>0$ and the digamma function is strictly increasing. Hence
$b(\alpha,\mu,\kappa,d)$ is strictly increasing in $\mu$. Similarly,
\[
\frac{\partial}{\partial\mu}\log g(\alpha,\mu,\kappa)
=\frac1p\{\psi(\mu+p)-\psi(\mu)\}>0,
\]
so the ${\cal GW}$ bridge support is also strictly increasing in $\mu$.

The validity boundary is $\mu=1$ for $\mathcal H$ and, when $d\ge2$ or
$d=1$ with $\kappa\ge0$, it is
$\mu_{\rm GW}=\kappa+(d+1)/2$ for ${\cal GW}$. Therefore
\begin{equation}\label{eq:bridge-ratio-proof}
\left[
\frac{b(\alpha,1,\kappa,d)}
     {g(\alpha,\mu_{\rm GW},\kappa)}
\right]^p
=
\frac{2^p\Gamma(\kappa+1)\Gamma(d/2+2\kappa+1)}
     {\sqrt\pi\,\Gamma(d/2+3\kappa+3/2)}.
\end{equation}
The duplication formula yields
\[
\frac{2^p\Gamma(\kappa+1)}{\sqrt\pi}
=
\frac{\Gamma(2\kappa+2)}{\Gamma(\kappa+3/2)}.
\]
Let $x=d/2+2\kappa+1$ and $y=\kappa+3/2$. Since
$q=\kappa+1/2>0$, the function
$t\mapsto\Gamma(t+q)/\Gamma(t)$ is strictly increasing. Moreover,
\[
x-y=\frac d2+\kappa-\frac12\ge0
\]
under the stated dimensional--smoothness conditions. Consequently,
\[
\frac{\Gamma(y+q)/\Gamma(y)}
     {\Gamma(x+q)/\Gamma(x)}\le1,
\]
which is exactly the right-hand side of~\eqref{eq:bridge-ratio-proof}. This
proves~\eqref{eq:bridge-support-order}. Equality requires $x=y$, which in the
stated parameter region occurs only for $d=1$ and $\kappa=0$.

Finally, the standard Gamma-ratio expansion
$\Gamma(z+a)/\Gamma(z+b)\sim z^{a-b}$ gives
\begin{align*}
 b(\alpha,1,\kappa,d)^p
&=
\alpha^p\frac{2^p\Gamma(\kappa+1)}{\sqrt\pi}
\frac{\Gamma(d/2+2\kappa+1)}
     {\Gamma(d/2+\kappa+1/2)}\\
&\sim
\alpha^p\frac{2^p\Gamma(\kappa+1)}{\sqrt\pi}
\left(\frac d2\right)^{p/2},
\end{align*}
which proves~\eqref{eq:H-bridge-asymptotic}. Likewise,
\[
 g(\alpha,\mu_{\rm GW},\kappa)^p
=
\alpha^p
\frac{\Gamma(d/2+3\kappa+3/2)}
     {\Gamma(d/2+\kappa+1/2)}
\sim
\alpha^p\left(\frac d2\right)^p,
\]
proving~\eqref{eq:GW-bridge-asymptotic} and the order $d^{-1/2}$ for the
support ratio.
\end{proof}
}

\subsection{Proofs of Theorems~\ref{theo10}--\ref{theo12_new}}

We use a direct spectral sandwich for the profiled quadratic forms.  Put
\[
p:=2\kappa+1=2\delta-d,
\qquad
G_{\mu,a}(x):=
\frac{a^p}{F(\kappa,\mu,d)}\,
\mathcal H_{\kappa,\mu,a,d}(x),
\]
and let $g_{\mu,a}$ denote the isotropic spectral density of
$G_{\mu,a}$.  Thus
\[
\sigma^2\mathcal H_{\kappa,\mu,a,d}
=\theta G_{\mu,a},
\qquad
\theta=\frac{\sigma^2F(\kappa,\mu,d)}{a^p}.
\]
For $\eta=(\mu,a)$, write $G_n(\eta)$ for the covariance matrix obtained
by evaluating $G_{\mu,a}$ at the observation locations.  Then the profiled estimator can be written as
\begin{equation}\label{eq:theta-normalized-qf}
\widehat\theta_n(\eta)
=\frac1n\mathbf Z_n^\top G_n(\eta)^{-1}\mathbf Z_n.
\end{equation}

\begin{lemma}\label{lem:uniform-spectral-sandwich}
Let $d\in\{1,2,3\}$, let $I=[a_L,a_U]\subset(0,\infty)$, and let
$J=[\mu_L,\mu_U]$ with $\mu_L>1+d/2$.  Set
$\Theta=J\times I$, fix $\eta_0=(\mu_0,a_0)\in\Theta$, and suppose that
\[
\operatorname{Cov}_0(\mathbf Z_n)=\theta_0G_n(\eta_0).
\]
Then
\begin{equation}\label{eq:uniform-sandwich-rootn}
\sqrt n\,
\sup_{\eta\in\Theta}
\left|\widehat\theta_n(\eta)-\widehat\theta_n(\eta_0)\right|
\xrightarrow{p}0.
\end{equation}
The same conclusion holds when $J$ is reduced to a singleton
$\{\mu\}$ with $\mu>1+d/2$.
\end{lemma}

\begin{proof}
We first obtain a uniform high-frequency comparison.  The scale-mixture
representation in Theorem~\ref{emery-antologico} implies that
$g_{\mu,a}(\boldsymbol\omega)>0$ for every $\boldsymbol\omega$ whenever
$\mu>1$.  Since $g_{\mu,a}$ is continuous in
$(\mu,a,\boldsymbol\omega)$, compactness of $\Theta$ gives, for every
fixed $R<\infty$,
\begin{equation}\label{eq:sandwich-compact-positive}
0<c_R\le g_{\mu,a}(\boldsymbol\omega)\le C_R<\infty,
\qquad
\eta\in\Theta,\quad \|\boldsymbol\omega\|\le R.
\end{equation}

Let $\rho=\|\boldsymbol\omega\|$.  In the $\mathcal H$ parameterization,
Lemma~\ref{the3} and $F(\kappa,\mu,d)=L(\beta,\gamma)$ give
\[
g_{\mu,a}(\rho)
=K_{\delta,d}^{\star}\rho^{-2\delta}
\left\{1+\mathcal O(\rho^{-2})
      +\mathcal O(\rho^{1-\mu})\right\},
\qquad \rho\to\infty.
\]
The remainder bounds can be chosen uniformly for $\eta=(\mu,a)\in\Theta$.
Indeed, $a$ is bounded away from zero and infinity, while the lower
hypergeometric parameters and all Gamma-function arguments appearing in the
coefficients range over compact subsets of $(0,\infty)$.  The finite-order
remainder bounds in the standard large-argument expansion of ${}_1F_2$
used in Lemma~\ref{the3} can therefore be chosen uniformly on $\Theta$;
see \citet{Luke:1969}.
The leading term is therefore the same for every nuisance value.  Hence,
with
\[
\xi:=\min\{2,\mu_L-1\}>\frac d2,
\]
there exists $C<\infty$ such that
\begin{equation}\label{eq:spectral-envelope-decay}
w(\boldsymbol\omega)
:=\sup_{\eta\in\Theta}
\left|
\frac{g_{\mu,a}(\boldsymbol\omega)}
     {g_{\mu_0,a_0}(\boldsymbol\omega)}-1
\right|
\le C(1+\rho)^{-\xi}.
\end{equation}
In particular $w\in L^2(\mathbb R^d)$.  Combining
\eqref{eq:sandwich-compact-positive} and
\eqref{eq:spectral-envelope-decay} also yields constants
$0<m<M<\infty$ such that
\begin{equation}\label{eq:sandwich-uniform-comparison}
m\le
\frac{g_{\mu,a}(\boldsymbol\omega)}
     {g_{\mu_0,a_0}(\boldsymbol\omega)}
\le M,
\qquad
\boldsymbol\omega\in\mathbb R^d,\quad \eta\in\Theta.
\end{equation}

Define the lower and upper spectral envelopes
\[
g_-(\boldsymbol\omega):=\inf_{\eta\in\Theta}g_{\mu,a}(\boldsymbol\omega),
\qquad
g_+(\boldsymbol\omega):=\sup_{\eta\in\Theta}g_{\mu,a}(\boldsymbol\omega).
\]
By~\eqref{eq:sandwich-uniform-comparison}, they are integrable and satisfy
$mg_{\mu_0,a_0}\le g_-\le g_+\le Mg_{\mu_0,a_0}$; hence they define
positive definite comparison covariances.  Let $G_n^-$ and $G_n^+$ be their
covariance matrices at the observation locations.  The spectral
representation gives the Loewner ordering
\begin{equation}\label{eq:matrix-spectral-sandwich}
G_n^-\preceq G_n(\eta)\preceq G_n^+,
\qquad \eta\in\Theta,
\end{equation}
so that
\[
(G_n^+)^{-1}\preceq G_n(\eta)^{-1}\preceq(G_n^-)^{-1}.
\]
Consequently, with
\[
\widehat\theta_{n,+}
:=\frac1n\mathbf Z_n^\top(G_n^+)^{-1}\mathbf Z_n,
\qquad
\widehat\theta_{n,-}
:=\frac1n\mathbf Z_n^\top(G_n^-)^{-1}\mathbf Z_n,
\]
we have the finite-sample sandwich
\begin{equation}\label{eq:theta-finite-sandwich}
\widehat\theta_{n,+}
\le \widehat\theta_n(\eta)
\le \widehat\theta_{n,-},
\qquad \eta\in\Theta.
\end{equation}
Since $\eta_0\in\Theta$,
\begin{equation}\label{eq:uniform-width-bound}
\sup_{\eta\in\Theta}
\left|\widehat\theta_n(\eta)-\widehat\theta_n(\eta_0)\right|
\le \widehat\theta_{n,-}-\widehat\theta_{n,+}.
\end{equation}

It remains to show that the width of this sandwich is negligible at the
root-$n$ scale.  Put $G_{0n}=G_n(\eta_0)$ and
\[
B_n^-:=G_{0n}^{-1/2}G_n^-G_{0n}^{-1/2},
\qquad
B_n^+:=G_{0n}^{-1/2}G_n^+G_{0n}^{-1/2}.
\]
Then~\eqref{eq:sandwich-uniform-comparison} implies
\[
mI_n\preceq B_n^-\preceq I_n\preceq B_n^+\preceq MI_n.
\]
For $\boldsymbol\omega\in\mathbb R^d$, let
\[
\mathbf e_n(\boldsymbol\omega)
:=\left(
 e^{i\boldsymbol\omega^\top\mathbf s_1},\ldots,
 e^{i\boldsymbol\omega^\top\mathbf s_n}
\right)^\top
\]
and define the finite-sample spectral leverage
\begin{equation}\label{eq:spectral-leverage}
\Lambda_n(\boldsymbol\omega)
:=g_{\mu_0,a_0}(\boldsymbol\omega)\,
\mathbf e_n(\boldsymbol\omega)^*
G_{0n}^{-1}\mathbf e_n(\boldsymbol\omega).
\end{equation}
The spectral representation gives
\begin{equation}\label{eq:leverage-mass}
\int_{\mathbb R^d}\Lambda_n(\boldsymbol\omega)\,d\boldsymbol\omega=n.
\end{equation}
We next record a bound that is uniform in both $n$ and frequency.  Choose
$\chi\in C_c^\infty(\mathbb R^d)$ such that $\chi=1$ on a neighborhood of
$D$.  For an exponential polynomial
\[
p(\boldsymbol\lambda)=\sum_{j=1}^n c_j
 e^{i\boldsymbol\lambda^\top\mathbf s_j},
\]
Fourier inversion applied to $\chi$ gives, up to the fixed Fourier-normalization
constant,
\[
p(\boldsymbol\omega)
=\int_{\mathbb R^d}\widehat\chi(\boldsymbol\omega-\boldsymbol\lambda)
 p(\boldsymbol\lambda)\,d\boldsymbol\lambda.
\]
By~\eqref{eq:sandwich-compact-positive} and the common spectral tail,
$g_{\mu_0,a_0}(\boldsymbol\omega)\asymp
(1+\|\boldsymbol\omega\|)^{-2\delta}$ on $\mathbb R^d$.  Hence
\[
\frac{g_{\mu_0,a_0}(\boldsymbol\omega)}
     {g_{\mu_0,a_0}(\boldsymbol\lambda)}
\le C(1+\|\boldsymbol\omega-\boldsymbol\lambda\|)^{2\delta}.
\]
Cauchy--Schwarz and the rapid decay of $\widehat\chi$ therefore yield
\[
g_{\mu_0,a_0}(\boldsymbol\omega)|p(\boldsymbol\omega)|^2
\le C_D\int_{\mathbb R^d}|p(\boldsymbol\lambda)|^2
 g_{\mu_0,a_0}(\boldsymbol\lambda)\,d\boldsymbol\lambda,
\]
where $C_D$ is independent of $n$, $\boldsymbol\omega$, and the coefficients
$c_j$.  Taking the supremum over the coefficient vector gives
\begin{equation}\label{eq:leverage-bound}
0\le \Lambda_n(\boldsymbol\omega)\le C_D,
\qquad \boldsymbol\omega\in\mathbb R^d.
\end{equation}
This is the finite-dimensional spectral-projection estimate underlying the
Gaussian-measure equivalence criterion; see also \citet{Stein:2004}.

Using~\eqref{eq:spectral-leverage},
\eqref{eq:spectral-envelope-decay}, and the definition of $g_\pm$,
\begin{align}
\operatorname{tr}(B_n^+-I_n)
&\le \int_{\mathbb R^d}w(\boldsymbol\omega)
\Lambda_n(\boldsymbol\omega)\,d\boldsymbol\omega,\label{eq:trace-plus}\\
\operatorname{tr}(I_n-B_n^-)
&\le \int_{\mathbb R^d}w(\boldsymbol\omega)
\Lambda_n(\boldsymbol\omega)\,d\boldsymbol\omega.\label{eq:trace-minus}
\end{align}
For any fixed $R>0$, \eqref{eq:leverage-mass}--\eqref{eq:leverage-bound}
and Cauchy--Schwarz give
\begin{align*}
\frac1{\sqrt n}
\int_{\mathbb R^d}w\Lambda_n
&\le
\frac{C_D}{\sqrt n}
\int_{\|\boldsymbol\omega\|\le R}w(\boldsymbol\omega)\,d\boldsymbol\omega\\
&\quad+
\sqrt{C_D}
\left\{
\int_{\|\boldsymbol\omega\|>R}w(\boldsymbol\omega)^2
\,d\boldsymbol\omega
\right\}^{1/2}.
\end{align*}
First letting $n\to\infty$ and then $R\to\infty$ yields
\begin{equation}\label{eq:trace-envelope-small}
\frac{\operatorname{tr}(B_n^+-I_n)}{\sqrt n}
\longrightarrow0,
\qquad
\frac{\operatorname{tr}(I_n-B_n^-)}{\sqrt n}
\longrightarrow0.
\end{equation}

Now set
\[
\mathbf Y_n
:=\theta_0^{-1/2}G_{0n}^{-1/2}\mathbf Z_n
\sim\mathcal N(0,I_n).
\]
Then
\[
\widehat\theta_{n,-}-\widehat\theta_{n,+}
=\frac{\theta_0}{n}\mathbf Y_n^\top A_n\mathbf Y_n,
\qquad
A_n:=(B_n^-)^{-1}-(B_n^+)^{-1}\succeq0.
\]
Since $B_n^-\succeq mI_n$ and $B_n^+\succeq I_n$,
\[
\operatorname{tr}(A_n)
\le m^{-1}\operatorname{tr}(I_n-B_n^-)
   +\operatorname{tr}(B_n^+-I_n)
=o(\sqrt n)
\]
by~\eqref{eq:trace-envelope-small}.  Therefore
\[
\mathbb E_0\!\left[
\sqrt n\{\widehat\theta_{n,-}-\widehat\theta_{n,+}\}
\right]
=\frac{\theta_0}{\sqrt n}\operatorname{tr}(A_n)
\longrightarrow0.
\]
The width is nonnegative, so Markov's inequality together with
\eqref{eq:uniform-width-bound} proves~\eqref{eq:uniform-sandwich-rootn}.
\end{proof}

\begin{proof}[Proof of Theorem~\ref{theo10}]
Fix $\mu>1+d/2$ and a fixed $a>0$, and put
$\eta_0=(\mu,a_0)$ and $\eta=(\mu,a)$.  Consider the Gaussian measure
with covariance $\theta_0G_{\mu,a}$.  It has the same microergodic
parameter $\theta_0$ as the true covariance, so Theorem~\ref{Wv_vs_Wv}
applied to the $\mathcal H$ subfamily implies that the two Gaussian
measures are equivalent on the bounded domain $D$.

Under the measure with covariance $\theta_0G_{\mu,a}$,
\[
\frac{n\widehat\theta_n(\mu,a)}{\theta_0}\sim\chi_n^2.
\]
The standard chi-square exponential tail bound and Borel--Cantelli therefore
give $\widehat\theta_n(\mu,a)\to\theta_0$ almost surely under that measure,
and equivalence transfers the same almost-sure statement to the true
measure.  This proves part~(i).

For part~(ii), choose a compact interval $I_\star\subset(0,\infty)$
containing $a_0$ and $a$, and apply
Lemma~\ref{lem:uniform-spectral-sandwich} with
$\Theta=\{\mu\}\times I_\star$.  Then
\[
\sqrt n\{
\widehat\theta_n(\mu,a)-\widehat\theta_n(\mu,a_0)
\}=o_p(1).
\]
At the true nuisance parameters,
\[
\frac{n\widehat\theta_n(\mu,a_0)}{\theta_0}\sim\chi_n^2,
\]
and hence
\[
\sqrt n\{
\widehat\theta_n(\mu,a_0)-\theta_0
\}
\xrightarrow{\mathcal D}\mathcal N(0,2\theta_0^2).
\]
Slutsky's theorem proves part~(ii).
\end{proof}

\begin{proof}[Proof of Theorem~\ref{theo11_new}]
Apply Lemma~\ref{lem:uniform-spectral-sandwich} with
$\Theta=\{\mu\}\times I$.  Since $\widehat a_n\in I$,
\[
\sqrt n\{
\widehat\theta_n(\mu,\widehat a_n)
-\widehat\theta_n(\mu,a_0)
\}=o_p(1).
\]
The same relation without the factor $\sqrt n$ gives consistency in
probability.  At $(\mu,a_0)$ the normalized estimator is exactly
$\chi_n^2/n$, so the central limit theorem and Slutsky's theorem yield
\[
\sqrt n\{
\widehat\theta_n(\mu,\widehat a_n)-\theta_0
\}
\xrightarrow{\mathcal D}\mathcal N(0,2\theta_0^2).
\]
\end{proof}

\begin{proof}[Proof of Theorem~\ref{theo12_new}]
Apply Lemma~\ref{lem:uniform-spectral-sandwich} with
$\Theta=J\times I$ and $\eta_0=(\mu_0,a_0)$.  Since
$(\widehat\mu_n,\widehat a_n)\in J\times I$,
\[
\sqrt n\{
\widehat\theta_n(\widehat\mu_n,\widehat a_n)
-\widehat\theta_n(\mu_0,a_0)
\}=o_p(1).
\]
Again, this implies consistency in probability after dropping the factor
$\sqrt n$.  Finally,
\[
\frac{n\widehat\theta_n(\mu_0,a_0)}{\theta_0}\sim\chi_n^2,
\]
so
\[
\sqrt n\{
\widehat\theta_n(\widehat\mu_n,\widehat a_n)-\theta_0
\}
\xrightarrow{\mathcal D}\mathcal N(0,2\theta_0^2)
\]
by Slutsky's theorem.
\end{proof}
\section{Closed-form expressions of the hypergeometric model}
\label{closedforms}

\textbf{In Euclidean spaces of odd dimension.}
When $d$ is an odd integer, $\kappa=k$ is a non-negative integer and $\mu$ is a positive integer, the hypergeometric model yields a truncated polynomial \citep[Section 2.5.3]{emery2022gauss}. Specifically, if $\mu$ is odd, then
\begin{equation*}
\begin{split}
    \mathcal{H}_{k,\mu,a,d}(x) = \begin{cases}
    \frac{\Gamma(\frac{1}{2}-k)\Gamma(\frac{\mu+1}{2}+k)}{\Gamma(\frac{1-\mu-d}{2}-2k)} \bigg[\sum_{n=0}^{k+\frac{\mu-1}{2}} \frac{(-1)^n \Gamma(\frac{1-\mu-d}{2}-2k+n)}{\Gamma(\frac{1}{2}-k+n) \Gamma(\frac{\mu+1}{2}+k-n) n!} \left(\frac{x}{a}\right)^{2n}\\
    + \sum_{n=0}^{k-1+\frac{\mu+d}{2}} \frac{(-1)^n \Gamma(n+1-\frac{\mu}{2})}{\Gamma(\frac{3}{2}+k+n) \Gamma(\frac{\mu+d}{2}+k-n) n!} \left(\frac{x}{a}\right)^{2n+2k+1}\bigg] \text{ if $x < a$}\\
    0 \text{ if $x \geq a$,}
    \end{cases}
\end{split}
\end{equation*}
while, if $\mu$ is even,
\begin{equation*}
\begin{split}
    \mathcal{H}_{k,\mu,a,d}(x) = \begin{cases}
    \frac{\Gamma(\frac{1}{2}-k)\Gamma(\frac{\mu+d+1}{2}+2k)}{\Gamma(\frac{1-\mu}{2}-k)} \bigg[\sum_{n=0}^{2k+\frac{\mu+d-1}{2}} \frac{(-1)^n \Gamma(\frac{1-\mu}{2}-k+n)}{\Gamma(\frac{1}{2}-k+n) \Gamma(\frac{\mu+d+1}{2}+2k-n) n!} \left(\frac{x}{a}\right)^{2n}\\
    + \sum_{n=0}^{\frac{\mu}{2}-1} \frac{(-1)^n \Gamma(n+1-\frac{\mu+d}{2}-k)}{\Gamma(\frac{3}{2}+k+n) \Gamma(\frac{\mu}{2}-n) n!} \left(\frac{x}{a}\right)^{2n+2k+1}\bigg] \text{ if $x < a$}\\
    0 \text{ if $x \geq a$.}
    \end{cases}
\end{split}
\end{equation*}

The cases $k=0$, $\mu=1$ and $d=1$, $3$ and $5$ correspond to the well-known triangular, spherical and pentaspherical models \citep{Matern}, which behave linearly near the origin: 
\begin{equation*}  
 \mathcal{H}_{0,1,a,1}(x)=1-\frac{x}{a},  \quad x  \leq a,
\end{equation*}
\begin{equation*}  
 \mathcal{H}_{0,1,a,3}(x)=1-\frac{3x}{2a}+ \frac{x^3}{2a^3},  \quad x  \leq a,
\end{equation*}
\begin{equation*}  
 \mathcal{H}_{0,1,a,5}(x)=1-\frac{15x}{8a} + \frac{5x^3}{4a^3} - \frac{3x^5}{8a^5},  \quad x  \leq a,
\end{equation*}
whereas the cases $d=3$, $\mu=1$ and $k=1$ or $2$ correspond to the so-called cubic and penta models \citep{Chiles:Delfiner:2012}, which are associated with Gaussian RFs that are once and twice mean-squared differentiable, respectively: 
\begin{equation*}  
 \mathcal{H}_{1,1,a,3}(x)=1-\frac{7x^2}{a^2} + \frac{35x^3}{4a^3} - \frac{7x^5}{2a^5} + \frac{3x^7}{4a^7},  \quad x  \leq a,
\end{equation*}
\begin{equation*}  
 \mathcal{H}_{2,1,a,3}(x)=1-\frac{22x^2}{3a^2} + \frac{33x^4}{a^4} - \frac{77x^5}{2a^5} + \frac{33x^7}{2a^7} - \frac{11x^9}{2a^9} + \frac{5x^{11}}{6a^{11}},  \quad x  \leq a.
\end{equation*}\\

\noindent \textbf{In the 2D plane.} 
If $d=2$ and $\mu=1$, the following closed-form expressions are found.
\begin{enumerate}
    \item When $\kappa = 0$, formula 7.3.2.92 of \cite{prud} gives:
  \begin{equation}   \label{popoi}
  \mathcal{H}_{0,1,a,2}(x)=  \frac{2}{ \pi} \left( \arcsin   \sqrt{\left(1-\frac{x^2}{a^2}\right)_+}  -\frac{x}{a} \left(1-\frac{x^2}{a^2}\right)_+^{\frac{1}{2}} \right),
 \end{equation}
which corresponds to the circular correlation model \citep[2.46]{Chiles:Delfiner:2012}, which behaves linearly near the origin.

\item When $\kappa=\frac{1}{2}$, formula 7.3.2.99 of \cite{prud} gives, with
$r=x/a$,
\[
\mathcal{H}_{\frac12,1,a,2}(x)=
\begin{cases}
\displaystyle
\frac12\left[
(2+r^2)\sqrt{1-r^2}
-r^2(4-r^2)\arctanh\sqrt{1-r^2}
\right], & 0\leq x<a,\\[0.8em]
0, & x\geq a.
\end{cases}
\]
This is the mont\'ee (upgrading) \citep[I.4.18]{matheron65} of order $1$ of
the spherical correlation model in $\mathbb{R}^3$.

\item When $\kappa=1$, one obtains  \citep{weisstein}:
\begin{eqnarray}  \label{popoi2}
 \mathcal{H}_{1,1,a,2}(x)&=& \frac{2}{3 \pi} \Bigg\{\frac{x}{a}  \left(1-\frac{x^2}{a^2} \right)_+^{\frac{1}{2 }}  \left[15-4\left(1-\frac{x^2}{a^2}\right)_+\left(3-\frac{x^2}{a^2}\right)\right] \nonumber \\
&& +
 3\left[6\left(1-\frac{x^2}{a^2}\right)_+ -5\right] \arcsin   \sqrt{\left(1-\frac{x^2}{a^2} \right)_+}\Bigg\},
 \end{eqnarray}
which is the mont\'ee of order $2$ of the Euclid's hat model in $\mathbb{R}^4$ and is associated with a Gaussian RF that is mean-squared differentiable.


\end{enumerate}

\end{document}